%% file: main.tex
\documentclass[sigconf, nonacm]{acmart}
\usepackage{subfig}
\usepackage[T1]{fontenc}
\usepackage{color}
\usepackage{booktabs}
\usepackage{amstext}
\usepackage{relsize}
\usepackage{lipsum}
\usepackage{array}
\usepackage{subfig}
\usepackage{multirow}
\usepackage{float}
\usepackage{bm}
\usepackage{bbding}
\usepackage{hyperref}
\usepackage{float}
\usepackage{amsmath,amsthm}
\usepackage{enumitem}
\usepackage{algorithm}
\usepackage{algorithmic}
\usepackage{tabularx}
\usepackage[skip=4pt]{caption}

\newtheorem{myDef}{Definition}

\newtheorem{myEx}{Example}

\newcommand\vldbdoi{10.14778/3705829.3705858}
\newcommand\vldbpages{453 - 465}
\newcommand\vldbvolume{18}
\newcommand\vldbissue{2}
\newcommand\vldbyear{2024}
\newcommand\vldbtitle{\shorttitle} 
\newcommand\vldbavailabilityurl{https://github.com/yumengs-exp/MLSimp}
\newcommand\vldbpagestyle{empty} 

\begin{document}
\title{Quantifying Point Contributions: A Lightweight Framework for Efficient and Effective Query-Driven Trajectory Simplification}

%%
%% The "author" command and its associated commands are used to define the authors and their affiliations.

% \author{
% Yumeng Song$^{1}$,
% Yu Gu$^{1}$, 
% Tianyi Li$^{2}$,
% Yushuai Li$^{2}$,  
% Christian S. Jensen$^{2}$,
% Ge Yu$^{1}$}
% \affiliation{%
%   \institution{
%   {\large$^{1 }$}School of Computer Science and Engineering, Northeastern University, Shenyang, China \\
%   {\large$^{2}$} Department of Computer Science, Aalborg University, Aalborg, Denmark   \\
% {\large$^{1 }$}songym@stumail.neu.edu.cn, \{guyu, yuge\}@mail.neu.edu.cn \\ {\large$^{2}$}\{tianyi, yusli, csj\}@cs.aau.dk
% \\ %{\large$^{3 }$}\{wanganni.wan,aiao.cj,lidimeng.ldm,lifeifei\}@alibaba-inc.com
%   }
% }

% \begin{comment}
\author{Yumeng Song}
\affiliation{%
  \institution{Northeastern Univ., China}
}
\email{songym@stumail.neu.edu.cn}

\author{Yu Gu}
\authornote{Corresponding authors}
\affiliation{%
  \institution{Northeastern Univ., China}
}
\email{guyu@mail.neu.edu.cn}

\author{Tianyi Li}
\authornotemark[1]
\affiliation{%
  \institution{Aalborg Univ., Denmark}
}
\email{tianyi@cs.aau.dk}

\author{Yushuai Li}
\affiliation{%
  \institution{Aalborg Univ., Denmark}
}
\email{yusli@cs.aau.dk}

\author{Christian S. Jensen}
\affiliation{%
  \institution{Aalborg Univ., Denmark}
}
\email{csj@cs.aau.dk}

\author{Ge Yu}
\affiliation{%
    \institution{Northeastern Univ., China}
}
\email{yuge@mail.neu.edu.cn}
% \end{comment}
% \setlength{\marginparwidth}{1.1cm}
\begin{abstract}
%With geo-positioning capabilities such as GPS being readily available, increasingly large volumes of trajectory data are being accumulated to serve important applications. We aim to reduce the associated storage and querying costs by simplifying trajectories while retaining valuable information.

As large volumes of trajectory data accumulate, simplifying trajectories to reduce storage and querying costs is increasingly studied. 
{\color{black}Existing proposals face three main problems.  First, they require numerous iterations to decide which GPS points to delete. Second, they focus only on the relationships between neighboring points (local information) while neglecting the overall structure (global information), reducing the global similarity between the simplified and original trajectories and making it difficult to maintain consistency in query results, especially for similarity-based queries. Finally, they fail to differentiate the importance of points with similar features, leading to suboptimal selection of points to retain the original trajectory information.}

% Existing proposals (i) require numerous iterations to decide on which GPS points to delete, (ii) struggle to capture both the overall structure (global information) and neighboring point relationships (local information) of trajectories, and (iii) struggle to determine the relative importance of points with similar features.

We propose \textsc{MLSimp}, a novel Mutual Learning query-driven trajectory simplification framework that integrates two distinct models: GNN-TS, based on graph neural networks, and Diff-TS, based on diffusion models.  GNN-TS evaluates the importance of a point according to its globality, capturing its correlation with the entire trajectory, and its uniqueness, capturing its differences from neighboring points. It also incorporates attention mechanisms in the GNN layers, enabling simultaneous data integration from all points within the same trajectory and refining representations, thus avoiding iterative processes. Diff-TS generates amplified signals to enable the retention of the most important points at low compression rates. Experiments involving eight baselines on three databases show that \textsc{MLSimp} reduces the simplification time by 42\%--70\% and improves query accuracy over simplified trajectories by up to 34.6\%.

%outperforms eight baselines by 44\%-70\%  in simplification time and 
%The results show that \textsc{MLSimp} not only reduces simplification time by 44\%-70\% but also enhances query effectiveness  up to 21\%.
%reduces simplification time and improves query accuracy on the simplified trajectory database compared to state-of-the-art methods.

%Moreover, the model employs only GNN-TS for inference and adjusts the final simplification probability through query-level information.
%Experiments on three databases demonstrate that \textsc{MLSimp} reduces simplification time and improves query accuracy on the simplified trajectory database compared to state-of-the-art methods.
\vspace{-3mm}
\end{abstract}

%that directly quantifies each trajectory point's contribution to the simplified trajectory dataset, thereby circumventing the iterative process and substantially reducing simplification time. 

%\textsc{MLSimp} introduces two distinct trajectory simplification models GNN-TS and Diff-TS: GNN-TS based on graph neural networks and Diff-TS based on the diffusion model. 
%learns 
%both global and local information of trajectories via a trajectory graph  without the need of performing iterations and quantify trajectory-level point importance.
\maketitle

\pagestyle{\vldbpagestyle}
\begingroup\small\noindent\raggedright\textbf{PVLDB Reference Format:}\\
% \vldbauthors. 
Yumeng Song, Yu Gu, Tianyi Li, Yushuai Li, Christian S. Jensen, and Ge Yu.
\vldbtitle. PVLDB, \vldbvolume(\vldbissue): \vldbpages, \vldbyear.\\
\href{https://doi.org/\vldbdoi}{doi:\vldbdoi}
\endgroup
\begingroup
\renewcommand\thefootnote{}\footnote{
\noindent
This work is licensed under the Creative Commons BY-NC-ND 4.0 International License. Visit \url{https://creativecommons.org/licenses/by-nc-nd/4.0/} to view a copy of this license. For any use beyond those covered by this license, obtain permission by emailing \href{mailto:info@vldb.org}{info@vldb.org}. Copyright is held by the owner/author(s). Publication rights licensed to the VLDB Endowment. \\
\raggedright Proceedings of the VLDB Endowment, Vol. \vldbvolume, No. \vldbissue\ %
ISSN 2150-8097. \\
\href{https://doi.org/\vldbdoi}{doi:\vldbdoi} \\
}\addtocounter{footnote}{-1}
\endgroup

\vspace{-1mm}
\ifdefempty{\vldbavailabilityurl}{}{
\vspace{.1cm}
\begingroup\small\noindent\raggedright\textbf{PVLDB Artifact Availability:}\\
The source code, data, and/or other artifacts have been made available at \url{\vldbavailabilityurl}.
\endgroup
}

\input{src/introduction}

\input{src/relatedwork}
\input{src/preliminaries}

\input{src/model}

\input{src/experiments}

\input{src/conclusion}
\vspace{-1mm}
\begin{acks}
 % This work was supported by the [...] Research Fund of [...] (Number [...]). Additional funding was provided by [...] and [...]. We also thank [...] for contributing [...].
 % This work is supported by the National Natural Science Foundation of China (U23B2019, 62072083) and the Fundamental Research Funds of the Central Universities (N2216017).
 This research is supported in part by the National Natural Science Foundation of China (U23B2019, 62072083), the Fundamental Research
Funds of the Central Universities (N2216017) and the Innovation Fund Denmark project DIREC (9142-00001B).
\end{acks}

%\clearpage
\balance
\bibliographystyle{ACM-Reference-Format}
\bibliography{main}

\end{document}

%% file: src/introduction.tex
\section{Introduction}
\label{sec:1}
% Trajectories, representing the paths followed by objects over time, have found wide applications across various domains such as transportation, wildlife monitoring, and human behavior analysis. The significance of trajectories lies in their ability to capture spatiotemporal movement patterns, enabling valuable insights and decision-making in numerous fields.

The widespread use of mobile and location-aware devices generates large volumes of GPS trajectory data. Efficient compression of this data~\cite{han2017compress,nibali2015trajic,li2021trace, li2020compression} is attractive as it reduces storage and transmission costs. 
% Lossless compression, though effective, often achieves non-competitive efficiency and compression rates. In contrast, 
Trajectory simplification~\cite{wang2021survey}, which focuses on retaining only essential points, provides an efficient space reduction with acceptable information loss.

\begin{figure*}[t]
\centering
\includegraphics[width=0.9\textwidth]{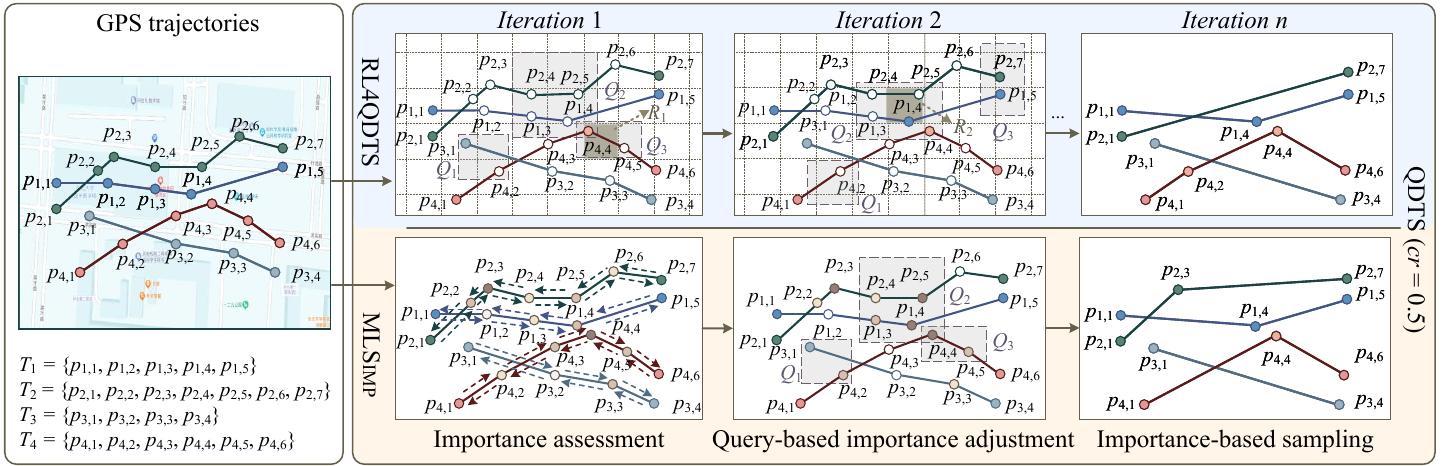}
\caption{Circles denote GPS points, with solid lines connecting them to form trajectories.  Colored circles indicate points retained in simplified trajectories, while white circles show points that are not  retained by the current iteration. %(iv) Blue dashed squares represent simulated range query regions, brown dashed points represent the importance assessment of candidate points, with darker colors indicating higher importance.
}
\label{intro}
\vspace{-5mm} 
\end{figure*}

\begin{comment}
\begin{myEx}
In Fig.~\ref{intro} (a) and (b), we show an example of the simplification process of the representative error-driven trajectory query method in a top-down manner, DP~\cite{DP},  with a compression rate of 0.5 involving four trajectories $T_i\,(1\leq i \leq 4)$. Taking $T_1$ as an example, $T_1$ contains five GPS points $p_1,j\,(1\leq j \leq 5)$, and the simplified trajectory $T_1^*$ should contain three points. Firstly, the starting point $p_{1,1}$ and the ending point $p_{1,5}$ are retained. Then, the errors for $p_{1,j}\,(2\leq j \leq 4)$,  are calculated, where we demonstrate PED in the example. The PED increases with the length of the Euclidean distance between the point and the segment $\overline{{p_{1,1} p_{1,5}}}$. In each step, the point with the maximum error, in this case $p_{1,4}$, is selected and added to the simplified trajectory. Finally, the simplified trajectory $T_1^*= \{p_{1,1}, p_{1,4}, p_{1,5}\}$. The simplification processes for $T_2$, $T_3$, and$T_4$  are similar to $T_1$, requiring 2, 0, and 1 steps, respectively.
\end{myEx}
\end{comment}

Most trajectory simplification methods~\cite{DPhull,TD-TR,RLTS}, such as those using Synchronized Euclidean Distance (SED)~\cite{SED1,SED2,SED3SAD1}, Perpendicular Euclidean Distance (PED)~\cite{PED1,TD-TR,PED3}, Direction-aware Distance (DAD)~\cite{DAD1,DAD2}, and Speed-aware Distance (SAD)~\cite{SED3SAD1}, evaluate the importance of GPS points based on 
 error margins. They are Error-Driven Trajectory Simplification (EDTS) methods. Such methods often fail to accommodate common queries. 
 % In particular, they do not take into account the relationships between trajectories, which is crucial for similarity-based queries, and they do not consider query distributions, as is necessary for range queries. 
{\color{black}
% Existing EDTS research utilizes different distance-based errors to guide the simplification. However, these methods overlook the usability of simplified trajectory data when supporting query and mining tasks in trajectory management systems. 
 An experimental report~\cite{TSsurvey} 
% highlights the performance of these methods on range, $k$NN, clustering, and similarity queries. The report 
finds that simplification methods using the DAD error, e.g., \cite{Error-search,dpts}, 
% perform poorly across all queries,
result in nearly a 50\% drop in performance on range queries.
% with the performance on range queries dropping by nearly 50\%. 
% This reduced performance is attributed to their emphasis on direction that disregards critical location and time information.
Simplification methods using the PED and SED errors, e.g., \cite{DP,DPhull,TD-TR}, 
% show competitive performance on range and clustering queries but 
% underperform on similarity and $k$NN queries, 
 lead to a performance decline of up to 57\% on $k$NN queries and up to 33\% on similarity queries.
% with the performance dropping by up to 57\% on $k$NN queries and up to 33\% on similarity queries. 
% This is because these methods simplify trajectories as individual entities without preserving inter-trajectory similarity.
}

Query-Driven Trajectory Simplification (QDTS)~\cite{QDTS} has been proposed to address the above limitations. QDTS aims to ensure {\color{black} that the} results of queries on simplified trajectories match those on the original trajectories closely across various query types, while also achieving low compression rates. 
Motivated by these advantages, 
{\color{black}we focus on batch mode QDTS. In this mode~\cite{QDTS,TD-TR,DPhull,batchmode}, all trajectories are stored in the database and simplified offline once. The resulting simplified trajectory database is then available for online querying.}
The state-of-the-art QDTS model, named RL4QDTS~\cite{QDTS}, employs a reinforcement learning model and takes into account the relationship between trajectories and queries during simplification. 
\vspace{-8mm} 
\begin{myEx}\label{ex:RL4QDTS}
Fig.~\ref{intro} provides an example of 
% trajectory simplification using 
RL4QDTS. Given four trajectories $T_i\,(1\leq i \leq 4)$ comprising a set of points $P\,(|P|=22)$ and a compression rate $cr=\textit{0.5}$, RL4QDTS has two major steps. First, RL4QDTS includes the start and end points of each trajectory in the simplified set $S=\{p_{1,1}, p_{2,1}, p_{3,1}, p_{4,1}, p_{1,5},  p_{2,7}, p_{3,4}, p_{4,6}\}$. 
Second, RL4QDTS iteratively simulates queries $Q_1$, $Q_2$, and $Q_3$ (blue rectangulars in Fig.~\ref{intro}). In the first iteration, it selects a region $R_1$ based on $Q_i$ ($1\leq i \leq 3)$ and adds $p_{4,4}$ that falls within $R_1$ (grey grid cells in Fig.~\ref{intro}) to $S$. In the second iteration, it updates the states based on the selected region $R_1$, selects $R_2$, and adds $p_{1,4}$ to $S$. {\color{black}This procedure is repeated until $\frac{|S|}{|P|}=\textit{cr}$, with a higher $\textit{cr}$ indicating a higher compression rate. } The final simplified trajectory set is  $S=\{T_1^*=\langle p_{1,1}, p_{1,4}, p_{1,5}\rangle,  T_2^*=\langle p_{2,1}, p_{2,7}\rangle, T_3^*=\langle p_{3,1}, p_{3,4}\rangle, T_4^*=\langle p_{4,1}, p_{4,4}, p_{4,6}\rangle\}$.
\vspace{-1mm}
%In each iteration, it selects regions (shaded grid cells in Fig.~\ref{intro}) based on $Q_1$, $Q_2$, and $Q_3$ and adds the points falling within the regions to $S$. In the first iteration, $p_{4,4}$ is added; followed by $p_{1,4}$ in the second iteration until $\frac{|S|}{|P|}=\textit{cr}$. The final simplified trajectory set is  $S=\{T_1^*=\{p_{1,1}, p_{1,4}, p_{1,5}\},  T_2^*=\{p_{2,1}, p_{2,7}\}, T_3^*=\{p_{3,1}, p_{3,4}\}, T_4^*=\{p_{4,1}, p_{4,4}, p_{4,6}\}\}$.

\end{myEx}

While existing studies demonstrate effectiveness in trajectory simplification, they face three major challenges:

\noindent 1) \textbf{\textit{How to compute the importance of GPS points non-iteratively?}} Trajectory simplification typically calculates the importance of GPS points iteratively, as their significance changes with updates to the simplified trajectory~\cite{DP,DPhull, bottomup,RLTS,S3}. For example in Fig.~\ref{intro}, when $p_{4,4}$ is included, the importances of its neighboring points $p_{4,3}$ and $p_{4,5}$, as determined by existing EDTS methods~\cite{DP,DPhull}, are likely to decrease. 
{\color{black} Moreover, EDTS methods~\cite{DP,DPhull,TD-TR,dpts,Error-search}  
 typically rely on dynamic programming or binary search, which are time-consuming.
 % Many error measures have  cubic time complexity, making them impractical for real-world applications. 
The recent methods RLTS~\cite{RLTS} and RL4QDTS~\cite{QDTS}, as shown in Example~\ref{ex:RL4QDTS}, implement an iterative process due to its use of reinforcement learning, resulting in substantial time overhead (as discussed in Sec.~\ref{sec:5.3}).}

\noindent 2) \textbf{\textit{How to retain ``global information'' in simplified trajectories for adapting to diverse queries?}}  The effectiveness of range-search queries (e.g., point, range, and window) is affected significantly  by the specific distribution of individual locations, referred to as \textit{local information}~\cite{berzal2002data}.  Conversely, similarity-based queries (e.g., $k$NN queries, similarity search, and clustering) depend on the overall structure of a trajectory, referred to as \textit{global information}~\cite{berzal2002data}. Existing studies~\cite{QDTS, DP,bottomup,DPhull,TD-TR,RLTS,SED1,SED2,PED1,PED3,DAD1,DAD2,SED3SAD1} typically simplify GPS points while focusing solely on local information, ignoring global information. %On the other hand, EDTS~\cite{DP,bottomup,DPhull,TD-TR,RLTS,SED1,SED2,PED1,PED2,PED3,DAD1,DAD2,SED3SAD1} methods typically evaluate point importance considering only information from neighboring points. 
For example, for trajectory $T_1$, an existing study~\cite{bottomup} measures the importance of $p_{1,3}$ based on its distance to the line segment $\overline{p_{1,4} p_{1,5}}$. Recent studies~\cite{zhou2023grlstm,jiang2023self,timecomplexity1,chang2023contrastive} employing trajectory embeddings 
% with LSTM or Transformer models 
aim to preserve similarity in the embedding space, but often fail to  represent accurately the distribution space of simplified trajectories, especially when having to capture global information across extended trajectories.

 \noindent 3) \textbf{\textit{How to identify the most important point for simplified trajectories from points with similar importance at low compression rates?}}  Existing studies~\cite{DP,bottomup,DPhull,TD-TR} typically select either the most important point to retain or the least important point to drop.  However,  candidate points often have equal importance and the ``most important'' one is chosen randomly.  
 At low compression rates—where few points are retained—this random selection can cause substantial deviations from the original trajectory. For example in Fig.~\ref{intro}, when simplifying trajectory $T_2$~\cite{DP},  initially, only the start and end points are retained. 
Next,  the importances of points $p_{2, i}\,(2 \leq i \leq 6)$ are measured  by their Euclidean distance to the segment $\overline{p_{2,1} p_{2,7}}$, and $p_{2,3}$ is selected. In the next iteration, $p_{2,5}$ and $p_{2,6}$ have equal distances to $\overline{p_{2,3} p_{2,7}}$, and $p_{2,6}$ is chosen randomly. If the iteration ends here due to the low compression rate, this choice results in more information loss than if $p_{2,5}$ was selected, as $p_{2,5}$ is better for maintaining $T_2$'s shape.  This highlights the need for means of more accurately discerning point importance, especially at low compression rates, ensuring the retention of the most important points.
 
To address the above challenges, we introduce a novel {\textbf{M}}utual {\textbf{L}}earning query-driven trajectory {\textbf{simp}}lification method (\textsc{MLSimp}). 
We define two key concepts to quantify point importance: \textit{globality} and \textit{uniqueness}. Globality captures the correlations of GPS points with their entire trajectory, while uniqueness captures the differences between a point and its neighboring points. %This allows us to take the global information of trajectories when evaluating the importance of a point, instead of just its neighbors. 
\textsc{MLSimp} incorporates a lightweight Graph Neural Network (GNN)-based trajectory simplification (GNN-TS) model.
For the first challenge, the GNN-TS models a trajectory as a graph with points as nodes. The GNN-TS enables parallel generation of node embeddings, avoiding iterative updates dependent on a trajectory's current state. 

To address the second challenge, the GNN-TS measures point importance using both globality and uniqueness, which enables consideration of  both the global and local information of a trajectory. %The former enables the consideration of the trajectory's global information while the latter enables that of local point-to-point relationships. 
Moreover, %GNN-TS utilizes a GNN to mine global information from trajectory graphs, using contrastive learning for training. 
the GNN-TS incorporates attention mechanisms in its GNN layers. This allows points to dynamically integrate data from adjacent nodes and refine their representations, thereby preserving essential trajectory details. To address the third challenge, we introduce Diff-TS, a complex diffusion-based trajectory simplification model, and integrate it with the GNN-TS in the \textsc{MLSimp} mutual learning framework. In this framework, the GNN-TS provides simplified trajectories with high compression rates as soft labels for Diff-TS, which, in turn, offers low compression rate trajectories as feedback to enhance GNN-TS training. This facilitates a clearer distinction between the importance of GPS points, based on data inferred from the integrated model. Subsequently, \textsc{MLSimp} adjusts point importance based on simulated queries and generates simplified trajectories by performing sampling. 
\vspace{-3mm}
\begin{myEx}
  Fig.~\ref{intro} illustrates \textsc{MLSimp} applied to trajectories $T_i (1\leq i \leq 4)$.  The dashed arrows indicate the direction of  information passing.  Each point $p_i (1\leq i \leq 22)$ aggregates global and local information from other points in the same trajectory, allowing for simultaneous information exchange and  thus avoiding iterations.  In trajectory $T_2$, although $p_{2,5}$ and $p_{2,6}$ initially have similar importance due to comparable semantics, the Diff-TS model subsequently decreases $p_{2,6}$ importance and increases that of $p_{2,5}$ based on simplified point generation conditioned on $T_2$. Next, we adjust the importance based on the distribution of three generated queries $Q_i\,(1\leq i \leq 3)$. The importances of $p_{1,3}$, $p_{1,4}$, $p_{2,4}$, $p_{2,5}$, $p_{4,2}$, $p_{4,4}$, and $p_{4,5}$ are elevated as they are included in range queries, whereas the remaining points have their importance reduced as they are not  in the range queries. Finally, we sample three GPS points based on the importance and obtain the final simplified trajectory dataset $S= \{T_1^*=\langle p_{1,1}, p_{1,4}, p_{1,5}\rangle,  T_2^*=\langle p_{2,1}, p_{2,3}, p_{2,7}\rangle, T_3^*=\langle p_{3,1}, p_{3,4}\rangle, T_4^*=\langle p_{4,1}, p_{4,4}, p_{4,6}\rangle\}$.
% \vspace{-1mm}

\end{myEx}

\vspace{-2mm}
Our contributions are summarized as follows:
\vspace{-1mm}
\begin{itemize}[leftmargin=3mm]
\item We introduce \textsc{MLSimp}, a novel mutual learning framework for QDTS. It alternates training between two models, GNN-TS and Diff-TS, but employs only the GNN-TS for simplification. To the best of our knowledge, this is the first mutual learning framework and the first application of GNNs and diffusion models in trajectory simplification.

     \item The GNN-TS model evaluates the importance of GPS points using two new metrics: globality and uniqueness. It aggregates data from all points in the same trajectory simultaneously to eliminate iterations and  captures global information efficiently.
    
    \item The Diff-TS model processes simplified trajectories with high compression rates from the GNN-TS to train its diffusion model. This feedback, in the form of low compression trajectories, sharpens the distinction in the importance of GPS points through insights from the combined model.
    
    \item We report on experiments with eight state-of-the-art methods on three datasets. The results show that \textsc{MLSimp} not only  can reduce the simplification time by 44\%--70\% but also enhances query accuracy by up to 34.6\%.

\vspace{-1mm}
\end{itemize}

We review related work in Section~\ref{sec:2} and cover preliminaries in Section~\ref{sec:3}. Section~\ref{sec:4} presents \textsc{MLSimp}, while Section~\ref{sec:5} discusses experimental findings. Section~\ref{sec:6} concludes the paper and outlines research directions.

%% file: src/relatedwork.tex
\vspace{-2mm}
\section{Related Work}
\label{sec:2}

%可删
{\color{black}Trajectory simplification can be performed in two modes:  online and  batch. In online mode~\cite{SED1, SED2, SED3SAD1,RLTS,S3,ECM,yao2023simplets}, sensors continuously collect trajectory data, storing it  temporarily in a local buffer. This mode aims to select key trajectory points to be saved on the server.
This mode is suitable in scenarios that require real-time updates, such as vehicle tracking, live sports analytics, and dynamic route optimization. 
Conversely, in batch mode~\cite{QDTS,TD-TR,bottomup,DP,DPhull}, all trajectory data is pre-stored in the database, and there are no updates. This mode reduces simplification errors and optimizes storage space, making it suitable in scenarios requiring high precision and data usability, such as historical data analyses and offline route optimization.
We review the studies on the batch mode, which is the focus of this paper.}
%可删

\vspace{-3mm}
\subsection{Error-Driven Trajectory Simplification}

% Existing EDTS methods~\cite{DP,DPhull,TD-TR,bottomup,dpts, Error-search,S3, RLTS,EOB}  compress trajectories  by controlling the deviation between the simplified trajectory and the original trajectory.

%on EDTS aim to effectively compress data by controlling the error between simplified trajectory points and the original trajectory.% thereby reducing storage overheads and enhancing data processing efficiency. Herein, we discuss both non-learning and learning-based methods in the batch mode which is the focus of this paper.
\vspace{-1mm}
\subsubsection{Non-learning-based methods.}
Many methods simplify trajectories using a top-down approach~\cite{DP,DPhull, TD-TR}. The DP method~\cite{DP} simplifies trajectories by recursively splitting them based on an error threshold, using the point with the maximum perpendicular Euclidean distance to dictate splits. To improve efficiency, DPhull~\cite{DPhull} leverages convex hull properties to identify significant points, reducing the computational complexity while maintaining the same output. Another extension of DP, TD-TR~\cite{TD-TR}, incorporates time, using synchronized Euclidean distance to measure errors. 
% In contrast, the Bottom-Up method~\cite{bottomup} begins with individual GPS points. It merges GPS points into segments until a specified error threshold is met, focusing on error reduction while preserving essential trajectory features.

In contrast, Bottom-Up trajectory simplification~\cite{bottomup} begins with individual GPS points and gradually aggregates them into segments until a specified error threshold is satisfied. This method focuses on gradually reducing errors by merging segments while preserving critical trajectory characteristics. Despite its effectiveness at preserving location information, the method may omit critical points that are important for clustering and querying. To address this, the DPTS method~\cite{dpts} retains both directional and positional information by considering angular distances.

%Non-learning methods are mainly heuristic-based to iteratively remove or retain trajectory points. Some of these algorithms are sensitive to parameter settings, requiring manual adjustment to achieve optimal compression results. Additionally, these methods often exhibit high complexity, demanding significant computational resources. To address these challenges, researchers introduce deep learning for trajectory simplification as a replacement for manual adjustments.
\vspace{-2mm}
\subsubsection{Learning-based methods.}
RLTS~\cite{RLTS} uses reinforcement learning for trajectory simplification. It models trajectory simplification as a Markov decision process and employs policy gradient methods to learn simplification policies. Although it achieves minimal error, the high time cost of training  poses challenges. To address this, S3~\cite{S3} leverages two Seq2Seq models that utilize BiRNNs~\cite{BiRNN} to compress and then reconstruct trajectories, reducing both training and simplification times. Despite higher error rates compared to RLTS, S3 lowers the simplification time considerably. EB-OTS~\cite{EOB} also employs reinforcement learning for trajectory simplification, targeting a minimal compression rate given a fixed error threshold.  In contrast, \textsc{MLSimp} assumes a fixed compression rate and then performs simplification to meet that rate (see Example~\ref{intro}). Thus, \textsc{MLSimp} is not compared against EB-OTS.

In summary, while the majority of trajectory simplification studies~\cite{DP,DPhull,TD-TR,bottomup,RLTS} seeks to minimize error, they often overlook enhancing query correctness—a crucial simplification goal in real-world applications. 
In Section~\ref{sec:5}, we include  DPHull~\cite{DPhull}, Bottom-Up~\cite{bottomup}, RLTS~\cite{RLTS}, and S3~\cite{S3}  which are considered as baselines.

%Error-driven methods, despite preserving critical information under sufficient storage, struggle to maintain all essential details at lower compression

%These two methods respectively prioritize low error and low simplification time, but they do not simultaneously address both aspects. In contrast, \textsc{MLSimp} considers both effectiveness and efficiency simultaneously. Through a mutual learning approach, we enable the simple Graph-TS model to learn from a complex Diff-TS model, enhancing the performance of Graph-TS. During simplification, only the simple model is utilized, ensuring high efficiency.

%Additionally, EB-OTS~\cite{EOB} is another trajectory simplification method that utilizes reinforcement learning. However, its primary objective is to minimize compression rates under an error bound. This differs from the fixed compression rate simplification problem addressed in this paper. Therefore, \textsc{MLSimp} does not compare against it.

%In summary, most error-driven trajectory simplification research aims to minimize a given error metric while simplifying a trajectory. However, they fail to give sufficient attention to one of the main goals of simplification, which is enhancing query efficiency. While error-driven approaches can maintain essential information in the simplified trajectory with enough storage budget, it becomes challenging for the simplified trajectory to fully preserve all essential information in a low compression ratio. Therefore, error-driven approaches are not suitable for trajectory simplification tasks aimed at trajectory queries.

\vspace{-2mm} 
\subsection{Query-Driven Trajectory Simplification}
Error-driven trajectory simplification methods treat trajectory compression and storage as independent components. However, in a larger trajectory management system, trajectories are often stored and indexed to support query processing or pattern mining. Thus, Zhang et al.~\cite{TSsurvey} propose considering data availability as a simplification quality measure.  Inspired by this work, Wang et al.~\cite{QDTS}  introduce the Query-Driven Trajectory Simplification (QDTS) problem. They aim to find a simplified trajectory database within a given storage budget while preserving query accuracy as much as possible on the simplified database. They present a reinforcement learning-based method, RL4QDTS. 
This method works in a top-down manner. 
% initially adding the starting and ending points of all trajectories to the simplified trajectory database and then iteratively adding additional points from the original database until the storage space is exhausted. 
RL4QDTS establishes an octree index based on the trajectory database, defining the selection process as two decision tasks. 
% The first decision task selects a spatiotemporal cube constructed from 2D space and 1D time from an octree index. The second decision task selects a point from the spatiotemporal cube. The two decision tasks are performed by reinforcement learning agents.
The model learns the selection strategy from the difference between the query results of the original and the simplified database for a set of range queries. The simplified database not only supports range queries but also effectively supports $k$NN queries, similarity queries, and clustering.

However, RL4QDTS still has shortcomings: First, although it works in a top-down manner, it remains an iterative simplification method that requires multiple steps.
% Moreover, in each simplification step, it needs to update the reinforcement learning states. 
Second, RL4QDTS no longer considers whether the simplified trajectory still retains the key information of the original trajectory. It relies solely on training the model based on range queries, making it easy to lose critical information during simplification, thereby reducing the accuracy of similarity-based queries (such as $k$NN queries). In contrast, \textsc{MLSimp} mines the semantics of trajectories, predicts the importance of each GPS point based on the information contained in each  point, and then adjusts the importance based on the generated range query workload. \textsc{MLSimp}  considers  trajectory and query information simultaneously while avoiding iterations.

% \subsection{Deep Trajectory Learning}

% In recent years, deep learning has been successfully applied to trajectory data analysis and applications, including trajectory reconstruction and generation~\cite{generation1,generation2}, trajectory representation and similarity search~\cite{representation1,representation2, representation3}, and trajectory prediction~\cite{prediction1,prediction2}, etc. Among these applications, trajectory representation and similarity search are most relevant to the research in this paper. These studies~\cite{representation2, representation3} involve capturing hidden features in trajectories through representation learning to generate low-dimensional trajectory embeddings. The distribution of trajectory embeddings reflects the results of $k$NN queries between real trajectories. However, from a problem perspective, the goal of QDTS is to support various types of queries, while similarity search only supports $k$NN queries. From a technical perspective, although \textsc{MLSimp} also captures hidden features of trajectories and generates embedding spaces, our objective is to observe the features of trajectory point embeddings to ensure that the simplified trajectory retains key original information rather than preserving the similarity of hidden features with the original trajectory. Therefore, this problem fundamentally differs from both QDTS.

%% file: src/preliminaries.tex
\vspace{-2mm}
\section{Preliminaries}
\label{sec:3}
\begin{myDef}
A GPS \textbf{point} $(x,y,t)$ records the longitude $x$ and the latitude $y$ at time $t$.
\end{myDef}
\vspace{-4mm}
\begin{myDef}
A \textbf{trajectory} $T$ is a sequence of GPS points, i.e., $T=\langle p_1, p_2, \ldots, p_n \rangle$, where $p_i.t < p_{i+1}.t (1 \leq i \leq n-1)$.
\end{myDef}

% \begin{figure}
% \centering
% \subfloat[Spatial range query]{
% \includegraphics[width=3.5cm]{figures/query_1.pdf}}  
% \quad
% \subfloat[Temporal range query]{ 
% \includegraphics[width=3.5cm]{figures/query_2.pdf}} 
% \vspace{-4mm}

% \subfloat[$k$NN and similarity queries]{
% \includegraphics[width=3.5cm]{figures/query_3.pdf}}  
% \quad
% \subfloat[Clustering]{ 
% \includegraphics[width=3.5cm]{figures/query_4.pdf}}
% \caption{Examples of queries.}
% \label{query}
% \vspace{-5mm} 
% \end{figure}

\vspace{-3mm}
\begin{myDef}
Given a database $D$ of trajectories, a \textbf{range query}  $Q_\mathit{range}=(x_\mathit{min}, x_\mathit{max}, y_\mathit{min}, y_\mathit{max}, t_\mathit{min}, t_\mathit{max})$ finds all trajectories that contain at least one point $p_i = (x_i, y_i, t_i)$ such that $x_\mathit{min}\leq x_i \leq x_\mathit{max}$, ${ymin} \leq y_i \leq y_\mathit{max}$, and $t_\mathit{min} \leq t_i \leq t_\mathit{max}$.
\end{myDef}

% \vspace{-2mm}
% Figs.~\ref{query} (a) and (b) show the spatial and temporal distribution of the trajectory database $D$ in Fig.~\ref{intro}.   
% For a range query $Q_\mathit{range}=(4, 8, 2, 6, 0, 6)$, the points satisfying both spatial and temporal ranges include $p_{1,3}, p_{1,4}, p_{2,4}, p_{2,5}, p_{4,3}$, and $p_{4,4}$. Thus, the result of $Q_\mathit{range}$ is $\{T_1, T_2, T_4\}$.

\vspace{-2mm}
\begin{myDef}
Given a trajectory database $D$, a \textbf{\textit{k}NN query} $Q_\mathit{kNN}=(k, T_q, [t_s, t_e])$ finds a set $R$ of $k$ trajectories such that $\forall T_i \in R; \forall T_j \in D - R;(\Theta(T_q[t_s, t_e], T_i[t_s, t_e]) \leq \Theta(T_q[t_s, t_e], T_j[t_s, t_e]))$, where $\Theta(\cdot, \cdot)$ represents a dissimilarity measure for trajectories. 
\end{myDef}

In this paper, we use EDR~\cite{edr} to instantiate $\Theta(\cdot, \cdot)$. However, our proposals are orthogonal to the dissimilarity measure used.

% Performing a \textit{k}NN query $Q_\mathit{kNN}=(1, T_1, [2,6])$. The subtrajectories for this time window are shown in Fig.~\ref{query} (c).
% We compute the EDR between $T_q[2,6]$ and $T_i[2,6] (2\leq i \leq 4)$. The ranking of the EDR scores is  $\Theta(T_q[2, 6], T_2[2, 6]) < \Theta(T_q[2, 6], T_4[2, 6]) <\Theta(T_q[2, 6], T_3[2, 6])$. Therefore, the result of $Q_\mathit{kNN}$  is  $\{T2\}$.

\vspace{-1mm}
\begin{myDef}
Given a trajectory database $D$, a \textbf{similarity query} $Q_\text{sim}=(T_q, [t_s, t_e], \Delta)$ finds a set $R$ of trajectories defined as 
 $R = \{T \in D | \forall i \in [t_s,t_e] (d(T_q[i], T[i]) \leq \Delta)\}$ where $T_j \in R$, $d(\cdot, \cdot)$ denotes Euclidean distance.
\end{myDef}
\vspace{-3mm}

% Assuming a similarity query $Q_\mathit{sim} = (T_1, [2,6], 2)$ on $D$. The trajectories in the time window are the same for $Q_\mathit{kNN}$.  From Fig.~\ref{query} (c), we find that $d(T_q[t],T_2[t])<2$ where $t \in [2,6]$, while $d(T_q[2],T_3[2])>2$ and $d(T_q[2],T_4[2])>2$. Therefore, the result of $Q_\mathit{sim}$ is $\{T_2\}$.

\vspace{-1mm}
\begin{myDef}
Given a trajectory database, trajectory \textbf{clustering}~\cite{cluster} partitions each trajectory into subtrajectories and then clusters subtrajectories based on some notion of  trajectory distance.
\end{myDef}
\vspace{-2mm}

% Fig.~\ref{query} (d) illustrates clustering of $T_1$, $T_2$, $T_3$, and $T_4$. First, the four trajectories are divided into eight subtrajectories. Then, through line clustering algorithms, they are grouped into three clusters where $T_2 \in \{C_1\}$, $T_i \in \{C_2, C_3\} (t\in \{1,3,4\})$.

\vspace{-2mm}
\begin{myDef}
\textbf{Trajectory simplification} aims to eliminate points from a trajectory $T$ to obtain a simplified trajectory $T'$ of the form $T' = \langle p_{s_1}, p_{s_2}, \ldots, p_{s_m} \rangle$,
\end{myDef}
\vspace{-1mm}

\vspace{-2mm}
\begin{myDef}
Given a trajectory database $D$ and a storage budget $W$ indicating a fraction $r$ of the original points in $D$ to be retained, \textbf{Query-Driven Trajectory Simplification}~\cite{QDTS} aims to find a trajectory database $D'$ of simplified trajectories, such that the difference between query results on $D$ and $D'$ are minimized.
\end{myDef}
\vspace{-2mm}

%% file: src/model.tex
%gnn-ts framework 
\begin{figure*}[t]
\centering
\includegraphics[width=0.9\textwidth]{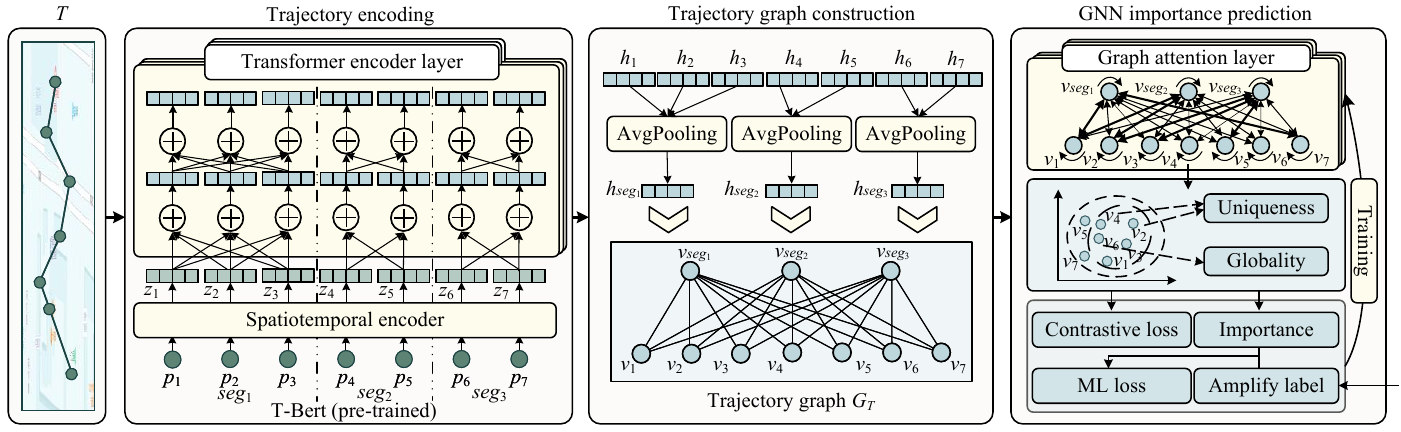}
\caption{GNN-TS model overview.}
\vspace{-6mm} 
\label{gnn-ts}
\end{figure*}

\vspace{-1mm}
\section{Mutual Learning Trajectory Simplification}
\label{sec:4}

% Given a trajectory database, our goal is to predict the importance of GPS points and determine which points to include in the simplified database. We adopt mutual learning that involves two models that are trained iteratively, with one model being used for simplification. We proceed to introduce the two models and then describe the mutual learning algorithm.

\subsection{GNN-based Trajectory Simplification Model}
% We first present the GNN-based trajectory simplification model. 
\vspace{-1mm}
\subsubsection{Framework}
% The Transformer~\cite{transformer} has shown excellent performance at trajectory representation learning by mining hidden information in trajectories. However, due to model framework and hardware constraints, the Transformer does not support encoding trajectories with more than 512 points. 
% % Long trajectories are common in real trajectory databases, such as Geolife and T-Drive, where the average trajectory length exceeds 1,000 points, as shown in Table~\ref{dataset}. 
% Therefore, 
GNN-TS consists of three steps: trajectory encoding, trajectory graph construction, and GNN importance prediction, as shown in  Fig.~\ref{gnn-ts}.

\noindent\textbf{Trajectory Encoding}. Given a trajectory $T$, a pre-trained Trajectory Bert (T-Bert) encodes each point, generating embeddings for 
 points. We detail T-Bert in Sec.~\ref{subsubsec:t-bert}.

\noindent\textbf{Trajectory Graph Construction}. Trajectory segment embeddings are generated from the embeddings of GPS points. A long trajectory is constructed into a graph $G_T$ using trajectory segments and points, as described in Sec.~\ref{subsubsec:graph construction}.

\noindent\textbf{GNN Importance Prediction}. The GNN aggregates hidden information from multiple segments to update node embeddings by $G_T$. It analyzes the uniqueness and globality of embeddings to predict their importance and construct a self-supervised contrastive learning loss. Importance is combined with amplification signals provided by Diff-TS to form a mutual learning loss (ML loss). Both contrastive and mutual learning losses are used to train the GNN. The prediction and training process is detailed in Sec.~\ref{subsubsec:gnn importance prediction}.

\subsubsection{Trajectory Encoding}
\label{subsubsec:t-bert}

\vspace{-2mm}
Given a trajectory $T=\langle p_1,p_2,\ldots,p_n\rangle$, we partition it into trajectory segments, each containing 
 $w$ points.  The trajectory segment sequence is denoted as $S_T=\langle\textit{seg}_1, \textit{seg}_2, \ldots, $\\$\textit{seg}_m\rangle$, where $\textit{seg}_k=\langle p_k^j, p^{j+1}_{k}, \ldots, p^{j+w}_k\rangle\, (1\leq k\leq m)$.  Each segment is encoded separately. If the last segment, $\textit{seg}_m$, has fewer than $w$ points, it is automatically padded to $w$ points during processing. Next, we detail the encoding of the points in a segment.

%$1\leq j\leq w-c$, and $c\leq w$

T-Bert first encodes the location and time of point $p_i=(x_i,y_i,t_i)$, where $p_i \in \textit{seg}_k$ and  $\textit{seg}_k \in S_T$,  as inputs to the Transformer layers through a spatiotemporal encoder. For location encoding, we employ node2vec~\cite{node2vec} to capture the location information of GPS points. To capture temporal differences of points, we follow an existing study~\cite{temporalencoder} and use a set of trainable parameters to encode time. 
Thus,  the initial encoding of $p_i$ is $ \textbf{z}_i=\textbf{z}_i^t+\textbf{z}_i^l$, 
% \begin{equation}
%   \textbf{z}_i=\textbf{z}_i^t+\textbf{z}_i^l.
% \end{equation}
where $\textbf{z}_i^t$ and $\textbf{z}_i^l$ denote the temporal and spatial embeddings of point $i$, respectively.
% denoted as
% \begin{equation}
%   \textbf{z}_i^l=\Omega(x_i,y_i),
% \end{equation}
% where $\Omega$ denotes the node2vec encoding. 

% To capture temporal differences of points, we follow an existing study~\cite{temporalencoder} and use a set of trainable parameters to encode time. Formally, 
% \begin{equation}
%   \textbf{z}_i^t=[\texttt{cos}(\omega_1 t_i),\texttt{sin}(\omega_1 t_i), \ldots, \texttt{cos}(\omega_d t_i), \texttt{sin}(\omega_d t_i)],
% \end{equation}
% where $\omega$ is the trainable parameters and $d$ is the encoding dimensionality. Thus,  the initial encoding of $p_i$ is:
% \begin{equation}
%   \textbf{z}_i=\textbf{z}_i^t+\textbf{z}_i^l.
% \end{equation}

{\color{black} The resulting sequence of vectors of each segment is then fed into a stacked Transformer encoder layer~\cite{transformer}, where each layer consists of a multi-head self-attention layer and a feed-forward neural network, generating embedding for each point based on its trajectory segment.} The $l$-th layer output embedding for $p_i$ of the Transformer is denoted as:
\vspace{-1mm}
\begin{equation}
  \textbf{h}_i^{(l)}=\texttt{FFN}(\texttt{MultiHeadAttention}(\bm{h}_i^{(l-1)})),
  \vspace{-1mm}
\end{equation}
where $\texttt{MultiHeadAttention}(\cdot)$ is the multi-head self-attention layer, $\texttt{FFN}(\cdot)$ is the feed-forward neural network and $\textbf{h}_i^{(0)}=\textbf{z}_i$.

Inspired by an existing study~\cite{bert}, we employ a Masked Language Model (MLM) to construct a self-supervised training task. Given a trajectory $T$, we randomly mask 20\% of the points  using a special token $[mask]$. For a masking set $M$, we train the model by predicting the value of its original token.
% using the following loss function:
% \begin{equation}
%  \begin{split}
%  \mathcal{L}_{\mathit{MLM}} &= \mathop{\texttt{argmin}}\limits_{\theta_{\mathit{MLM}}}\sum_{i \in M} p(l_i|\hat{l}_i) \\
%  \hat{l}_i&=\texttt{f}_{\mathit{MLM}_{l_i}}(\textbf{h}_i^{(L)})),
%  \end{split}
% \end{equation}
% where $l_m \in \langle x_m,y_m\rangle$, $\theta_{\mathit{MLM}}$ denotes the set of all learnable parameters, $\texttt{f}_{\mathit{MLM}_{l_i}}(\cdot)$ is a fully-connected forward network, and $L$ is the number of Transformer layers.
% After pre-training,  the resulting embedding vectors $\bm{h}_i^{(L)}$ from the $L$-layer Transformer are contextually relevant to the corresponding segment and contain specific information about spatial locations and time sequences.

\vspace{-2mm}
\subsubsection{Trajectory Graph Construction}
\label{subsubsec:graph construction}

Using T-Bert, we obtain the output of the last layer of the Transformer as the embedding for each point, i.e., $\bm{h}_i = \bm{h}_i^{(L)}$. For each segment $\textit{seg}_i = \langle p_i^1, ..., p_i^w \rangle$, we compute its embedding $\bm{h}_{\textit{seg}_i}$ through average pooling.
% : 
% \begin{equation}
% \bm{h}_{\textit{seg}_i} = \texttt{AvgPooling}(\bm{h}_{p_i^1}, \bm{h}_{p_i^2}, \ldots, \bm{h}_{p_i^w}),
% \end{equation}
% where $p_i^1$ represents the first point of the segment, and $p_i^w$ represents the last point of the segment. 
% Then $\bm{h}_{\textit{seg}_i}$ contains information about all points in the segment, and because we embed the time label of each point, $\bm{h}_{\textit{seg}_i}$ also captures the relative temporal relationship between segments.

To explore the relationships between GPS points and other segments in trajectory $T$, we construct a trajectory graph.  
\vspace{-2mm}
\begin{myDef}
The \textbf{trajectory graph}  $G_T = (V, E)$ of trajectory $T$ is an undirected graph, where $V$ and  $E$ are the set of nodes and edges, respectively.  The node set $V = V_{p} \cup V_\mathit{seg}$, where $V_p = \{v_1, v_2, \ldots, v_n\}$ is the set of all GPS points in $T$, and $V_{\textit{seg}} = \{v_{\textit{seg}_1}, v_{\textit{seg}_2}, \ldots, v_{\textit{seg}_m}\}$ is the segments of $T$. Each segment node is connected to each GPS node via an edge in $E$. The feature matrix $\bm{H} = [\bm{H}_p, \bm{H}_{\textit{seg}}]$  that contains the initial representation of the GPS point and segment nodes: $\bm{H}_p = [\bm{h}_1, \bm{h}_2, \ldots, \bm{h}_n]$ and  $\bm{H}_{\textit{seg}} = [\bm{h}_{\textit{seg}_1}, \bm{h}_{\textit{seg}_2}, \ldots, \bm{h}_{\textit{seg}_m}]$.
\vspace{-2mm}
\end{myDef}
The trajectory graph $G_T$  in Fig.~\ref{gnn-ts} has $V_p = \{v_1, v_2, v_3, \ldots, v_7\}$, $V_{\textit{seg}} = \{v_{\textit{seg}_1}, v_{\textit{seg}_2}, v_{\textit{seg}_3}\}$, and $E = \{(v_i, v_{\textit{seg}_j}) | 1 \leq i \leq 7, 1 \leq j \leq 3\}$. Next, we define a feature vectors of the point $p_i$ and segment $v_{\textit{seg}_j}$ node are  $\bm{h}_i$ and $\bm{h}_{\textit{seg}_j}$, respectively. 

% Compared to connecting all pairs of GPS points, our approach is computationally more efficient, avoiding unnecessary connections between unrelated nodes, and capturing the global information in a trajectory better.

{\color{black}The  trajectory partitioning and graph construction  methods we propose are designed to address the input limitations of the Transformer encoder. This is inspired by the encoding of long text, as discussed elsewhere~\cite{Longtrans,Longtrans2}. 
Unlike existing techniques that involve trajectory partitioning, we do not require goal-oriented segmentation. For example, the goal of one study~\cite{segtraj1} is to consider the homogeneity in the neighborhoods of space-time points, and the goal of another study  \cite{segtraj2} is to minimize the cost of multicut.}

\vspace{-2mm}
\subsubsection{GNN Importance Prediction}
\label{subsubsec:gnn importance prediction}

Using the trajectory graph $G_T$, we update the representation of trajectory nodes using a GNN. GNNs~\cite{gcn,gat,GNN1,GNN2,CHGNN} are neural networks that operate on graph-structured data, allowing for efficient aggregation of node information from edges. {\color{black} Here, we use a GNN consisting of  Graph Attention Network (GAT)~\cite{gat} layers.} GATs can learn complex relationships between nodes, and their attention mechanism allows the model to focus on specific nodes in a graph. Specifically, given a node $i$ and its neighbor node  set $\mathcal{N}_i$, we propagate the
features as:
% \vspace{-1mm}
\begin{equation}
\bm{g}_{i}^{(l)} = \texttt{GATLayer}(\bm{g}_i^{(l-1)},\bm{g}_{\mathcal{N}_i}^{(l-1)}),
\vspace{-1mm}
\end{equation}
where $\texttt{GATLayer}(\cdot)$ is the propagation function of a GAT layer, $1\leq l\leq L $, $L$ is the number of GAT layers, and  $\bm{g}_i^{(0)}=\bm{h}_i$.  
% Specifically, given a central node $i$ and a neighbor node $j$ with feature representations $\bm{h}_i$ and $\bm{h}_j$, respectively, the aggregation function for node $i$ is defined as
% \begin{equation}
% a_{ij} = \texttt{LeakyReLU}\left(\bm{W_a}^T [\texttt{concat}(\bm{h}_i, \bm{h}_j)]\right),
% \end{equation}
% where $\bm{W_a}$ is a learnable parameter vector, $\texttt{LeakyReLU}(\cdot) $is the leaky rectified linear unit activation function with a leaky coefficient, and $\texttt{concat}(\cdot, \cdot)$ denotes concatenation. Then, the attention coefficients are normalized using the softmax function:
% \begin{equation}
% \alpha_{ij} = \frac{\texttt{exp}(a_{ij})}{\sum_{k \in \mathcal{N}_i} \texttt{exp}(a_{ik})},
% \end{equation}
% where $\mathcal{N}_i$ is the set of neighboring nodes of node $i$. Finally, the aggregated representation of node $i$ is obtained as a weighted sum of the features of the neighboring nodes using the attention coefficients:
% \begin{equation}
% \bm{h}_i' = \sigma\left(\sum_{j \in \mathcal{N}_i} \alpha_{ij} \bm{h}_j\right),
% \end{equation}
% where $\sigma$ is an activation function. The aggregated representation $\bm{h}_i'$ of node $i$ can be updated as follows:
% \begin{equation}
% \bm{h}_i' = \texttt{ReLU}\left(\bm{W} \bm{h}_i' + \bm{b}\right),
% \end{equation}
% where $\bm{W}$ and $\bm{b}$ are learnable parameters for updating and $\texttt{ReLU}(\cdot)$ is the rectified linear unit activation function.

Using the $L$ layers of GAT, the node representation $\bm{g}_i = \bm{g}_i^{(L)}$ learns the representations of GPS points across the entire trajectory rather than just containing information within the current segment. To quantify the importance of each point, we analyze the distribution of  point representations and define two metrics for trajectory importance: uniqueness and globality.

\noindent\textbf{Uniqueness}. In a trajectory, due to the short sampling interval (usually only a few seconds), neighboring points are often highly similar, leading to redundancy, as illustrated by points $p_{2,2}$ and $p_{2,3}$ of $T_2$ in Fig.~\ref{intro}. To reduce this redundancy, we measure the uniqueness of points relative to the embeddings of neighboring points.
{\color{black} In the uniqueness calculation,  neighbors consist of the $k$ nearest trajectory points to $p_i $, where $k$ is a predefined parameter. These points are selected based on their cosine similarity to $p_i$'s representation vector; the higher the similarity, the closer the point to the target. This selection process ensures that the chosen neighbors are semantically similar to $p_i$, effectively filtering out points that significantly differ during the comparison. 
% Points with high uniqueness are valued more in the trajectory due to their distinct semantic characteristics; while those with low uniqueness, which contribute less to the global distinction and local representativeness of the trajectory, are prioritized for removal.
}
% We assess the importance of each point by evaluating its uniqueness and globality. Points with high uniqueness are valued more in the trajectory due to their distinct semantic characteristics; while those with low uniqueness, which contribute less to the global distinction and local representativeness of the trajectory, are prioritized for removal.
% Having higher uniqueness indicates that a GPS point may contain unique semantic information, making it more important. 
The uniqueness of a point $p_i$ represented by $\bm{g}^{(L)}_i$ is calculated as follows:
\vspace{-2mm}
\begin{equation}
 \mathcal{L}_{\mathit{uni}}(\bm{g}_i) = \frac{1}{|\mathcal{N}_i|} \sum_{j \in \mathcal{N}_i} \| \bm{g}_i - \bm{g}_j \|_2,
 \vspace{-2mm}
\end{equation}
where $\mathcal{N}_i$ is the neighbors of point $p_i$ in $T$.

% \begin{figure*}[t]
% \centering
% \includegraphics[width=\textwidth]{figures/diff-ts.pdf}
% \caption{Overview of the training and inference processes of Diff-TS.}
% \label{diff-ts}
% \vspace{-4mm} 
% \end{figure*}

\noindent\textbf{Globality}. The goal of trajectory simplification is to select points that best represent the original trajectory. Therefore, points that capture the semantics of the trajectory are important, such as the starting point $p_{4,1}$, end point $p_{4,6}$, and turning point $p_{4,4}$ of  $T_4$ in Fig.~\ref{intro}. Hence, we define globality to capture the similarity of a point with all other points in its trajectory, using the representations of the points. If a point's globality is high, the point is important. Conversely, if the similarity between a point and all other points is low, the GPS point may be noisy or simply wrong. Globality is defined as follows:
\vspace{-3mm}
\begin{equation}
 \mathcal{L}_{\mathit{glob}}(\bm{g}_i) = \texttt{log}\left(\frac{1}{|T|-1} \sum_{j=1,j \neq i}^{|T|} e^{-2 \| \bm{g}_i - \bm{g}_j \|^2_2}\right),
 \vspace{-2mm}
\end{equation}
where $|T|$ is the total number of  points in trajectory $T$.

% For example, in Figure 3, node $v6$ has high globality as it represents the smooth trajectory path in the latter part of the trajectory. $v3$ has high uniqueness because it is far from $v4$ and $v5$, representing a clear turn in the trajectory.

To better capture both semantic similarities and relationships between  points, we align uniqueness and globality using contrastive learning. Contrastive learning~\cite{contrastivelearning} is a self-supervised training method based on the alignment and uniformity of the distribution of all objects. 
% Specifically, an object $o$ has a positive example set $P$ and a negative example set $N$. The positive examples are objects of the same class as $o$, and the negative examples are objects belonging to different classes.
% Alignment measures the similarity between positive examples, while uniformity measures the distance between all objects. The target of contrastive learning is to minimize the distances between objects in the same class and to maximize the distances between all objects, ensuring that the objects are evenly distributed in the embedding space.

During training, we treat the neighboring points of a point $p_i$ as its positive examples to encourage similarity in the embedding space (i.e., minimizing $ \mathcal{L}_{\mathit{uni}}$). Conversely, we treat distant points from $p_i$ as negative examples to ensure separation (i.e. minimizing $ \mathcal{L}_{\mathit{glob}}$). Therefore, the loss function is:
\vspace{-1mm}
\begin{equation}
 \mathcal{L}_{\mathit{con}}(\bm{g}_i) = \mathcal{L}_{\mathit{uni}}(\bm{g}_i) + \lambda_1 \mathcal{L}_{\mathit{glob}}(\bm{g}_i),
 \label{gnnts-con loss}
 \vspace{-1mm}
\end{equation}
where $\lambda_1$ is a hyperparameter balancing the two terms. We fix it at 0.5 following the literature~\cite{contrastivelearning}.

When measuring the importance of point $p_i$, we consider  uniqueness and globality simultaneously and define the importance:
\vspace{-1mm}
\begin{equation}
I_{p_i}= \mathcal{L}_{\mathit{uni}}(\bm{g}_i)  \times \mathcal{L}_{\mathit{glob}}(\bm{g}_i) + \epsilon,
\label{importance}
\vspace{-1mm}
\end{equation}
where $\epsilon$ eliminates any zero values in the importance scores.
% Thus, the importance of a point increases only when both uniqueness and globality are relatively high, reducing the influence of redundant and noisy data. 
{\color{black}We assess   the importance of each point by evaluating its uniqueness and globality. Points with high uniqueness are valued more due to their distinct semantic characteristics; while those with low uniqueness, which contribute less to the global distinction and local representativeness of the trajectory, are prioritized for removal.}

We employ a mutual learning algorithm for training, where the results of the Diff-TS are used as amplified labels compared to the importance generated by GNN-TS, serving as the ML loss. This is covered in Sec.~\ref{mutallearning}.

{\color{black}\noindent\textbf{Discussion.} 
% GNN-TS includes a pre-trained T-Bert with an encoding layer that uses a Transformer encoder. Although this encoder is effective at learning sequence representations, its efficiency is affected by the sequence length. 
% The time and space complexity of the Transformer encoder are due to its multi-head attention modules and feed-forward neural networks. For a trajectory of length $|T|$ encoded into $d$ dimensions, the time complexity of a single-layer Transformer encoder is $O(|T|^2 d + |T|d^2)$, and the space complexity is $O(|T|^2 + |T|d)$. With a fixed $d$, the time and space complexity of encoding are quadratic in the trajectory length.
Since long trajectories are common in databases (see Table~\ref{dataset}), the space complexity of encoding an entire trajectory using a Transformer encoder is very costly and makes encoding inefficient.
For a trajectory of length $|T|$ encoded into $d$ dimensions, the time complexity of a single-layer Transformer encoder is $O(|T|^2 d + |T|d^2)$, and the space complexity is $O(|T|^2 + |T|d)$.  
% With a fixed $d$, the time and space complexity of encoding are quadratic in the trajectory length.
Therefore, GNN-TS  first segments long trajectories for encoding. Long trajectories are typically segmented into chunks of length $w$ based on the available computational capabilities. For a trajectory of length $|T|$, the number of segments is $k = \lceil |T|/w \rceil$. If all segments are encoded serially, their time complexity is $O(k w^2  d + k w d^2)$.  Due to serial encoding, the space complexity is $O(w^2 + w d) < O(|T|^2 + |T|d)$. 
Next, we construct a trajectory graph where 
% points from different segments are connected through segment nodes. 
the number of nodes is $|T| + k$ and the number of edges is $k|T|$. Finally, we use a GAT to fine-tune the encoding of nodes
% . in the trajectory graph, avoiding information loss introduced by the segmentation
, and predict the importance of trajectory points. Since the trajectory graph is sparse, with each trajectory node connected only to $k$ segment nodes, the encoding time complexity is approximately $O((|T| + k) d^2 + k|T| d)$. Therefore, the time complexity for GNN-TS to encode a trajectory is $O(k w^2  d + k w d^2) + O((|T| + k) d^2 + k |T| d) = O((kw^2+k|T|)d + (kw+T+k)d^2)$. When $\frac{w^2}{w-1} + \frac{w+1}{w-1}d<|T|$, the time complexity of GNN-TS is lower than when using only a Transformer encoder to generate a trajectory embedding. Thus, for long trajectories (e.g., the average trajectory length exceeds 1000), segmenting with a reasonable $w$ can improve encoding efficiency markedly.
%解释w虽然可以自由选择，以降低编码时间，但是不宜选择过小，因为过小会影响effectiveness。
However, although GNN-TS connects trajectory nodes from different segments through the trajectory graph, segmentation encoding introduces information loss. 
% When a Transformer encoder encodes an entire trajectory, each point's embedding learns information from all other points in the trajectory. GNN-TS, using segmented encoding, integrates and fine-tunes through GAT, making each point consider fine-grained information within its segment and coarse-grained information from other segments. 
To limit this loss, $w$ in GNN-TS is generally not very small (e.g., 500 in this paper).
Thus, compared to a trajectory graph containing only trajectory points with $|T|$ nodes, the number of additional segment nodes $\lceil |T|/w \rceil \ll |T|$ does not impact the GAT encoding efficiency substantially.
}

% \begin{figure}[t]
% \centering
% \includegraphics[width=0.47\textwidth]{figures/diff-ts-half.pdf}
% \caption{Overview of the training process of Diff-TS.}
% \label{diff-ts}
% \vspace{-4mm} 
% \end{figure}

\vspace{-2mm}
\subsection{Diffusion-based Trajectory Simplification Model}

% Although the GNN-TS measures the importance of each  point in a trajectory based on uniqueness and globality, similar points may have similar importance. 
% % While sampling based on importance can preserve key information of a trajectory and does not significantly affect the simplification performance at high sampling ratios, it can lead to information loss at low sampling ratios due to redundancy. 
% % This suggests that generative models are suitable for use at low sampling ratios. Generative models~\cite{generation1,generation2} use the hidden information of an original trajectory to generate simplified points. The diffusion model~\cite{generation2} is a state-of-the-art generative model that performs well on many tasks. Therefore, we propose Diff-TS, which transforms trajectory simplification into a conditional generation task, using an original trajectory as a condition to generate a simplified trajectory. However, the diffusion model's generation process has a high computational cost due to the model's iterative denoising process. Although it is possible to accelerate the processing by skipping iterations, the cost of the diffusion model still exceeds that of neural network models that do not require iteration. 
% Therefore, our goal is to generate low compression rate simplified trajectories using Diff-TS and then incorporate these as amplified labels in the training of GNN-TS to enhance the importance of prediction. Diff-TS is ignored during simplification. 
% We proceed to cover the training and inference processes of Diff-TS.

\subsubsection{Training Process}
% The training process of Diff-TS is shown in Fig.~\ref{diff-ts}. 
During training, since we lack ground truth optimal simplified trajectories, we instead employ a high compression rate $cr_\mathit{high}$ simplified trajectories $T*$, sampled based on the importance generated by GNN-TS. We set $0.5 \leq cr_\mathit{high} < 1$ and assume that $T^*$ retains a considerable amount of trajectory information and can serve as soft labels for training. We concatenate the original trajectory $T$ with the simplified trajectory $T^*$ and then encode the trajectory to generate trajectory embeddings. Next, the trajectory is fed into a diffusion module. Finally, the model parameters are updated using a joint diffusion and similarity loss.

\noindent\textbf{Trajectory Encoding}.
We use an Transformer to encode the concatenated trajectory $T_\mathit{concat}$ mapping the points into initial representation vectors $\bm{H}_{T_\mathit{concat}} = [\texttt{concat}(\bm{H}_{T}, \bm{H}_{T^*})]$, where $\bm{H}_{T} = [\bm{h}_1, \ldots, \bm{h}_n]$, $\bm{H}_{T^*} = [\bm{h}^*_1, \ldots, \bm{h}^*_{z}]$, $n=|T|$, and $z=|T^*|$.
% Here, for long trajectories, we can employ a segmentation strategy similar to that used  for question-answering in NLP~\cite{wolf-etal-2020-transformers}, partitioning the original trajectory into segments, with each segment concatenated with the simplified trajectory for individual training tasks. Due to space limitations, we assume that trajectories can be encoded all at once without partitioning.

\noindent\textbf{Diffusion Module}.
After obtaining the input encoding $\bm{H}_{T_\mathit{concat}}$, the continuous diffusion model conditionally generates embeddings for the simplified trajectory. The diffusion model consists of a forward and a reverse process. 

\noindent(i) \textit{Forward process}: In each forward step $q(\bm{H}_{\gamma}|(\bm{H}_{\gamma - 1})$, we gradually inject Gaussian noise $\epsilon \sim N(0,I)$ into the hidden state $\bm{H}_{\gamma - 1}$ from the previous step to obtain $\bm{H}_{\gamma}$. Inspired by an existing study~\cite{diffuseq}, we only apply noise to $\bm{H}_{T^*}$, allowing for conditional modeling by the diffusion model. After $\gamma$ forward steps, a noisy representation $\bm{H}_\gamma$ is obtained: 
\vspace{-3mm}
\begin{equation}
\begin{split}
\bm{H}_\gamma = [\texttt{concat}(\bm{H}_T, \bm{H}_{T^*}^\gamma)],  \quad\bm{H}_{T^*}^\gamma = N(\sqrt{1-\beta_\gamma}\bm{H}_{T^*}^{\gamma-1}, \beta_\gamma \bm{I}),
\end{split}
\vspace{-3mm}
\end{equation}
 where   $\gamma \in \{1,2,...,\Gamma \}$, $\Gamma$ is the total number of diffusion steps, $\bm{H}_{T^*}^{0} = N(\bm{H}_{T^*},\beta_0\bm{I})$,   and $\beta_\gamma \in \{\beta_0, \beta_1, \beta2, \ldots, \beta_\Gamma\}$, where $\beta_0<\beta_1 < \beta2< \ldots < \beta_\Gamma$, is a hyperparameter.

\noindent(ii) \textit{Reverse process}: During the training, the goal of the reverse process is to recover the original $\bm{H}_{T^*}$ through denoising. Once the noisy representation of the simplified trajectory part is obtained, we perform the reverse process to remove the noise conditioned on the previous step's trajectory representation:
\vspace{-1mm}
\begin{equation}
    p_\theta(\bm{H}_{T^*}^{\gamma-1}|\bm{H}_{T^*}^{\gamma}) = N(\bm{H}_{T^*}^{\gamma-1}; \mu_\theta(\bm{H}_{T^*}^{\gamma}), \sigma^2_\theta(\gamma)\bm{I}),
    \vspace{-1mm}
\end{equation}
where $\mu_\theta(\cdot)$ and $\sigma^2_\theta(\cdot)$ are  models for predicting the mean and standard deviation during the forward process $q((\bm{H}_{\gamma-1}|\bm{H}_{\gamma})$. Here we use stacked Transformer layers as the noise prediction model.

\noindent\textbf{Training Loss}. After $\Gamma$ rounds of the backward process, the recovered representation vectors of the simplified trajectory are $\hat{\bm{H}}_{T^*} = [\hat{\bm{h}}^*_1, \ldots, \hat{\bm{h}}^*_{z}]$. We train the diffusion model using the diffusion loss $\mathcal{L}_{\mathit{diff}}$ in an existing study~\cite{diffuseq}.

% \vspace{-2mm}
% \begin{equation}
%  \mathcal{L}_{\mathit{diff}} = \sum_{\gamma=2}^{\Gamma} ||\bm{H}_{T^*}^0 - \hat{f_\theta}(\bm{H}_{T^*}^\gamma,\gamma)||^2 + ||\bm{H}_{T^*} - \hat{f_\theta}(\bm{H}_{T^*}^\gamma,1)||^2 +  \mathcal{R}(\bm{H}_{T^*}^0),
%  \vspace{-2mm}
% \end{equation}
% where $\hat{f_\theta}(\cdot)$ is the function reconstructing $\bm{H}_{T^*}^0$ from the $\gamma$-th step, i.e., $\mu_\theta(\cdot)$ and $\sigma^2_\theta(\cdot)$. $\mathcal{R}(\cdot)$ is the L2 regularization term.

We also introduce a diversity loss to reduce the similarity among generated simplified  points, thus reducing redundant information in generated trajectories. We define the diversity loss as:
\begin{equation}
\mathcal{L}_{\mathit{div}}(\hat{\bm{h}}_i) = \texttt{log}\left(\frac{1}{|T^*|-1} \sum_{j=1,j \neq i}^{n} e^{-2 \|\hat{\bm{h}}_i - \hat{\bm{h}}_j \|^2_2}\right),
\end{equation}
where $|T^*|$ is the length of the simplified trajectory.
% Minimizing the diversity loss helps make  generated  points different.
Finally, we combine the diffusion loss and the diversity loss.
% , making a trade-off between learning the soft labels provided by the GNN-TS and obtaining diverse generated trajectory nodes. 
The training loss function is:
\vspace{-2mm}
\begin{equation}
\mathcal{L}_{\mathit{Diff-TS}} = \mathcal{L}_{\mathit{diff}} + \lambda_2\mathcal{L}_{\mathit{div}},
\label{diff-ts loss}
\vspace{-1mm}
\end{equation}
where $\lambda_2$ is a hyperparameter that balances the two losses.
\vspace{-1mm}
\subsubsection{Inference Process}
For inference, Diff-TS first obtains the representation $\bm{H}_T$ of the original trajectory $T$, and then adds Gaussian noise to obtain the initial simplified trajectory embedding through a Markov transition.
% \begin{equation}
% \bm{H}_T^{0} \sim N(\bm{H}_T, \beta_0 \bm{I})
% \end{equation}
Next, we randomly sample $\alpha$ ($\alpha \ll n$) points of Gaussian noise embeddings as the initial simplified trajectory embeddings input $\bm{H}_{T'}^{0} \in \mathbb{R}^{m \times d}$, where $d$ is the dimensionality of the embeddings.
The concatenated input to the diffusion model is $\bm{H}^{\#}_{0} = [\texttt{concat}(\bm{H}_T^{0}, \bm{H}_{T'}^{0})]$. 
The diffusion model utilizes the learned reverse denoising process (generation process) to iteratively remove Gaussian noise. 
After $\Gamma$ steps of diffusion, we obtain the output simplified trajectory representation $\tilde{\bm{H}}_{T'} = [\tilde{\bm{h}}_{s_1}, \tilde{\bm{h}}_{s_2}, \ldots, \tilde{\bm{h}}_{s_\alpha}]$. Then, Diff-TS computes the matching between the generated summary representation $\tilde{\bm{h}}_i$ and the original trajectory $\bm{H}_T$ representation, selects the point with the highest score, and adds it to the simplified trajectory. Finally, we obtain the simplified trajectory.

\vspace{-2mm}
\subsection{Mutual Learning and Simplification}
\label{mutallearning}

% \begin{figure}
% \centering
% \includegraphics[width=8.5cm]{figures/ml.pdf}
% \caption{The training and simplification process of \textsc{MLSimp}.}
% \label{ml}
% \vspace{-5mm} 
% \end{figure}

{\color{black} GNN-TS    can be viewed as an \textit{extractive} simplification model. 
 Similar to extractive summarization in NLP, GNN-TS selects the $l_s$ most important points as the simplified trajectory 
 by evaluating the importance of all points. 
 % The advantages of extractive simplification are that (i) the important trajectory points are guaranteed to exist in the original trajectory and that (ii) the retained points preserve some of the original trajectory information, such as shape and direction. 
 However, extractive simplification may retain redundant points, as the importance is fixed and is unaffected by the importance of already retained points.
Diff-TS can be considered a \textit{generative} simplification model. It generates a simplified trajectory by learning from the original one. 
% A well-trained generative model can produce a simplified trajectory that is semantically similar to the original and can reduce redundancy during generation.
However,  (i) it cannot guarantee that  generated points exist in the original trajectory, and (ii) the generated trajectory may deviate considerably from the original in terms of information---it may contain noise and may fail to maintain the original shape.

To address the shortcomings of GNN-TS, 
% inspired by the combination of extractive and generative methods in NLP for text summarization, 
we use ML to integrate the two models for trajectory simplification. The unsupervised GNN-TS provides training labels for the supervised Diff-TS, ensuring that the simplified trajectories generated by Diff-TS do not deviate substantially from the original trajectories. The trajectories generated by Diff-TS are regarded as amplified signals that prompt GNN-TS to retain important points that might be lost due to redundancy. 
% GNN-TS adaptively filters the information from Diff-TS based on importance scores, selectively encouraging the retention of important trajectory points within the amplified signals, thereby reducing redundant points under the same compression rate.
} Next, we present the framework for mutual learning between the two models and the process of simplification using the trained models.
% The training and simplification processes are illustrated in Fig.~\ref{ml}.
% During training, we employ two models for mutual learning, while during inference, we only use the GNN-TS. We adjust the importance based on queries and finally obtain the simplified database through sampling.

\vspace{-2mm}
\subsubsection{Mutual Learning Training}
We iteratively train two simplification models. This is inspired by an existing sudy~\cite{mutuallearning}, where multiple networks learn from each other in a supervised setting. 
% In our task, GNN-TS is an unsupervised model, while Diff-TS is a supervised model. Therefore, the key to training Diff-TS without labeled data is to generate training data. 
% As shown in Algorithm~\ref{train_algor}, 
We train two models with two stages. 

In the first stage, we train the GNN-TS using only the contrastive loss in a self-supervised manner, without using the mutual learning loss. The trained GNN-TS  is used to infer a high compression rate simplified database.
% ., retaining simplified data containing the original trajectory information.
In the second stage, we perform mutual learning for both models. We use the simplified database as the labels for Diff-TS training.
% Then, we train the Diff-TS model using these labels.
After training, Diff-TS infers the $\alpha$ most important points from trajectory $T$ to add these to the simplified trajectory $T^\#$, generating amplified labels $Y_T=\{y_1, ..., y_n\}$. If $p_i \in T^\#$, where $p_i \in T$, then $y_i=1$; otherwise $y_i=0$. We retrain the GNN-TS, introducing an ML loss: 
\vspace{-2mm}
\begin{equation}
    \mathcal{L}_{\mathit{ml}} = \sum_{i=1}^n -y_i \texttt{log}{I_{p_i}} - (1-y_i) \texttt{log}(1-I_{p_i}),
    \vspace{-2mm}
\end{equation}
where $I_{p_i}$ is the importance predicted by GNN-TS. The overall loss function of GNN-TS during the mutual learning stage is as follows:
\vspace{-1mm}
\begin{equation}
\mathcal{L}_{\mathit{GNN-TS}} = \mathcal{L}_{\mathit{con}} + \lambda_3\mathcal{L}_{\mathit{ml}},
\label{gnn-ts loss}
\vspace{-1mm}
\end{equation}
where $\lambda_3$ is a hyperparameter for balancing the two losses.

% \begin{algorithm}[H]
% \caption{Mutual Learning Training}
% \label{train_algor}
% \begin{algorithmic}[1]
% \STATE \textbf{Input:} Trajectory database $D$.
% \STATE Initialize GNN-TS and Diff-TS models.
% \STATE  \textit{\# Stage 1}
% \STATE Train GNN-TS using Eq.~\ref{gnnts-con loss}.
% \STATE Sample simplified trajectory database $D_{\mathit{simp}}$ based on the importance predicted by GNN-TS.
% \STATE \textit{ \# Stage 2}
% \FOR{$R$ epochs}
%     \STATE Train Diff-TS using $D$ and $D_{\mathit{simp}}$ using Eq.~\ref{diff-ts loss}.
%     \STATE Diff-TS infers amplified labels $Y_T$ for each trajectory.
%     \STATE Train GNN-TS using Eq.~\ref{gnn-ts loss}.
% \ENDFOR
% \end{algorithmic}
% \end{algorithm}

\vspace{-2mm}
\subsubsection{Simplification}
\label{sec:simp}
Upon training, we use the lightweight GNN-TS for inference. Given a dataset $D$ to be simplified, we first use GNN-TS to predict the importance $I_{p_i}$ of each point $p_i \in D$ and perform global normalization. However, the effectiveness of the simplified database on range queries is not ideal, as shown in Fig.~\ref{Ablation Study}. 
% This is because range queries depend on the distribution of queries. As shown in Figs.~\ref{query} (a) and (b), to ensure the performance of range queries, at least one point within the query range must be retained, which is not supported well in our simplification process.
Therefore, to support range queries better, we include range query-based importance adjustment.

\noindent\textbf{Importance Adjustment:} Since previous queries are not available, we synthesize a workload of range queries $Q^w$, where each query location is randomly sampled by following some distribution (e.g., a data distribution) as in the literature~\cite{QDTS}. Through the queries, we obtain a set of query results. We divide the spatial and temporal ranges into coarse-grained grid cells, where the cells hit by the query results are assigned an importance of 1, and the remaining cells are assigned 0. Finally, we normalize the query-based importance of each cell. The query-based importance $I_{p_i}^q$ of point $p_i$ in $D$ is consistent with the importance of the cell it belongs to. The adjusted importance of point $p_i$ is defined as follows:
\vspace{-1mm}
{\color{black}
\begin{equation}
    I_{p_i}^\mathit{adj} = (1-\delta) I_{p_i} + \delta I_{p_i}^q,
    \vspace{-1mm}
\end{equation}}
where $\delta$ is the adjustment ratio based on queries. 

Finally, we directly sample $m$  points with the adjusted importance to construct the simplified trajectory database, with compression rate $cr = m/n$.

\vspace{-2mm}
\subsubsection{Complexity}
\label{sec:4.3.3}
{\color{black}   Following existing learning-based trajectory  simplification models~\cite{RLTS, S3, QDTS}, we focus on analyzing the time complexity of simplification inference and exclude the training. 
 
%压缩了为什么不分析training time的原因
% for three primary reasons. (i) Only a small portion of the trajectories are typically used for training. For example, our experiments use only 4.02\% of the total data for training. (ii) The training of \textsc{MLSimp}  is conducted once in an offline process, and the resulting model is then applied to simplify all remaining trajectories. (iii) The time complexity of training varies according to factors such as the number of model parameters, which can be adjusted to meet specific requirements.

For a database with $m$ trajectories, where $w$ is the maximum input length of T-Bert and $|D'|$ is the storage budget of the simplified database, the simplification process includes the following steps: (i) \textit{\textbf{Trajectory encoding}} segments a trajectory, resulting in $\overline{n}/w$ encodings for a trajectory of length $\overline{n}$, with time complexity  $O(\overline{n}/w)$.
(ii) \textit{\textbf{Trajectory graph construction}} is a logical process where, in practice, it is not necessary to construct the actual graph structure.
% Instead, the feature matrix of the graph nodes and the edge index matrix of the graph are provided as input to the importance prediction. 
Thus, the time complexity of trajectory graph construction is $O(1)$. 
(iii) \textit{\textbf{GNN importance prediction}} consists of 
% the GAT layers//It is not good to say that prediction consists of layers, and layers are not a step, and layers are not matrix computations. Revise.
{\color{black}GAT's attention matrix computation and attention-based node feature aggregation,}
and importance calculation (uniqueness and globality computations). All steps are matrix computations on the feature matrix and the edge index matrix.
% Since these matrix operations can be parallelized, all points in a trajectory can be processed in a single prediction pass to calculate the importance of all points. 
Following learning-based trajectory studies~\cite{RLTS,S3,QDTS,timecomplexity1,timecomplexity2}, the time complexity of GNN importance prediction is $O(1)$.
(iv) \textit{\textbf{Importance adjustment}} of each trajectory point is calculated in parallel based on the adjustment values derived from the simulated range query results.
% since the importance adjustments of trajectory points do not affect each other.  
Thus, the time complexity of importance adjustment is $O(1)$.
(v) \textit{\textbf{Sampling}} is performed after determining the importance of all trajectories.  Thus, the sampling time complexity for the entire database is $O(1)$.
Therefore, the overall simplification time complexity is $m \cdot O(\overline{n}/w + 1 + 1 + 1) + O(1) = O(m \cdot \overline{n}/w)$.}

% Based on the above, we observe that \textsc{MLSimp} is independent of the compressed storage space, i.e., it is not directly affected by the compression rate. Instead, its performance is affected mainly by the number and length of trajectories, as well as the maximum sequence length acceptable to T-Bert. The maximum sequence length acceptable to T-Bert, in turn, depends on hardware constraints and the settings of the temporal encoding.

%% file: src/experiments.tex
\section{Experimental Study}
\label{sec:5}
\subsection{Experimental Setup}
\subsubsection{Datasets}

 We evaluate \textsc{MLSimp} on three publicly available real-world datasets. Table~\ref{dataset} provides statistics of the datasets.
 
 \noindent\textbf{Geolife\footnote{https://www.microsoft.com/en-us/research/publication/geolife-gps-trajectory-dataset-user-guide/}} is a GPS trajectory dataset collected in the Geolife project from 182 users over a period of more than three years. 
 % These trajectories are recorded by  GPS loggers and GPS-phones and have a variety of sampling rates.
 
 \noindent\textbf{T-Drive\footnote{https://www.microsoft.com/en-us/research/publication/t-drive-trajectory-data-sample/}} contains one-week trajectories from 10,357 taxis. The dataset contains about 17 million points.
 % , and the total distance of the trajectories is 9 million km.
 
\noindent\textbf{OSM\footnote{https://star.cs.ucr.edu/?OSM/GPS\#center=43.6,-56.1\&zoom=2}} contains 500 million GPS trajectories with 3 billion points shared by the community on OpenStreetMap over 9 years.
 
\begin{table}[t]
\centering
\caption{Trajectory dataset statistics.}
\renewcommand\tabcolsep{5pt}
\begin{tabular}{lccc}
\toprule
\textbf{Statistic} & \textbf{Geolife} & \textbf{T-Drive} & \textbf{OSM} \\
\midrule
\# of trajectories & 17,621 & 10,359 & 513,380 \\
Total \# of points & 24,876,978 & 17,740,902 & 2,913,478,785 \\
Ave. \# of pts per traj & 1,433 & 1,713 & 5,675 \\
Sampling rate & 1s--5s & 177s & 53.5s \\
Average length & 9.96m & 623m & 180m \\
\bottomrule
\end{tabular}
\label{dataset}
\vspace{-7mm}
\end{table}

\subsubsection{Competitors}
Since the existing algorithms for the QDTS problem only include RL4QDTS, the existing EDTS algorithm is also considered as a potential baseline. Guided by the skyline study in an existing study~\cite{QDTS}, we select the most competitive existing EDTS method as the target baseline. 

\noindent\textbf{Top-Down(E, PED) and Top-Down(W, PED)}~\cite{DPhull} start with the entire trajectory and iteratively divides it into segments until each segment meets an error threshold. In the following section, we use ``TD'' to represent ``Top-Down.''

\noindent\textbf{Bottom-Up(E, PED), Bottom-Up(E, SED), and Bottom-Up(E, DAD)}~\cite{bottomup} start with the individual points of a trajectory and merge them iteratively to meet the compression rate. In the following section, we use ``BU'' to represent ``Bottom-Up.''

\noindent\textbf{RLTS(E, SED)}~\cite{RLTS} employs reinforcement learning to iteratively simplify trajectories. It formulates trajectory simplification as a sequential decision-making problem, where an agent decides which points to retain or remove based on rewards associated with simplification decisions. 

\noindent\textbf{S3}~\cite{S3} uses two Seq2Seq models for training and generates simplified trajectories by feeding the original trajectory into an encoder-decoder consisting of Bi-LSTM units as a compressor.

\begin{figure}[t]
{
\includegraphics[width=5cm]{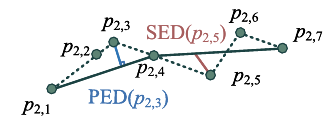}}

 \caption{\color{black}An example of the SED and PED errors, where (i) circles denote trajectory points, (ii) green solid and dash lines denote the original and simplification trajectory, respectively, and (iii) the red and blue solid lines denote the SED of $p_{2,5}$ and the PED  of $p_{2,3}$, respectively. }
 \label{fig:anchor}
   \vspace{-6mm}
   \end{figure}
   
\noindent\textbf{RL4QDTS}~\cite{QDTS} performs QDTS by reinforcement learning. It considers range queries and simplifies trajectories 
% accordingly to ensure that the simplified data remains suitable 
for queries.

 Top-Down(E, PED), Top-Down(W, PED, Bottom-Up(E, PED), Bottom-Up(E, SED), Bottom-Up(E, DAD), RLTS(E, SED), and S3 are EDTS baselines. RL4QDTS is a  QDTS baseline. ``E'' and ``W'' denote simplifying each trajectory in the database and considering the entire database as a whole by inserting or deleting points between all points in the database, respectively. 
 % PED~\cite{PED1,PED2,PED3}, SED~\cite{SED1,SED2,SED3SAD1}, and DAD~\cite{DAD1,DAD2} denote the error metrics used by the methods.
 % Since S3 is a trajectory generation framework that simplifies each trajectory without errors, it is not suitable for the above settings.

  \begin{figure*}
\centering
         \subfloat[ {Range query}]{
          \includegraphics[width=3.9 cm]{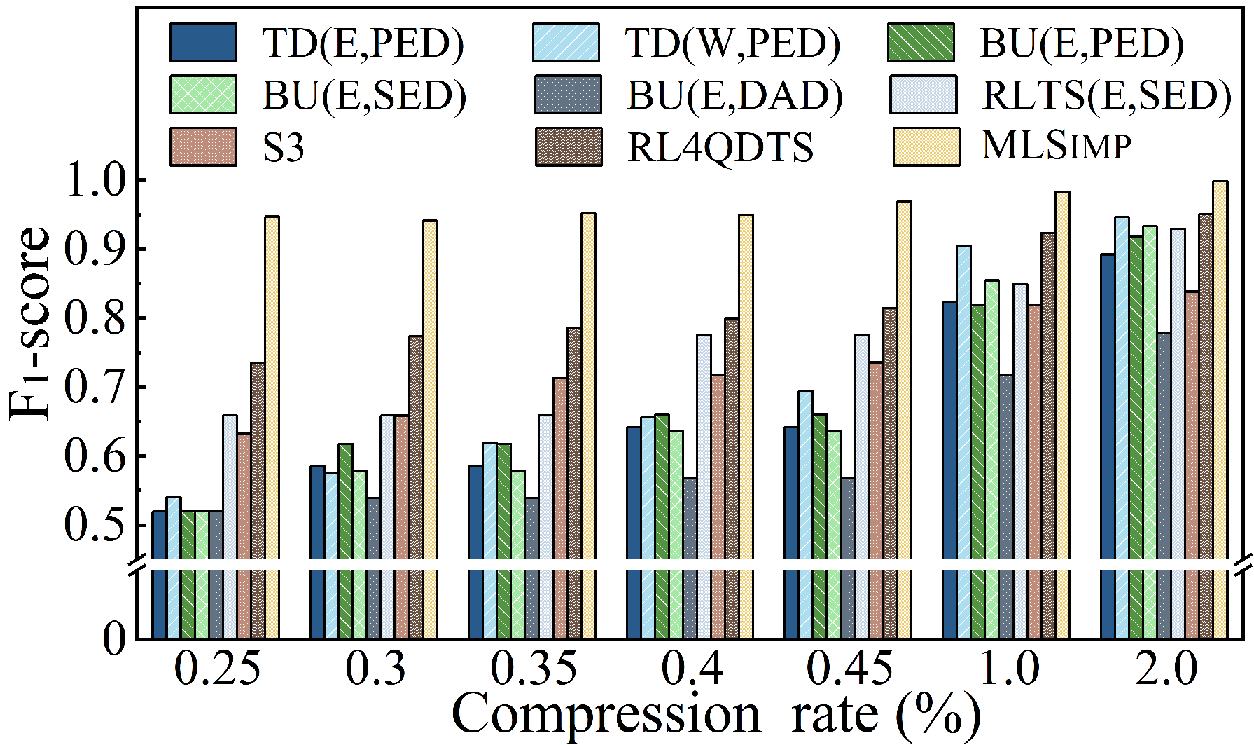}}   
           \quad
         \subfloat[ {$k$NN query}]{
          \includegraphics[width=3.9cm]{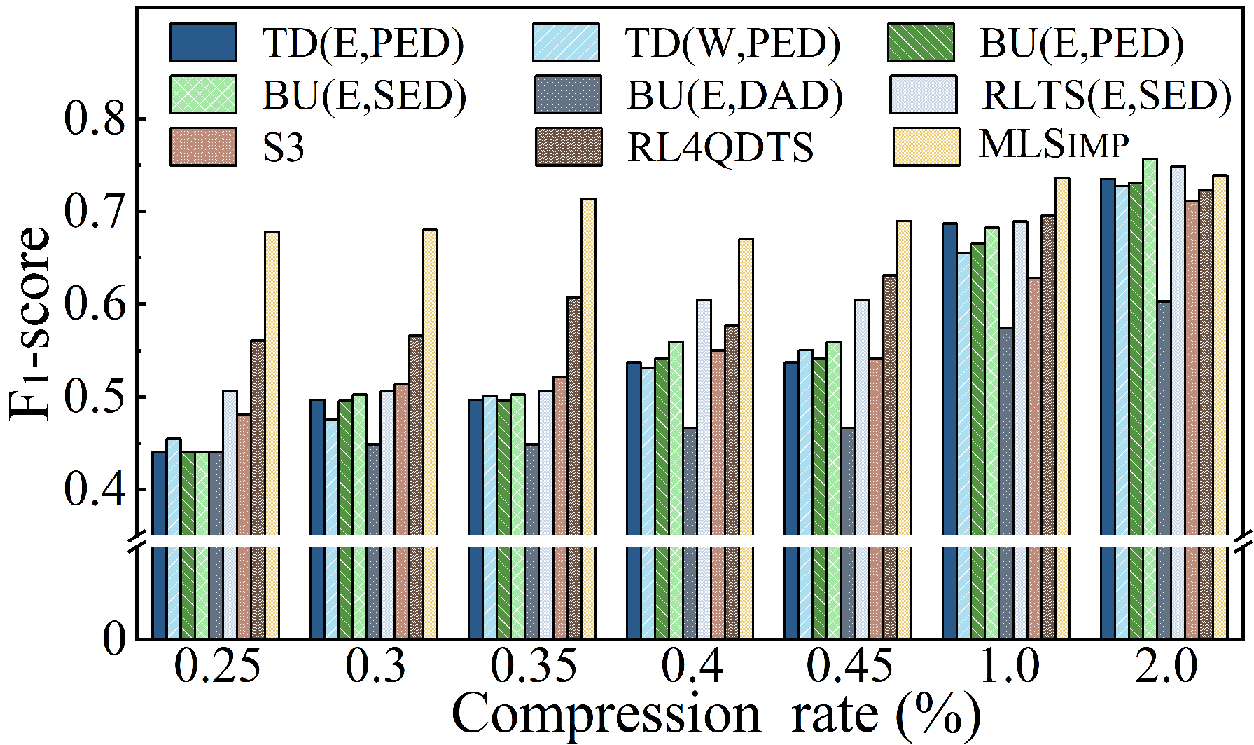}}
          \quad
                   \subfloat[ {Similarity query}]{
          \includegraphics[width=3.9 cm]{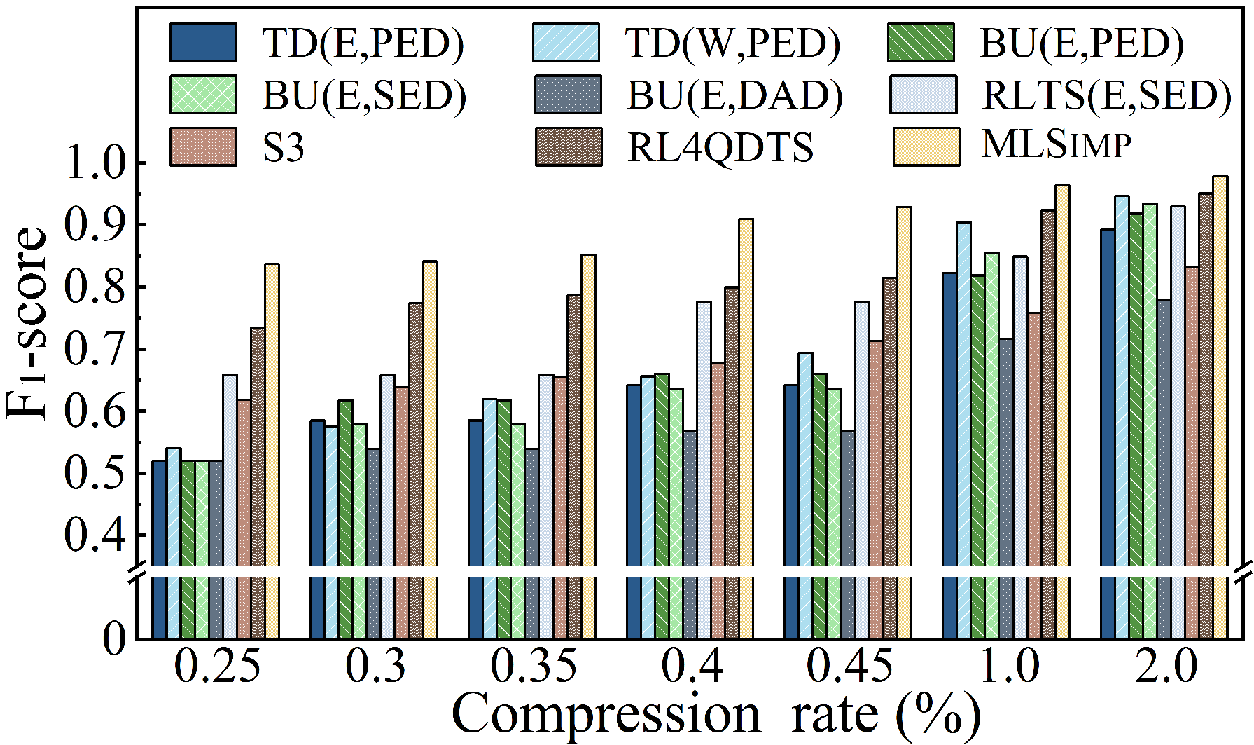}} 
          \quad
                   \subfloat[ {Clustering}]{
                   \includegraphics[width=3.9cm]{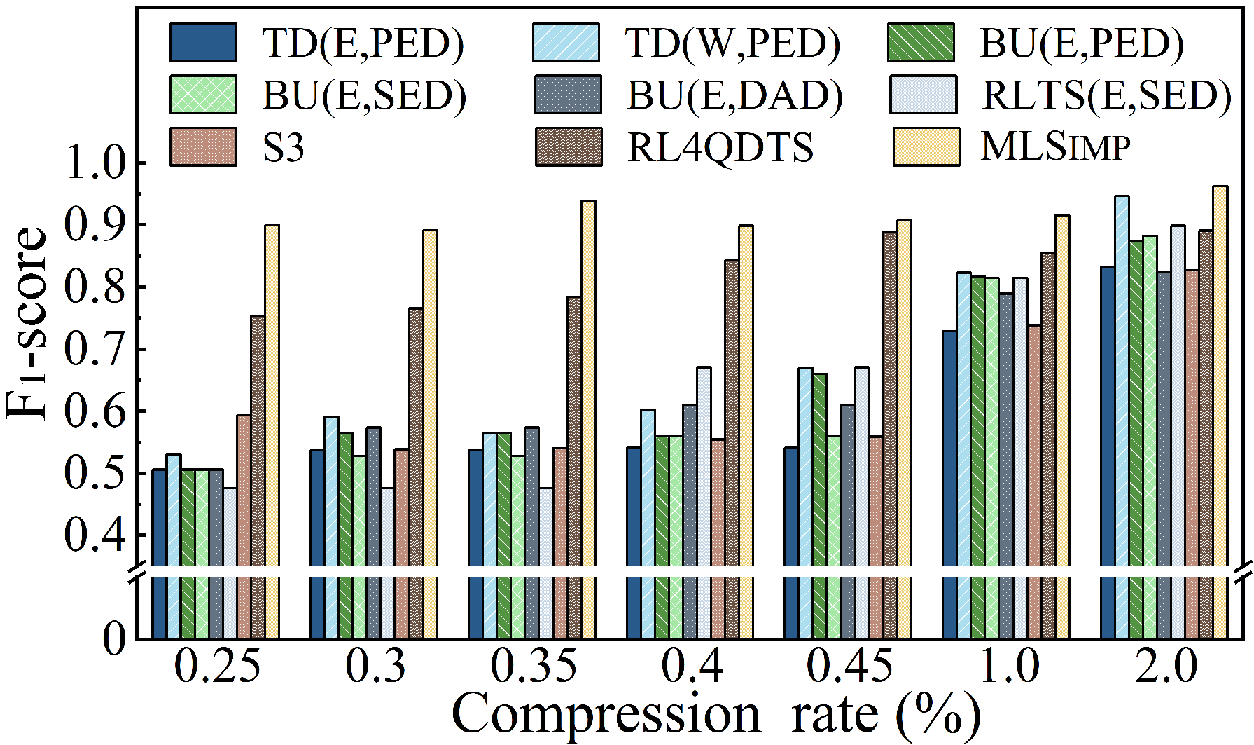}}
          \vspace{-3mm}

                   \subfloat[ {Range query}]{
                   \includegraphics[width=3.9cm]{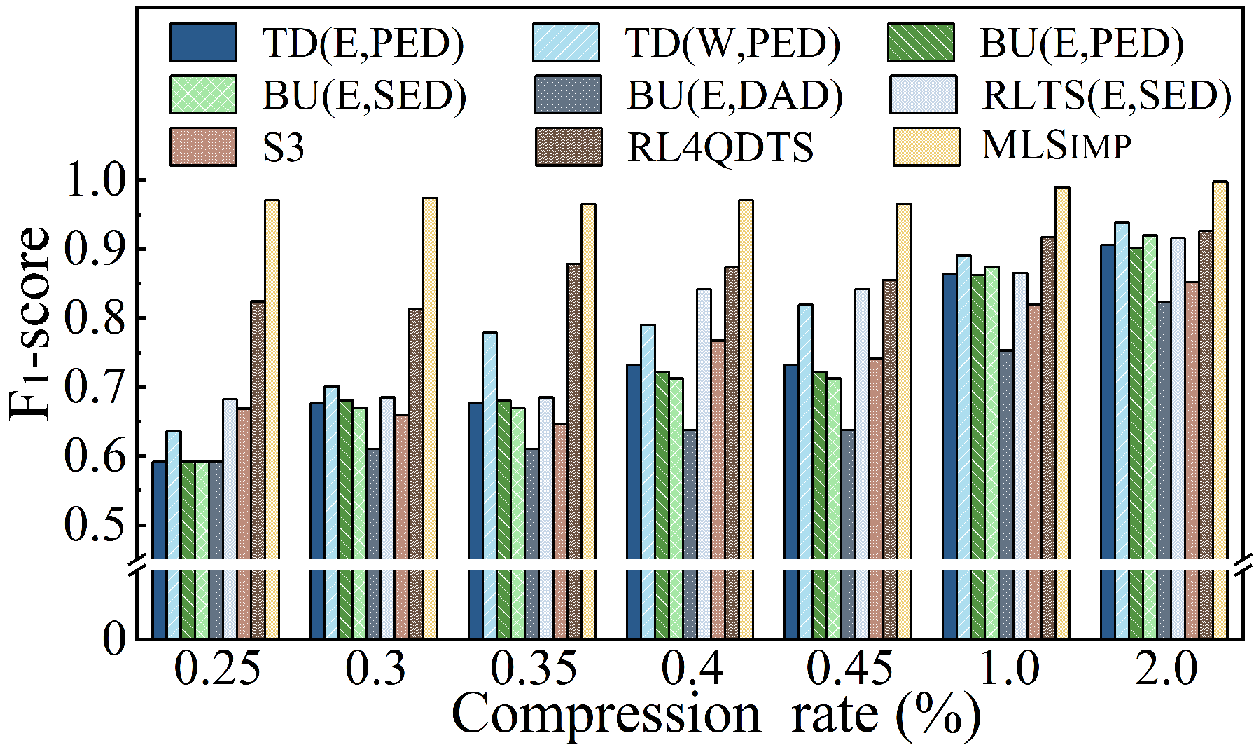}}
           \quad
         \subfloat[ {$k$NN query}]{
         \includegraphics[width=3.9cm]{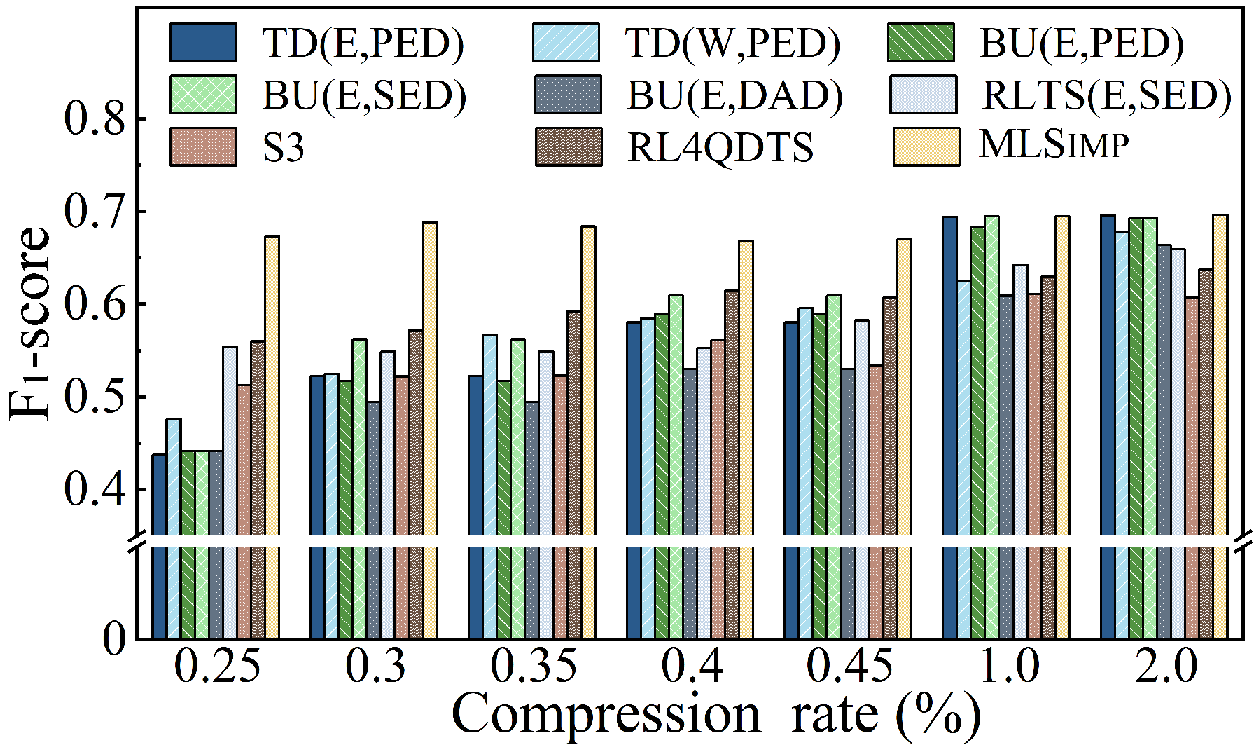}}      
          \quad
                   \subfloat[ {Similarity query}]{
          \includegraphics[width=3.9cm]{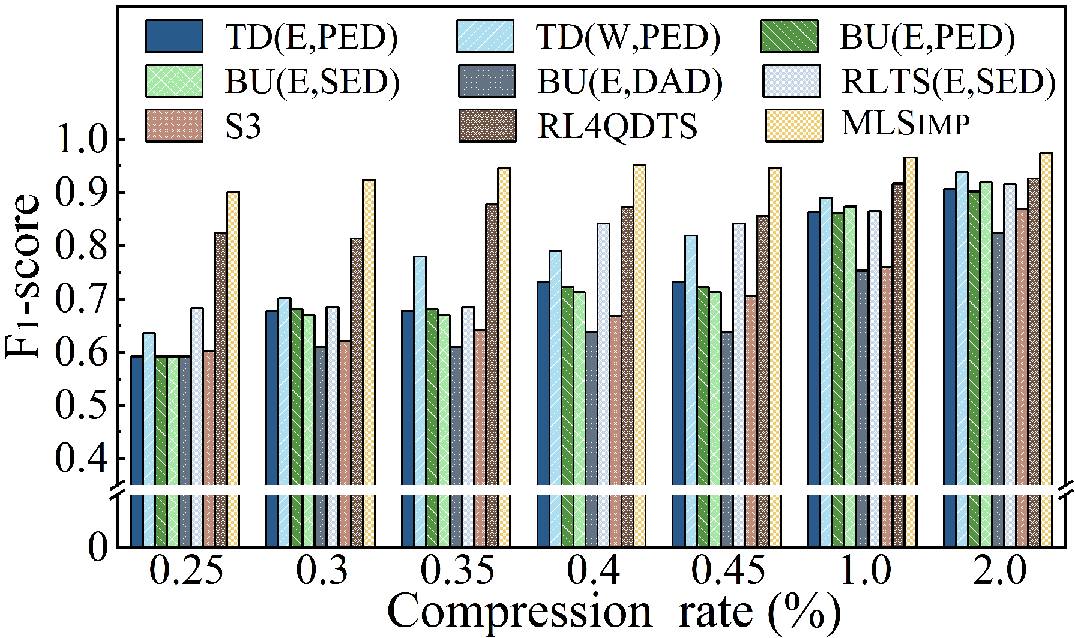}}
          \quad
                   \subfloat[ {Clustering}]{
          \includegraphics[width=3.9cm]{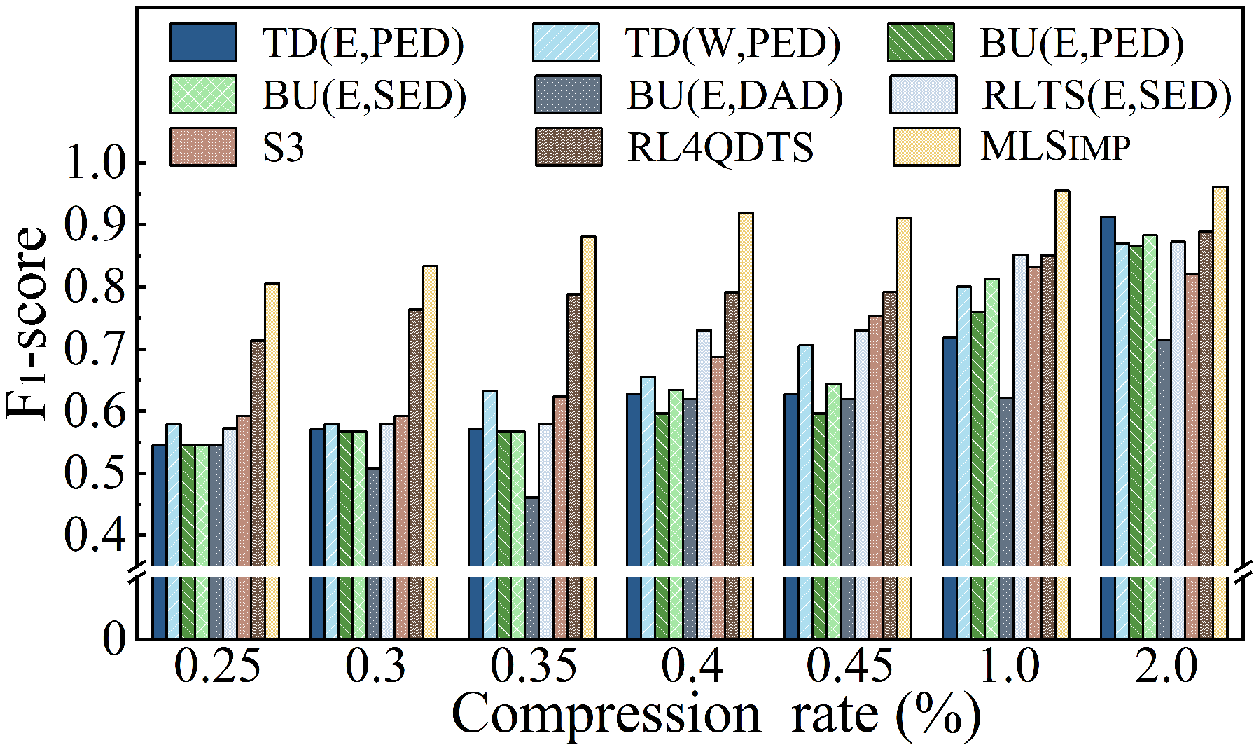}}
           \caption{\color{black}$F_1$-score with different compression rate (\%) on Geolife, where (a)--(d) are queries following the data distribution and (e)--(h) are queries following a Gaussian distribution. }

           \vspace{-6mm} 
     \label{geolife}
   \end{figure*}

\begin{figure*}
\centering
         \subfloat[ {Range query}]{
          \includegraphics[width=3.9cm]{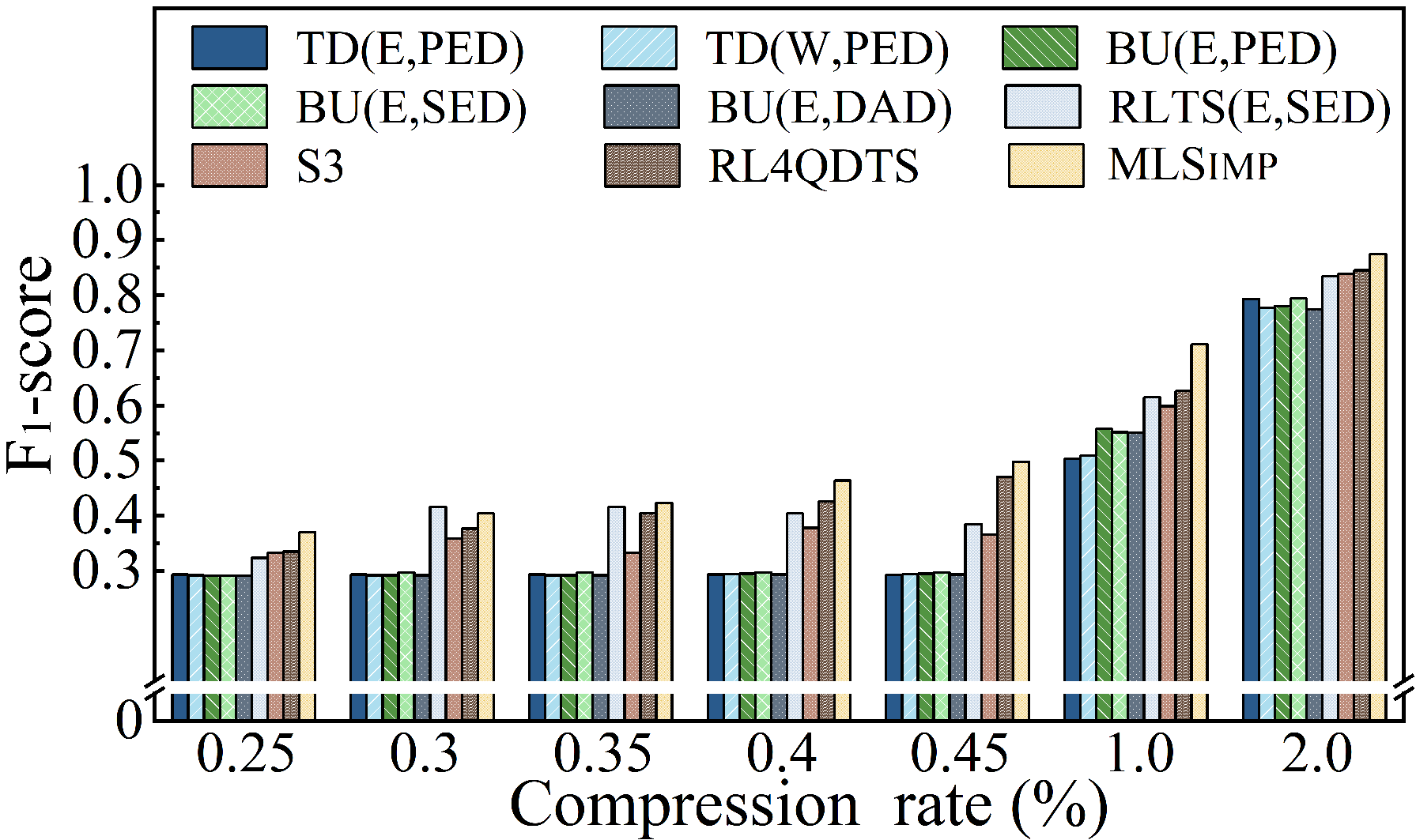}}   
           \quad
         \subfloat[ {$k$NN query}]{
          \includegraphics[width=3.9cm]{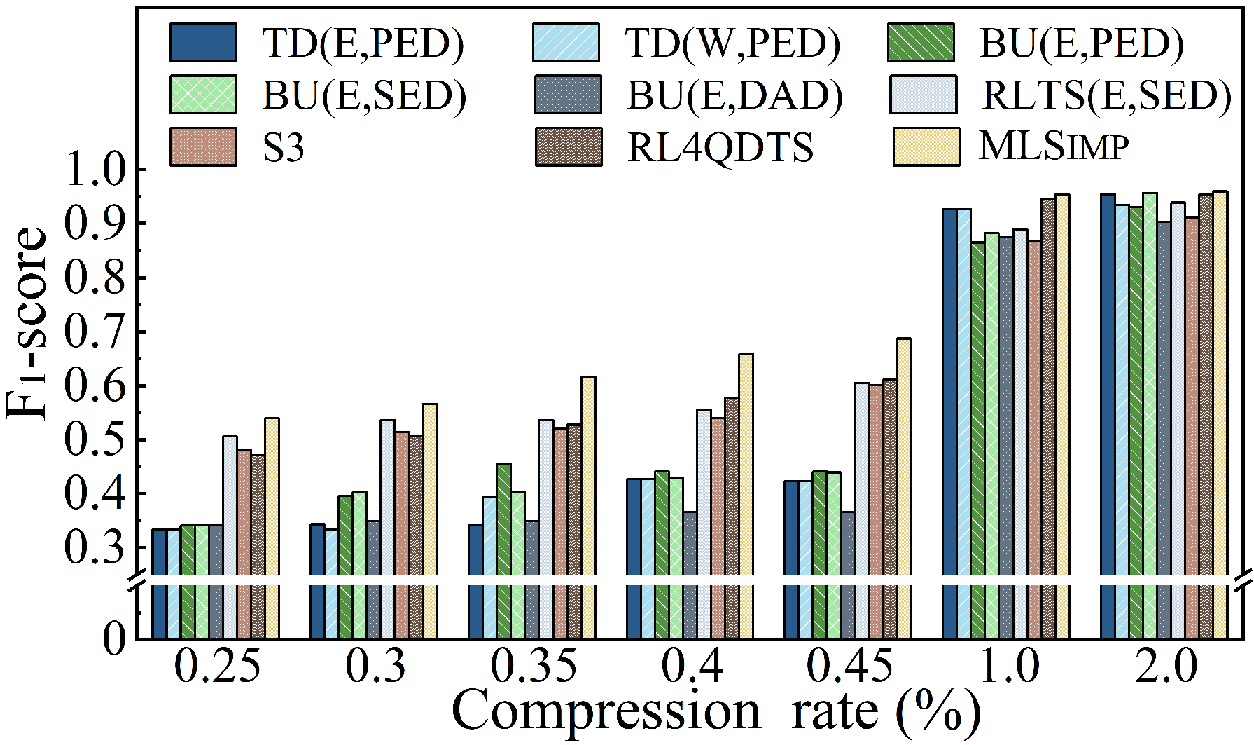}}
          \quad
                   \subfloat[ {Similarity query}]{
          \includegraphics[width=3.9 cm]{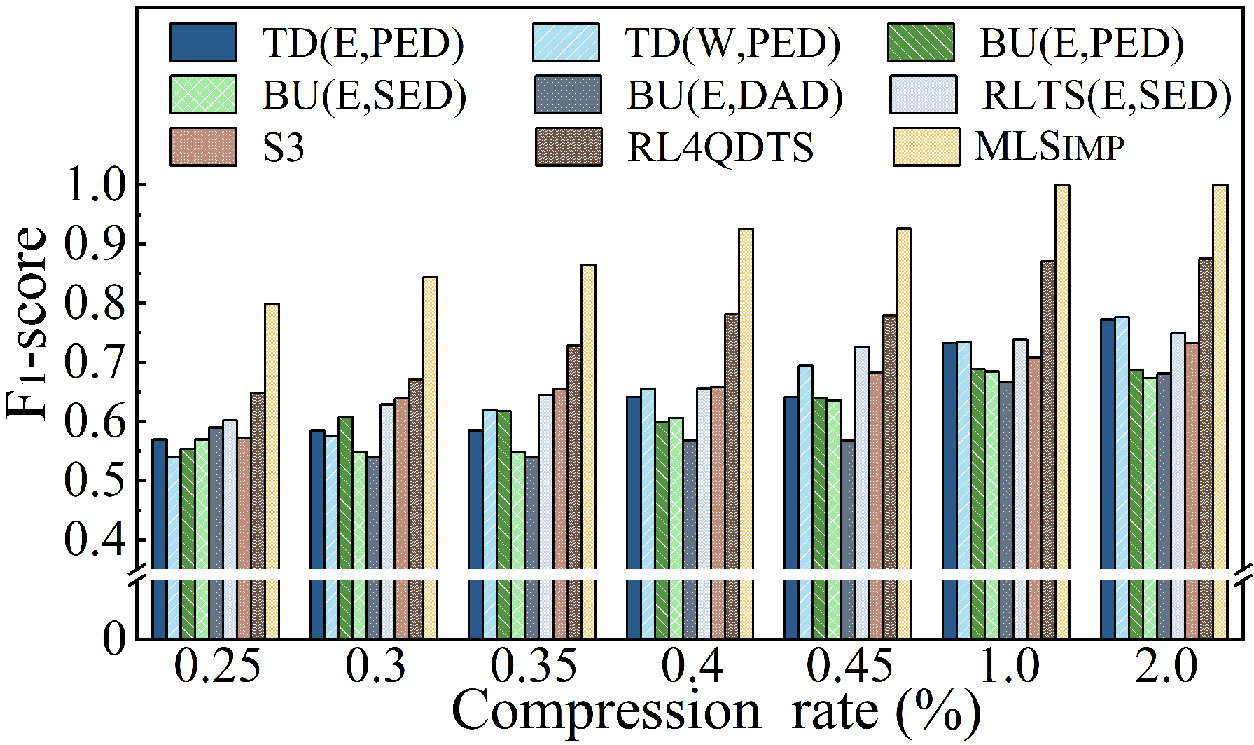}} 
          \quad
                   \subfloat[ {Clustering}]{
                   \includegraphics[width=3.9cm]{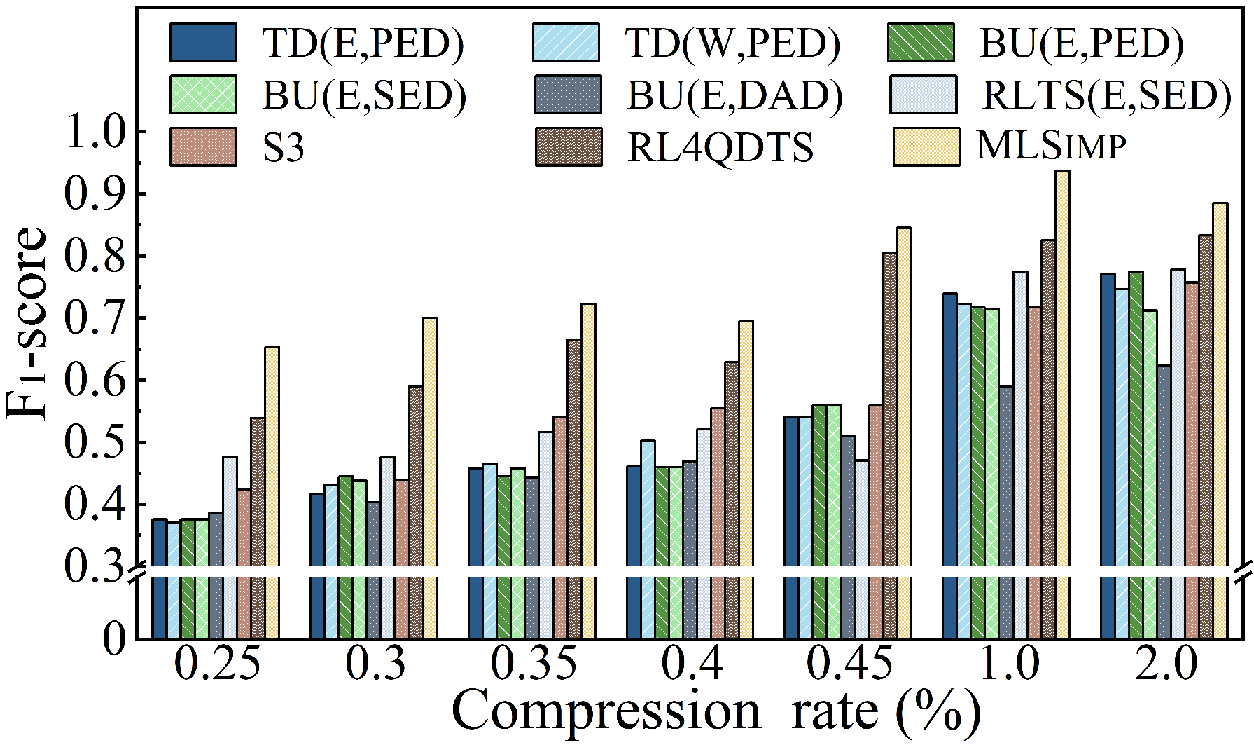}}
            \vspace{-3mm}

                   \subfloat[ {Range query}]{
                   \includegraphics[width=3.9cm]{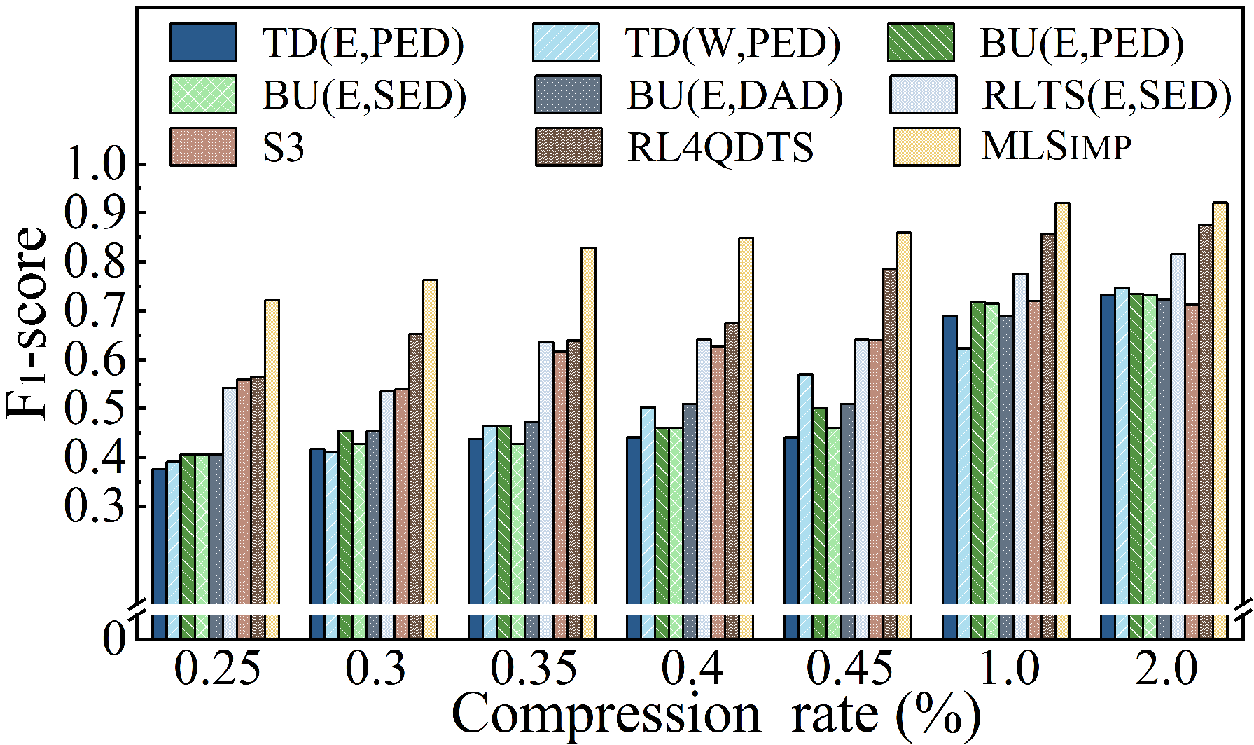}}
           \quad
         \subfloat[ {$k$NN query}]{
         \includegraphics[width=3.9cm]{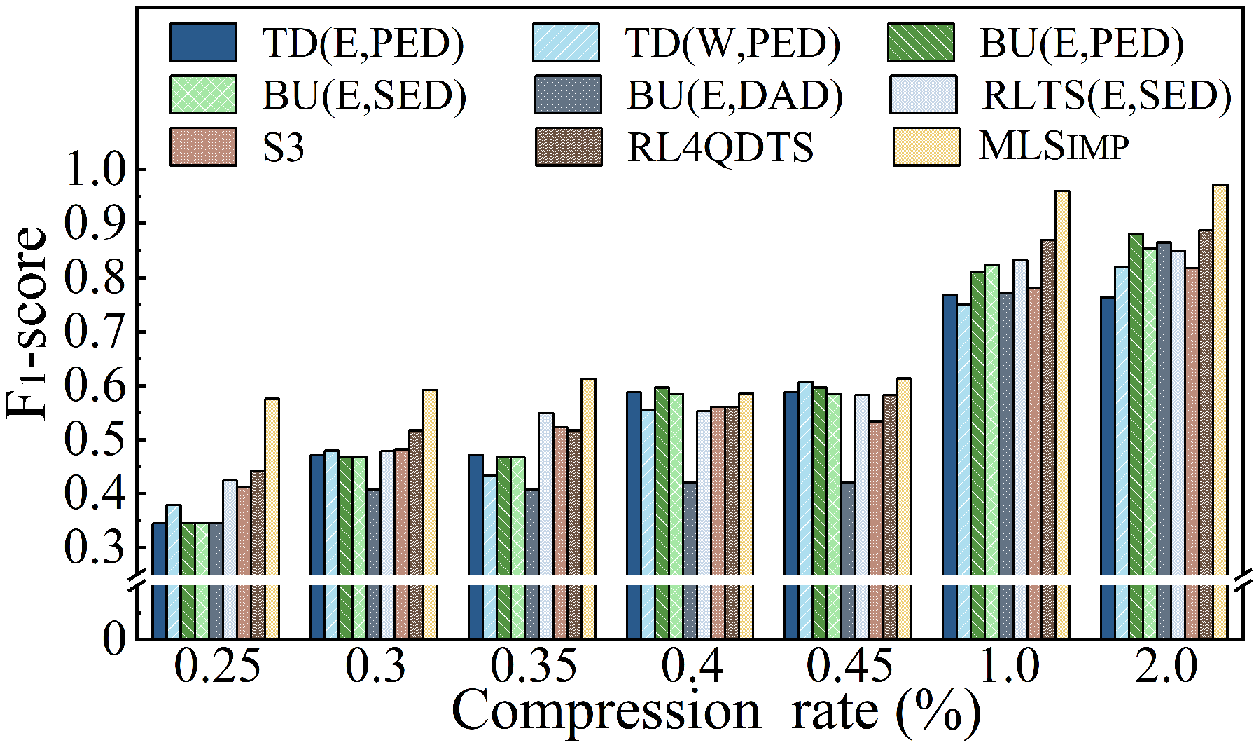}}      
          \quad
                   \subfloat[ {Similarity query}]{
          \includegraphics[width=3.9cm]{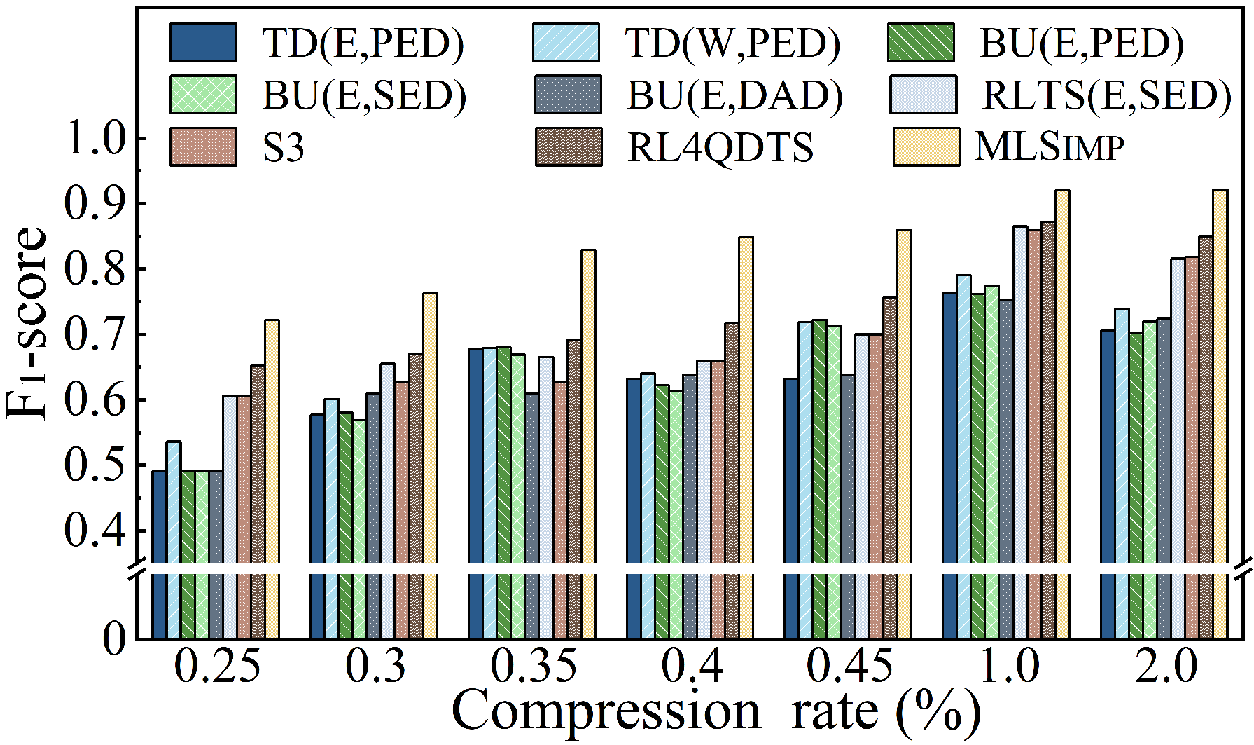}}
          \quad
                   \subfloat[ {Clustering}]{
          \includegraphics[width=3.9cm]{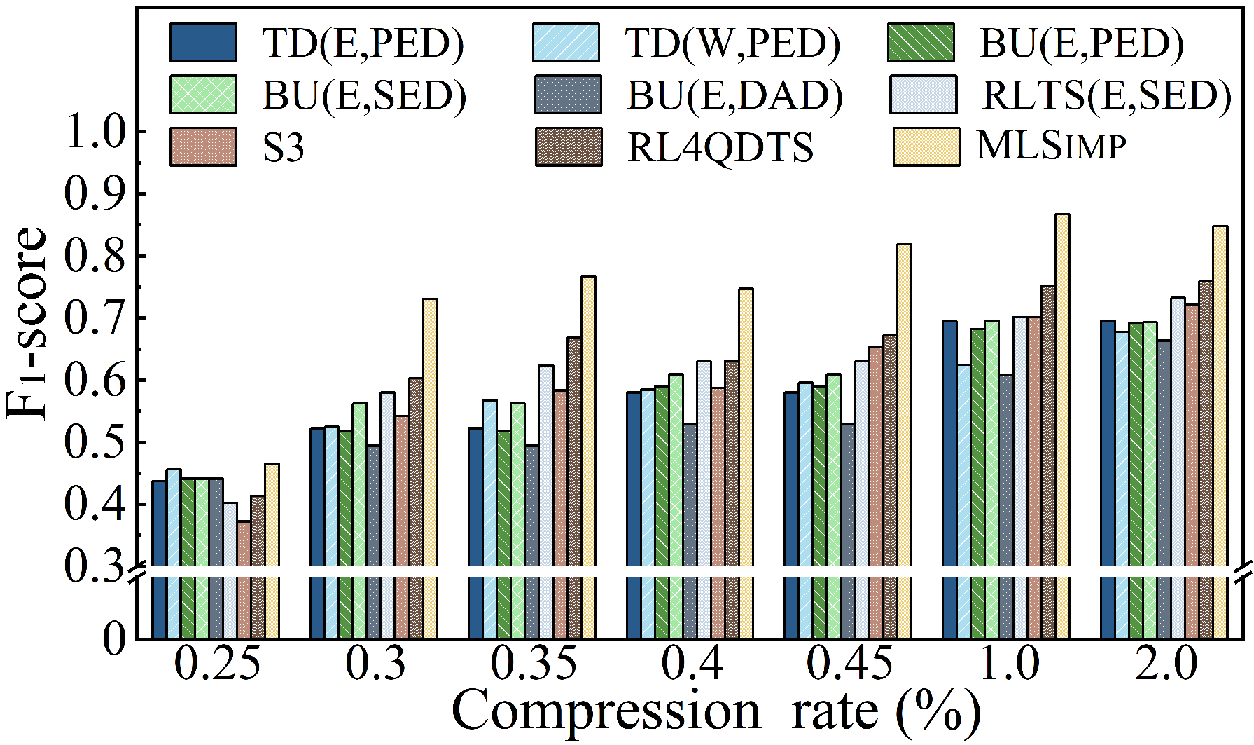}}
           \caption{\color{black}$F_1$-score with different compression rate (\%) on T-Drive, where (a)--(d) are queries following the data distribution and (e)--(h) are queries following a Gaussian distribution.}

     \label{T-drive}
 \vspace{-5mm} 
   \end{figure*}

\vspace{-2mm}
\subsubsection{Hyperparameters and Experimental Settings}

For training, we randomly sample 1,000 trajectories and fix each $|T|$ at 1,000. The remaining trajectories are used for testing. For testing,    
{\color{black} the compression rate ($cr$)  is defined as  shown in Example~\ref{ex:RL4QDTS}.
% $cr = \frac{|S|}{|P|}$, where $|S|$ is the number of points in the simplified database and $|P|$ is the number of points in the original database. 
The lower the $cr$, the fewer points remain. For example, when $cr = 0.25\%$, the amount of data after simplification is reduced by 400 (i.e., $100 / 0.25$) times compared to the original data.} We follow a previous study~\cite{QDTS} and generate range queries, $k$NN queries, similarity queries, and clustering. 
Queries are generated based on two distributions: the data distribution and a Gaussian distribution (with parameters $\mu = 0.5$ and $\sigma = 0.25$). 
The spatial window size for range queries is set to $2 km \times 2 km$. For all queries, the temporal window size is 7 days for Geolife, 3 hours for T-Drive, and 1 day for OSM. The similarity metric of $k$NN queries is  EDR~\cite{edr} with a threshold of 2 km. The value of  $k$ is set to 3. The distance threshold for similarity queries is set to 5 km. For clustering, we adopt the tracking algorithm~\cite{cluster}.

% We measure the difference between query results on an original database and those on a simplified database using the $F_1$-score. A larger $F_1$-score indicates a smaller difference between the query results. 

We set the hyperparameters of the baselines as suggested in the respective papers.  
In GNN-TS, the pre-trained T-Bert employs a transformer with 4 layers and 8 attention heads as the embedding layer, trained for 500 epochs. The embedding dimensionality is 128, and the maximum embedding sequence length is set to 500. The GNN consists of 2 layers of GATs with 4 attention heads, each producing an output with dimensionality 32. During training, the GNN-TS generates coarse-grained simplified trajectories with a compression rate of 0.5 for training the Diff-TS. The Diff-TS's encoder utilizes a transformer with 2 layers and 2 attention heads, producing output with a dimensionality of 128. The diffusion model employs a transformer with 2 layers and 2 attention heads, with an output dimensionality of 128. The diffusion steps are set to 500. Diff-TS generates amplified labels for simplified trajectories containing only 20 trajectory points. The hyperparameter $\lambda_1$ in $\mathcal{L}_{\mathit{con}}$ is fixed at 0.5~\cite{contrastivelearning}. The hyperparameter $\lambda_2$ is used to balance the two terms in $\mathcal{L}_{\mathit{Diff-TS}}$. The hyperparameter $\lambda_3$, set to 0.5.
% , is used to adjust the proportion of mutual learning in  $\mathcal{L}_{\mathit{GNN-TS}}$. 
The parameter for query-based adjustment $\delta$ is set to 0.5 (0.7 for T-Drive). 
% The models are trained using the Adam optimizer, with each model trained for 200 epochs per round and 10 rounds of mutual learning.
{\color{black}We  perform 100 range queries for the query-based importance adjustment. 
% These queries are only used to adjust the sampling for trajectories. 
% The queries are generated consistently with the test queries based on both the  data distribution and a Gaussian distribution.
}
All experiments are conducted on a server with an Intel(R) Xeon(R) W-2155 CPU, 128GB memory, and an NVIDIA TITAN RTX GPU. 
% The source code is available online\footnote{https://github.com/yumengs-exp/MLSimp}. 

\vspace{-2mm}
\subsubsection{Metrics}
 {\color{black}We use  the $F_1$-score to evaluate the performance of queries on the simplified database. 
 For each query distribution, we generate 100 queries and report the average  $F_1$-score. Given a query $Q \in \{Q_\textit{range}, Q_\textit{kNN}, Q_\textit{sim}\}$, where $Q_\textit{range}$ is a range query, $Q_\textit{kNN}$ is a $k$NN query, and $Q_\textit{sim}$ is a similarity query, and given the query results $R_s$ on the simplified database and $R_o$ on the original database, the $F_1$-score is:
 \vspace{-2mm}
\begin{equation}
    \begin{split}
    F_1\text{-score}(Q) = \frac{2P(Q)R(Q)}{P(Q) + R(Q)},
    % P(Q) = \frac{|R_s \cap R_o|}{|R_s|},
    % R(Q) = \frac{|R_s \cap R_o|}{|R_o|},
    \end{split}
     \vspace{-3mm}
\end{equation}
where $P(Q) = \frac{|R_s \cap R_o|}{|R_s|}$, $R(Q) = \frac{|R_s \cap R_o|}{|R_o|}$, and $|\cdot|$ returns the cardinality of its argument set.

For clustering queries, following the literature~\cite{TSsurvey}, we introduce $C$ to represent the clustering results on a database $D$. If $T_i$ and $T_j$ ($T_i, T_j \in D$) belong to the same cluster then $(T_i, T_j) \in C$. Given a clustering $Q_{c}$, the clustering results  $C_s$ on the simplified database, and the clustering results  $C_o$ on the original database, the $F_1$-score is defined as follows:
 \vspace{-2mm}
\begin{equation}
    \begin{split}
        F_1\text{-score}(Q_c) = \frac{2P(Q_c)R(Q_c)}{P(Q_c) + R(Q_c)},
      % P(Q_c) = \frac{|C_s \cap C_o|}{|C_s|}
      %   R(Q_c) = \frac{|C_s \cap C_o|}{|C_o|},
    \end{split}
     \vspace{-2mm}
\end{equation}
where $P(Q_c) = \frac{|C_s \cap C_o|}{|C_s|}$ and $R(Q_c) = \frac{|C_s \cap C_o|}{|C_o|}$.}

{\color{black}We use  SED and PED to evaluate the simplification errors for simplification methods.
We first define the anchor segment of each point in the original trajectory. 
\vspace{-2mm}
\begin{myDef} Given an original trajectory $T=\langle p_1, p_2, \ldots, p_n \rangle$ and a simplified trajectory $T' = \langle p_{s_1}, p_{s_2}, \ldots, p_{s_m} \rangle$ ($m\ll n$ and $1=s_1<s_2<\cdots<s_m=n$), the \textbf{anchor segment} of a point $p_i$ ($1\leq i\leq n$) is $\textit{AncSeg}_{p_i}=\overline{p_{s_j}p_{s_{j+1}}}$, where $\overline{p_{s_j}p_{s_{j+1}}}$ is a trajectory segment of $T'$ and ${s_j}\leq i \leq {s_{j+1}-1}$.
\vspace{-2mm}
\end{myDef}

% Fig.~\ref{fig:anchor} shows an original trajectory $T_2 = \langle p_{2,1}, p_{2,2}, p_{2,3}, p_{2,4}, p_{2,5}, \\ p_{2,6}, p_{2,7}\rangle$ and its simplified trajectory  $T'_2 = \langle p_{2,1},  p_{2,4},  p_{2,7}\rangle$. The anchor segment of $p_{2,3}$ is $\textit{AncSeg}_{p_{2,3}}=\overline{p_{2,1}p_{{2,4}}}$. The anchor segment of $p_{2,5}$ is $\textit{AncSeg}_{p_{2,5}}=\overline{p_{2,4}p_{{2,7}}}$

The simplification error $\epsilon(T')$ of a simplified trajectory $T'$ is defined as follows:
 \vspace{-2mm}
\begin{equation}
\epsilon(T') = \max_{1 \leq j \leq m-1} \max_{s_j \leq i < s_{j+1}} \epsilon(p_i),
 \vspace{-2mm}
\end{equation}
where $\epsilon(p_i)$ is the error of the trajectory point $p_i$ in $T$. 

The similarity errors PED and SED for each point are shown in Fig.~\ref{fig:anchor}.
% \vspace{-1mm}
% \begin{itemize}[leftmargin=2pt]
PED is the shortest distance from  $p_i$ to the anchor segment $\textit{AncSeg}_{p_i} = \overline{p_{s_j} p_{s_{j+1}}}$.
% , i.e.:
% \vspace{-2mm}
% \begin{equation}
% \texttt{PED}(p_i) = \frac{|(y_{s_{j+1}} - y_{s_{j}}) \cdot x_i - (x_{s_{j+1}} - x_{s_{j}}) \cdot y_i + x_{s_{j+1}} y_{s_{j}} - y_{s_{j+1}} x_{s_{j}}|}{\sqrt{(x_{s_{j+1}} - x_{s_{j}})^2 + (y_{s_{j+1}} - y_{s_j})^2}}
% \vspace{-2mm}
% \end{equation}
SED is the Euclidean distance between the trajectory point $p_i = (x_i, y_i, t_i)$ and its mapped point $p'_i = (x'_i, y'_i, t_i)$ on the anchor segment $\textit{AncSeg}_{p_i} = \overline{p_{s_j} p_{s_{j+1}}}$ of $T'$ based on $t_i$:
\begin{equation}
    \begin{split}
    x'_i &= x_{s_j} + \frac{t_i - t_{s_j}}{t_{s_{j+1}} - t_{s_j}} \cdot (x_{s_{j+1}} - x_{s_j})\\
y'_i &= y_{s_j} + \frac{t_i - t_{s_j}}{t_{s_{j+1}} - t_{s_j}} \cdot (y_{s_{j+1}} - y_{s_j})\\
&\texttt{SED}(p_i) = \sqrt{(x_i - x'_i)^2 + (y_i - y'_i)^2},
    \end{split}
    \vspace{-2mm}
\end{equation}
where $p_{s_j} = (x_{s_j}, y_{s_j}, t_{s_j})$ and $p_{s_{j+1}} = (x_{s_{j+1}}, y_{s_{j+1}}, t_{s_{j+1}})$.

}

\subsection{Effectiveness Evaluation}
\label{sec:5.2}

\begin{figure*}
\centering
         \subfloat[ {Range query}]{
          \includegraphics[width=3.9 cm]{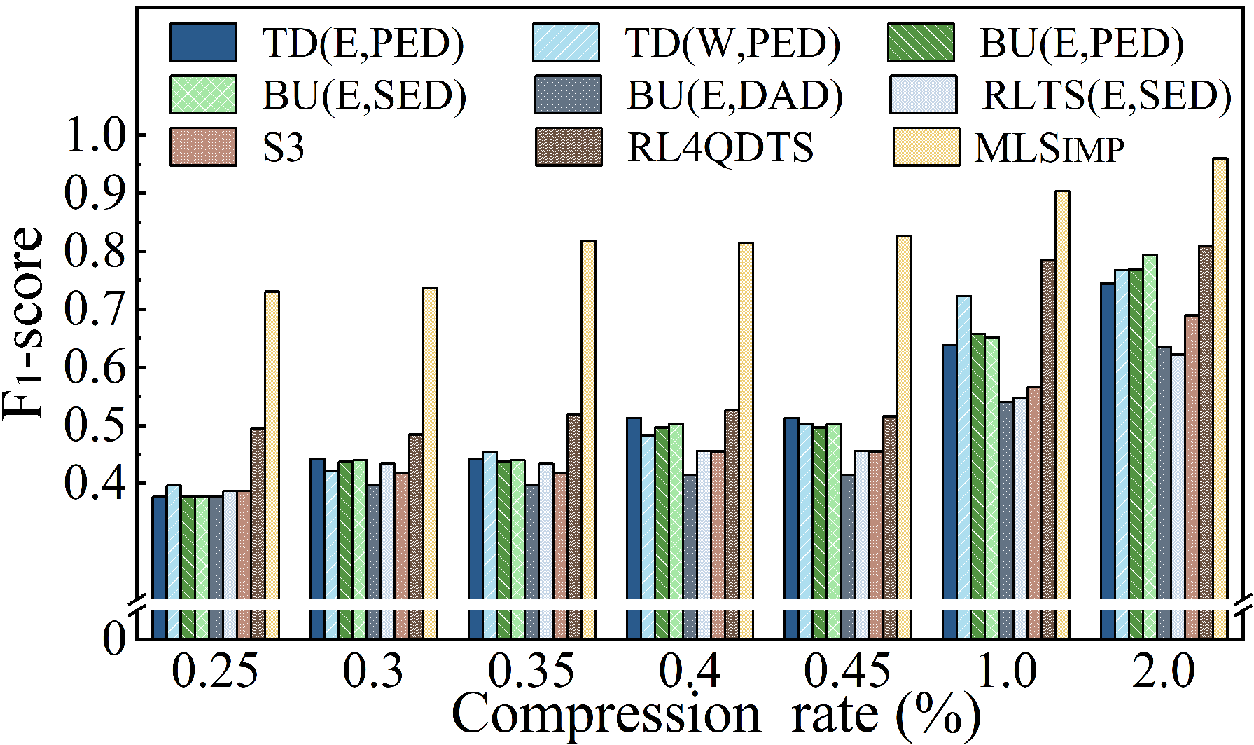}}   
           \quad
         \subfloat[ {$k$NN query}]{
          \includegraphics[width=3.9cm]{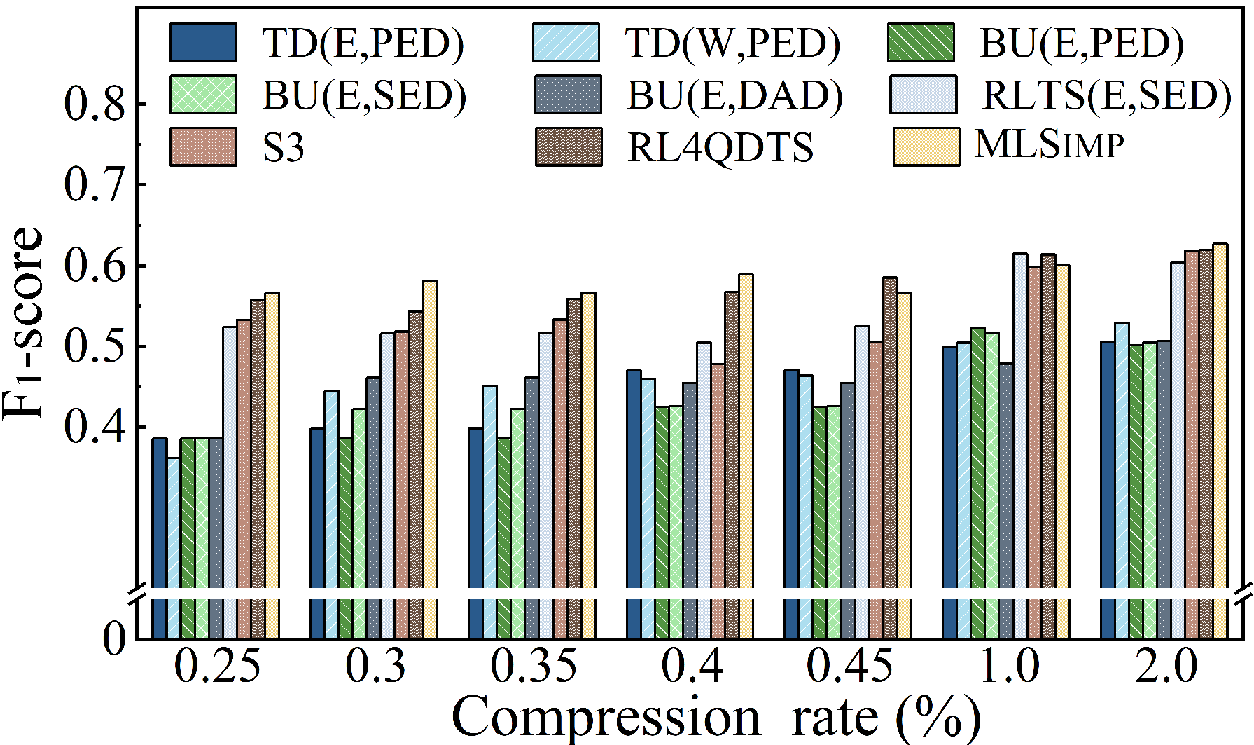}}
          \quad
                   \subfloat[ {Similarity query}]{
          \includegraphics[width=3.9 cm]{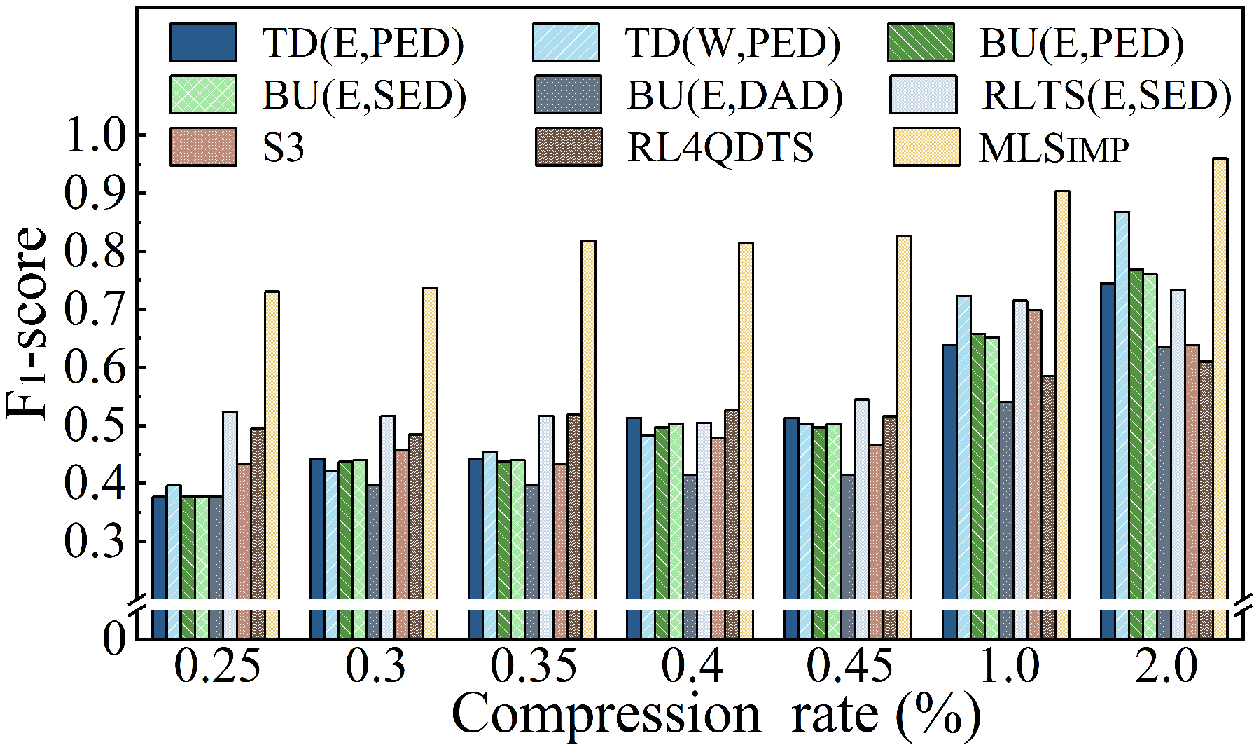}} 
          \quad
                   \subfloat[ {Clustering}]{
                   \includegraphics[width=3.9cm]{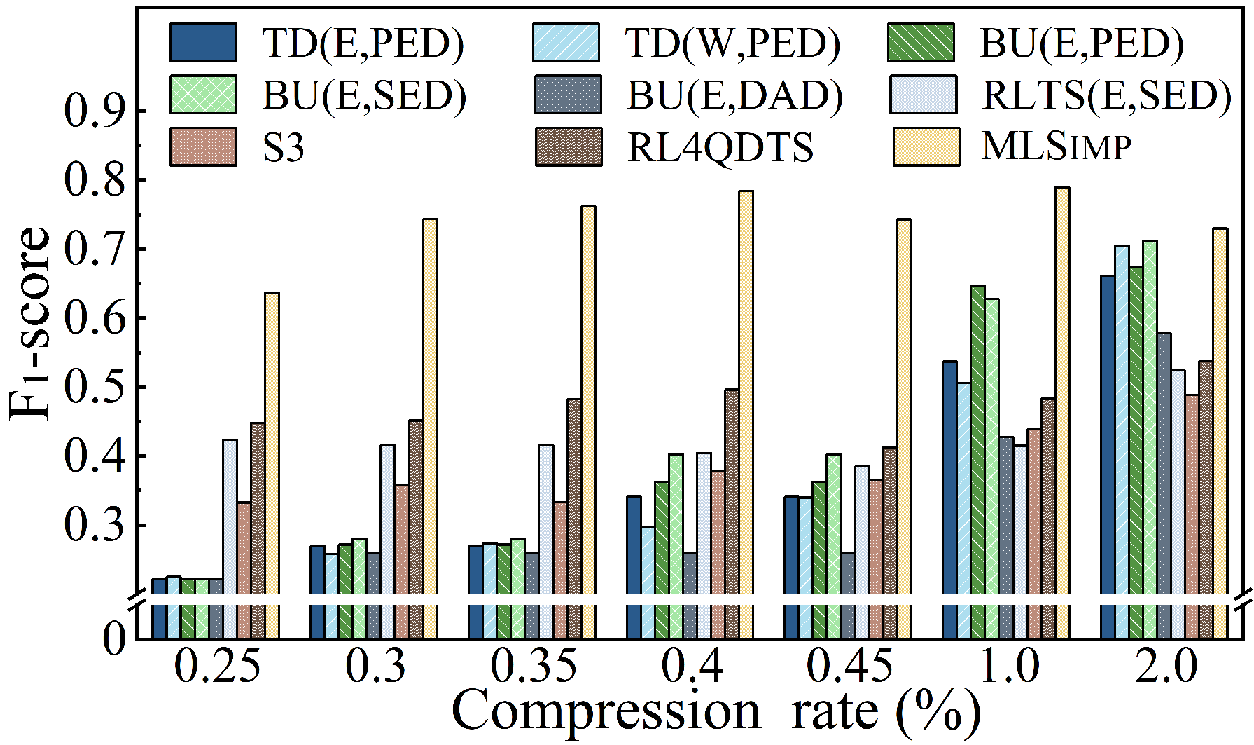}}
          \vspace{-3mm}

                   \subfloat[ {Range query}]{
                   \includegraphics[width=3.9cm]{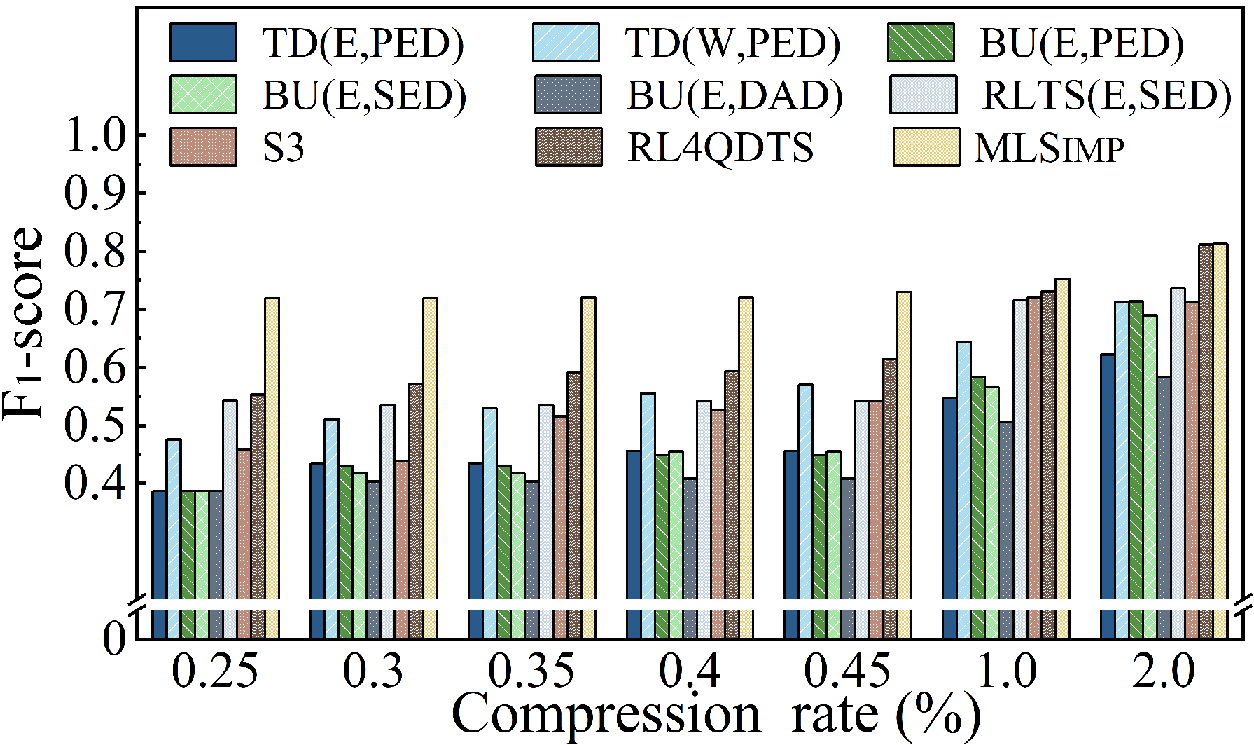}}
           \quad
         \subfloat[ {$k$NN query}]{
         \includegraphics[width=3.9cm]{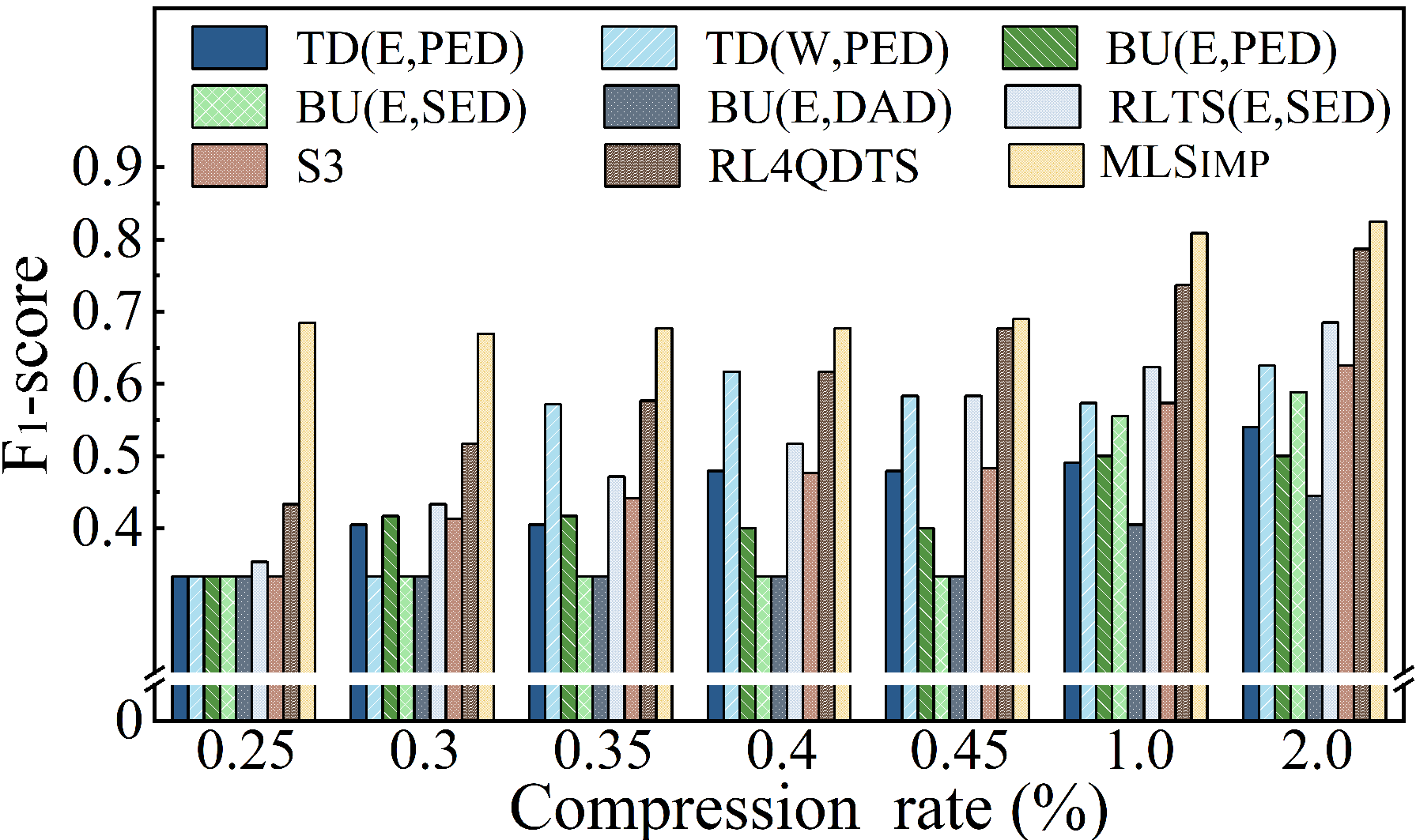}}      
          \quad
                   \subfloat[ {Similarity query}]{
          \includegraphics[width=3.9cm]{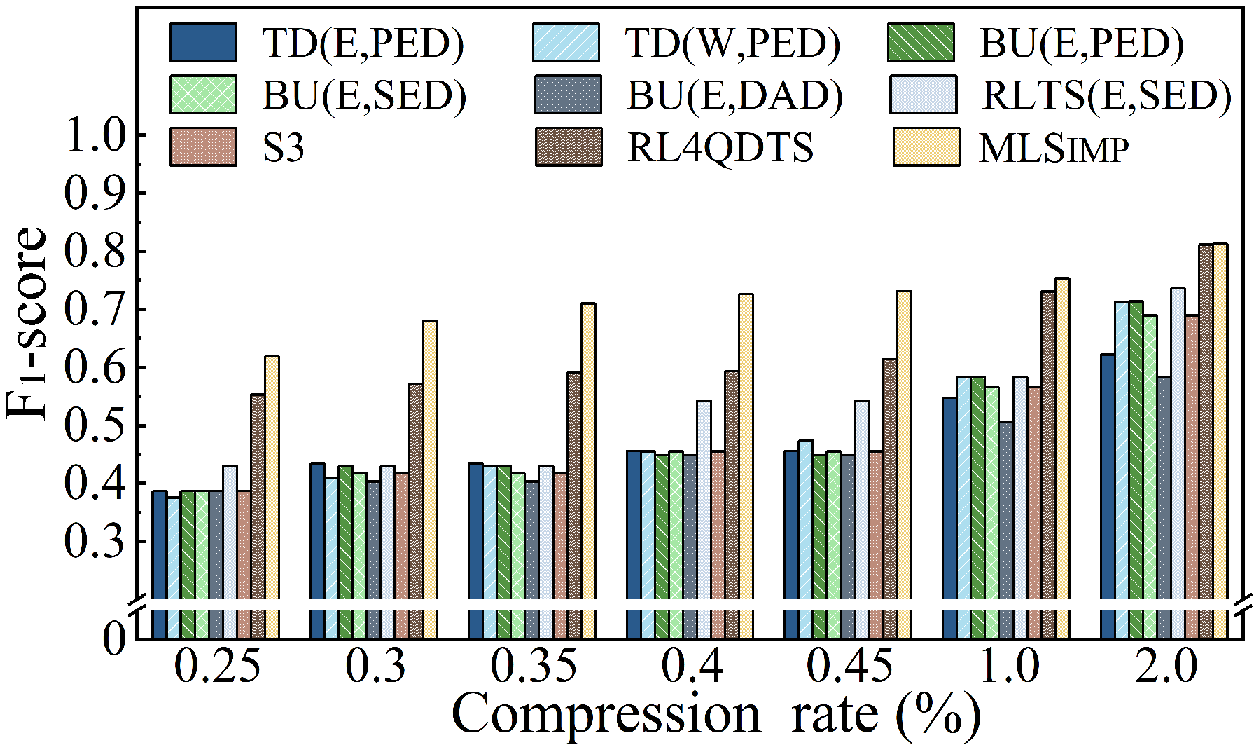}}
          \quad
                   \subfloat[ {Clustering}]{
          \includegraphics[width=3.9cm]{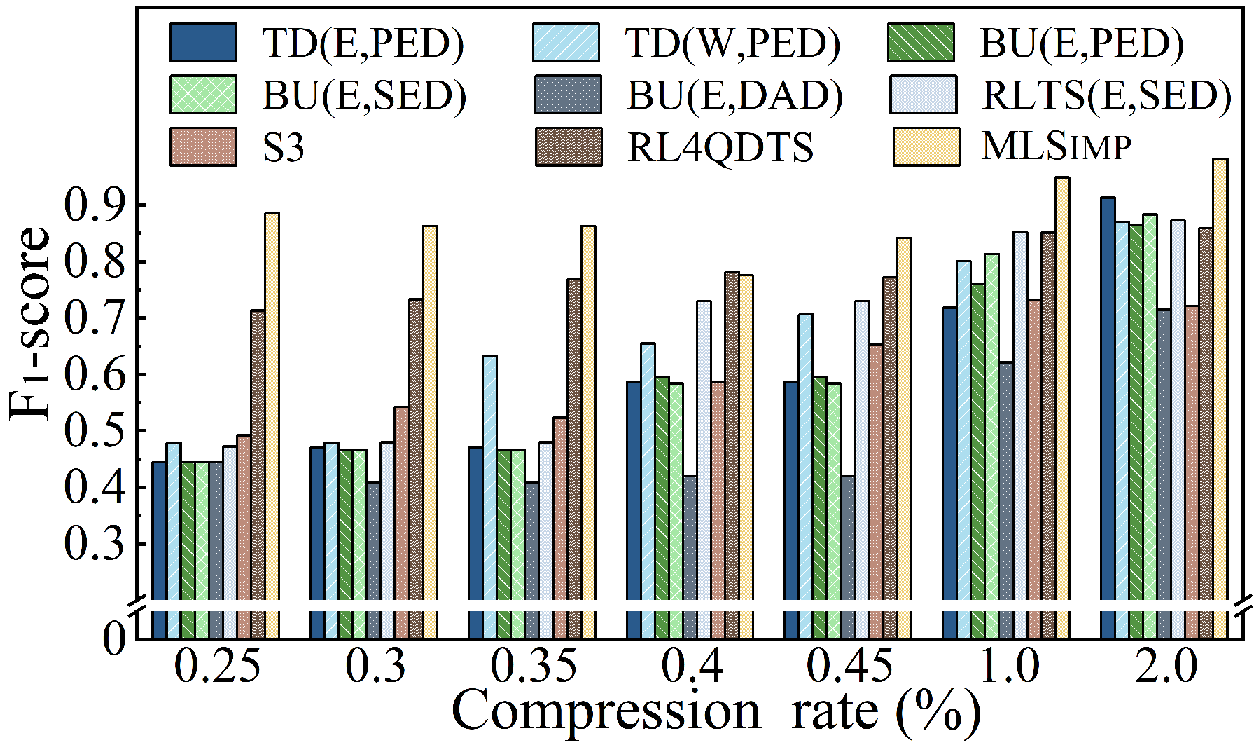}}
           \caption{\color{black}$F_1$-score with different compression rate (\%) on OSM, where (a)--(d) are queries following the data distribution and (e)--(h) are queries following a Gaussian distribution.}
\vspace{-7mm}
           
     \label{OSM}
   \end{figure*}

Figs.~\ref{geolife}, \ref{T-drive}, and \ref{OSM} report the $F_1$-scores of all baselines on the Geolife, T-drive, and OSM datasets with 2,000,000 trajectory points, respectively.
% RL4QDTS and \textsc{MLSimp} are trained based on the corresponding query distribution. 
\textsc{MLSimp} tops consistently across all datasets and queries. Even at a fairly low compression rate of 0.25\%, which means that we reduce the data by a factor of 400, most query scores exceed 0.5. Compared to the state-of-the-art method, RL4QDTS, \textsc{MLSimp} achieves up to a 21\% improvement in range query performance, up to an 11.8\% improvement in $k$NN query performance, up to a 34.6\% improvement in clustering, and up to a 20.3\% improvement in similarity queries on the three datasets. This is because \textsc{MLSimp} not only considers the impact of range queries on trajectory simplification but also, more importantly, preserves key points in the simplified database by mining global information, better fitting the trajectory similarity of the original database. 
% Moreover, we make two notable observations.

% First, \textsc{MLSimp} and RL4QDTS perform significantly better than other methods at low compression rates. This is because both methods involve global simplification and utilize simulated queries to give simplified trajectories relatively higher sample rates in query-hot areas. In high compression rate scenarios, although the EDTS methods can achieve query performance comparable to those of the QDTS methods, \textsc{MLSimp} still performs best because it considers the global information of trajectories, thereby mining the importance of trajectory points, and using Diff-TS to generate low compression rate amplified labels.

% Second, although TD(W, PED)  improves the performance of the EDTS methods via simplifying points considering the whole database, it still achieves lower performance that  do RL4QDTS and \textsc{MLSimp}.
%  This is because although simplification for the whole database selects the point with the maximum or minimum error in the database to keep or drop in each iteration,  the error gaps between GPS points are often very low, due to a large number of points and the existence of redundant and similar points. Therefore, the sampling results are similar to those obtained by sampling each trajectory individually. In contrast, \textsc{MLSimp} evaluates the importance of points by analyzing the representation distribution of points in the same embedding space, making the simplification points more representative.

\vspace{-3mm}
\subsection{Efficiency Evaluation}
\label{sec:5.3}
\begin{figure}
\centering
         \subfloat[ {Geolife}]{
          \includegraphics[width=2.7 cm]{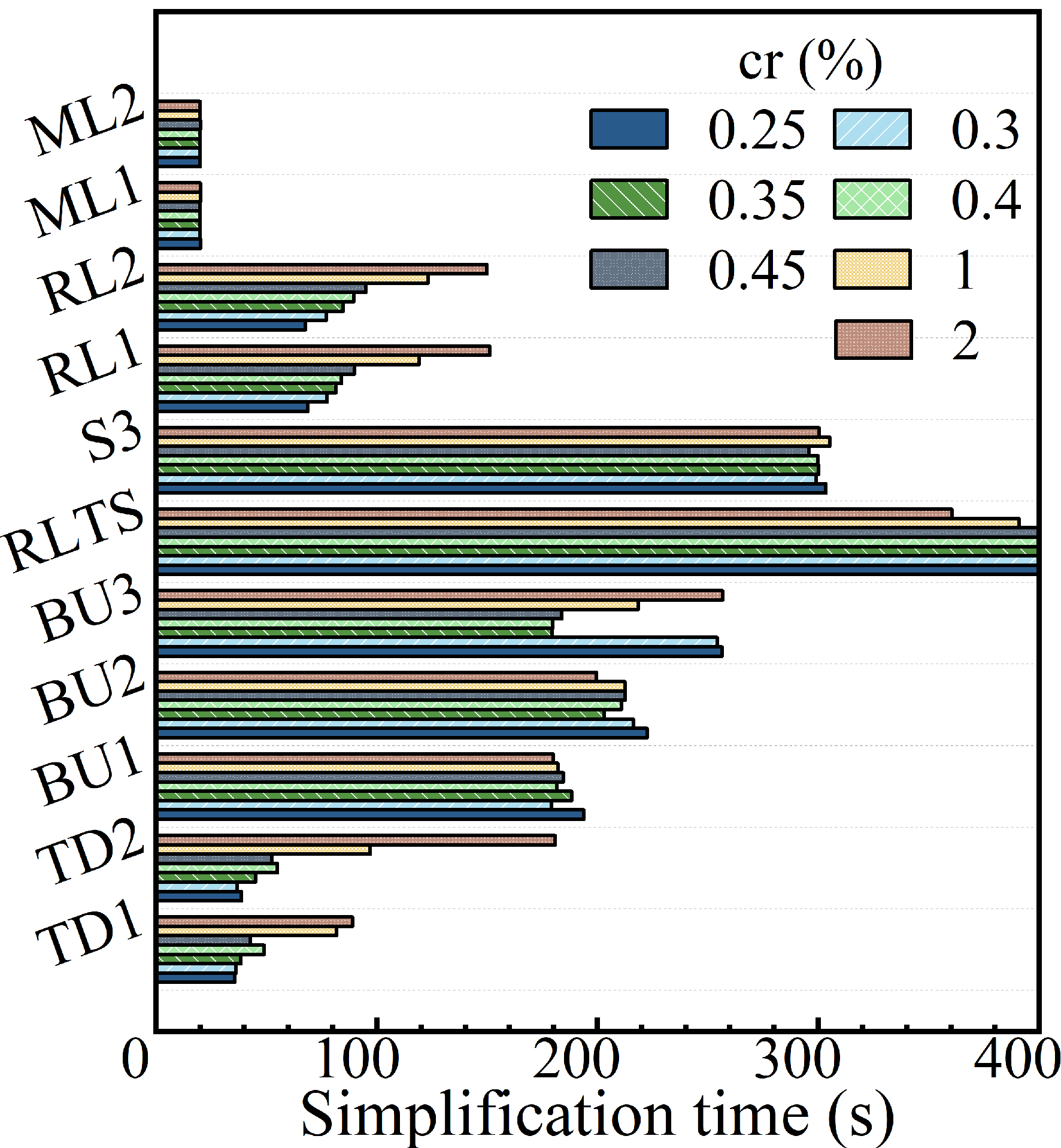}}   
         \subfloat[ {T-Drive}]{
          \includegraphics[width=2.7 cm]{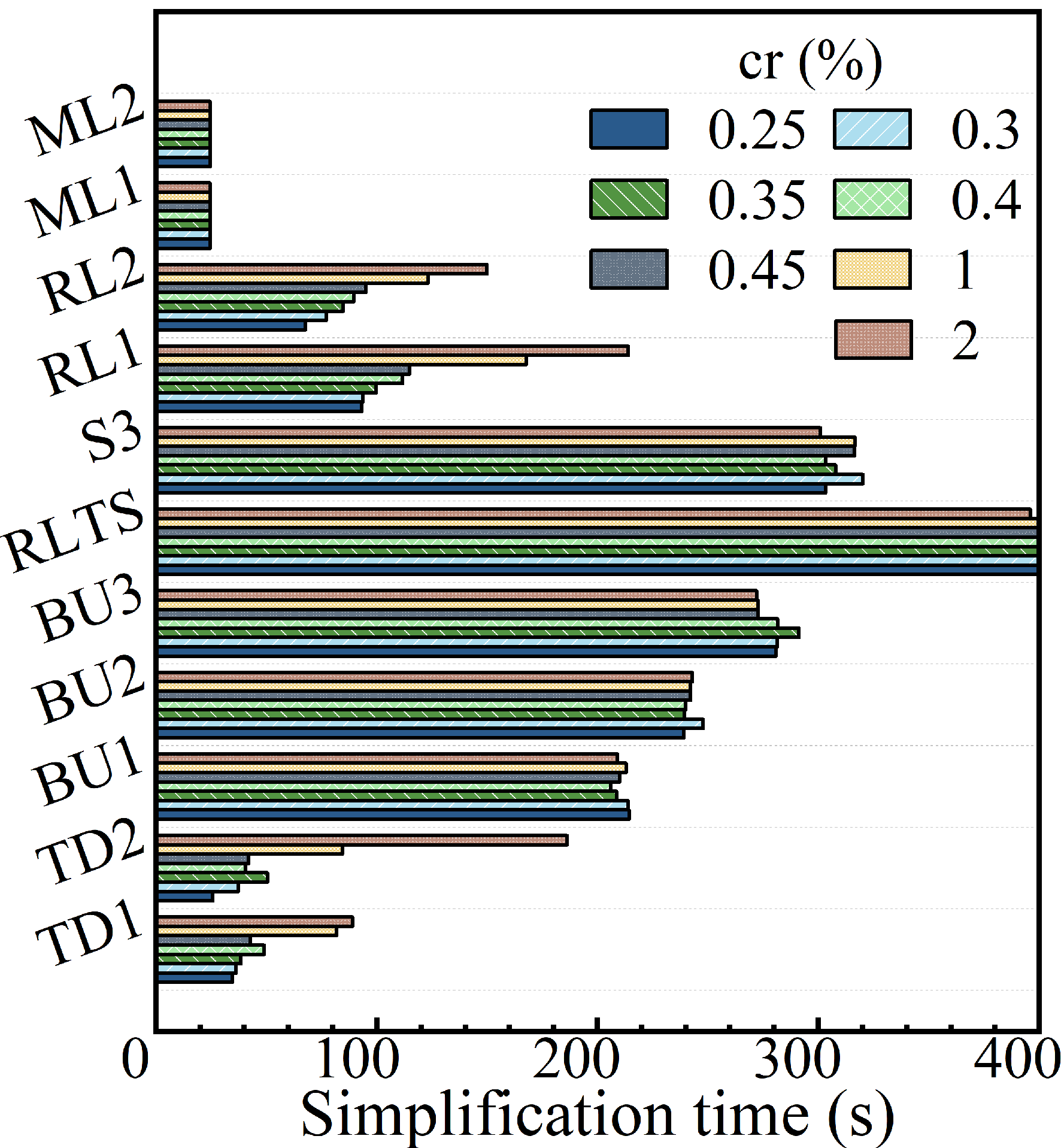}}
                   \subfloat[ {OSM}]{
          \includegraphics[width=2.7 cm]{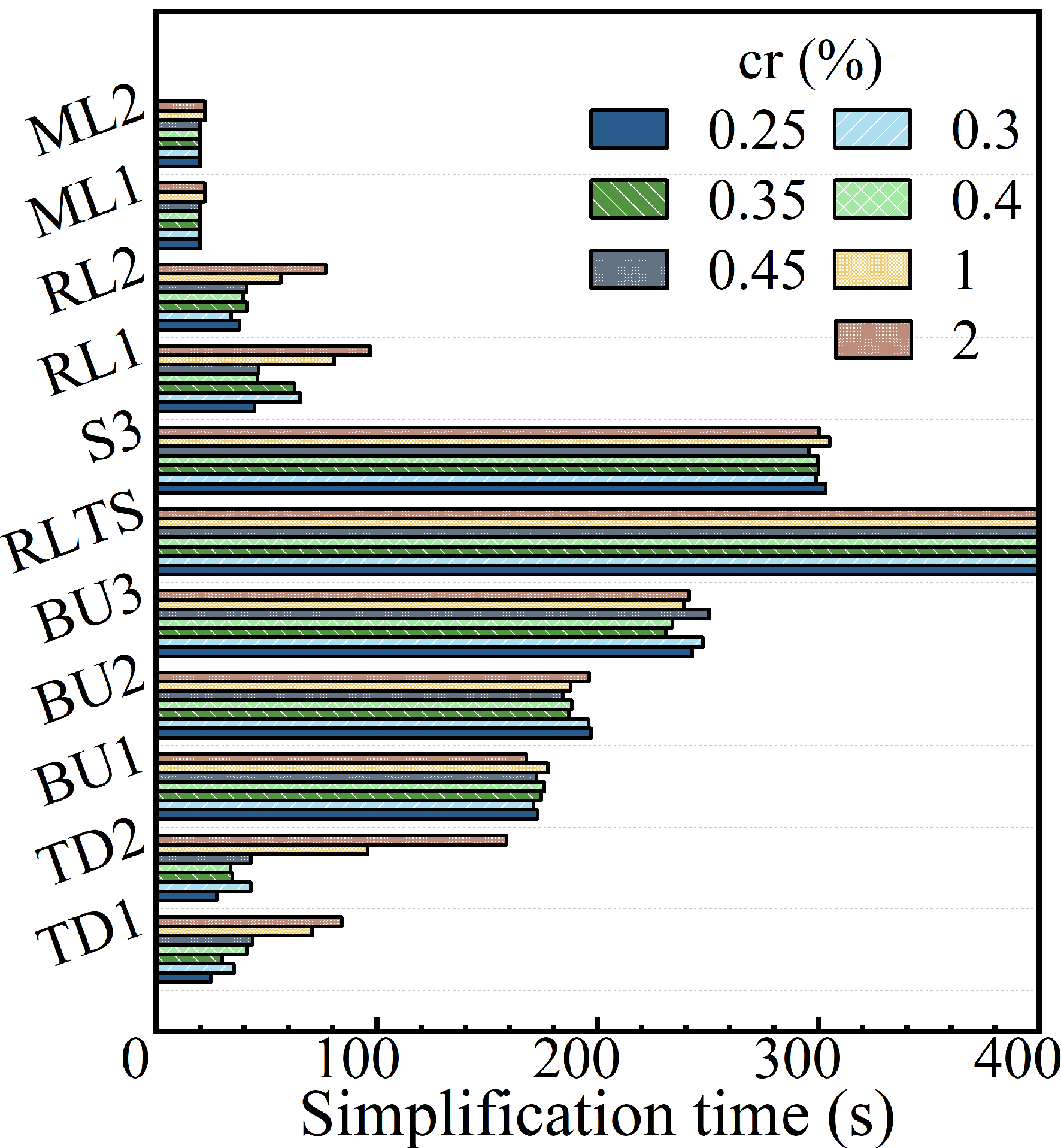}}

        %            \subfloat[ {ML1(Geolife)}]{
        %   \includegraphics[width=2.7 cm]{figures/ML1_geolife_time.png}}   
        %  \subfloat[ {ML1(T-Drive)}]{
        %     \includegraphics[width=2.7 cm]{figures/ML1_tdrive_time.png}}  
        %            \subfloat[ {ML1(OSM)}]{
        %    \includegraphics[width=2.7 cm]{figures/ML1_osm_time.png}}  

        %            \subfloat[ {ML2(Geolife)}]{
        %     \includegraphics[width=2.7 cm]{figures/ML2_geolife_time.png}}    
        %  \subfloat[ {ML2(T-Drive)}]{
        % \includegraphics[width=2.7 cm]{figures/ML2_tdrive_time.png}}  
        %            \subfloat[ {ML2(OSM)}]{
        %     \includegraphics[width=2.7 cm]{figures/ML2_osm_time.png}}  

           \caption{Simplification time with different compression rates (cr), where (i) TD1 and TD2 represent TD(E, PED) and TD(W, PED); (ii) BU1, BU2, and BU3 represent BU(E, DAD), BU(E, PED), and BU(E, SED); (iii) RLTS represents RLTS(E, SED); (iv) RL1 and RL2 represent  RL4QDTS for queries generated based on the data and  Gaussian distribution, respectively; (v) ML1 and ML2 represent  \textsc{MLSimp} for queries generated based on the data and Gaussian distribution, respectively.}
          \vspace{-6mm} 
     \label{efficiency}
   \end{figure}

\begin{figure}
\centering
                   \subfloat[ {\color{black}ML1 (Geolife)}]{
          \includegraphics[width=2.5 cm]{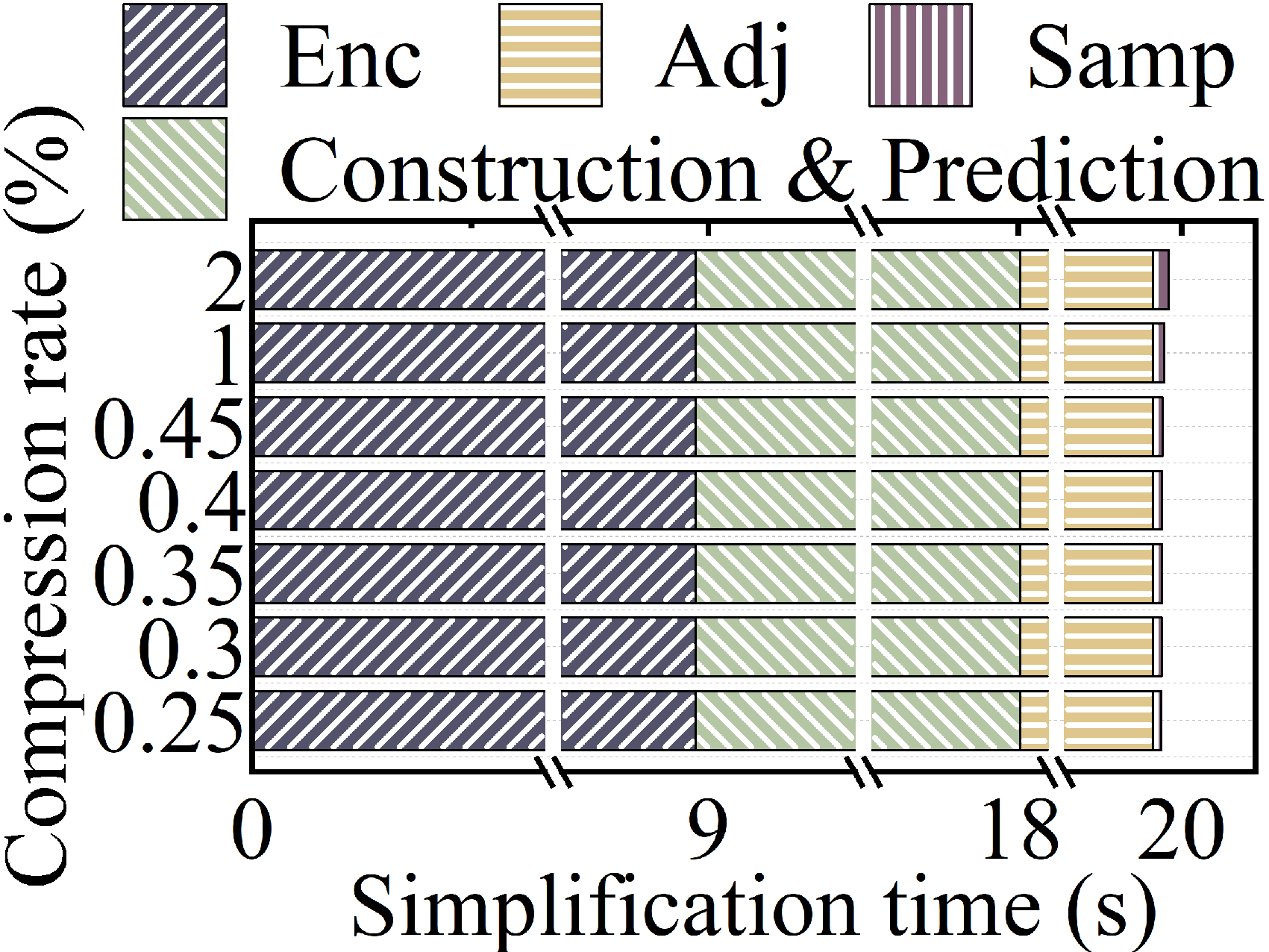}} 
\quad
            \subfloat[ {\color{black}ML1 (T-Drive)}]{
            \includegraphics[width=2.5  cm]{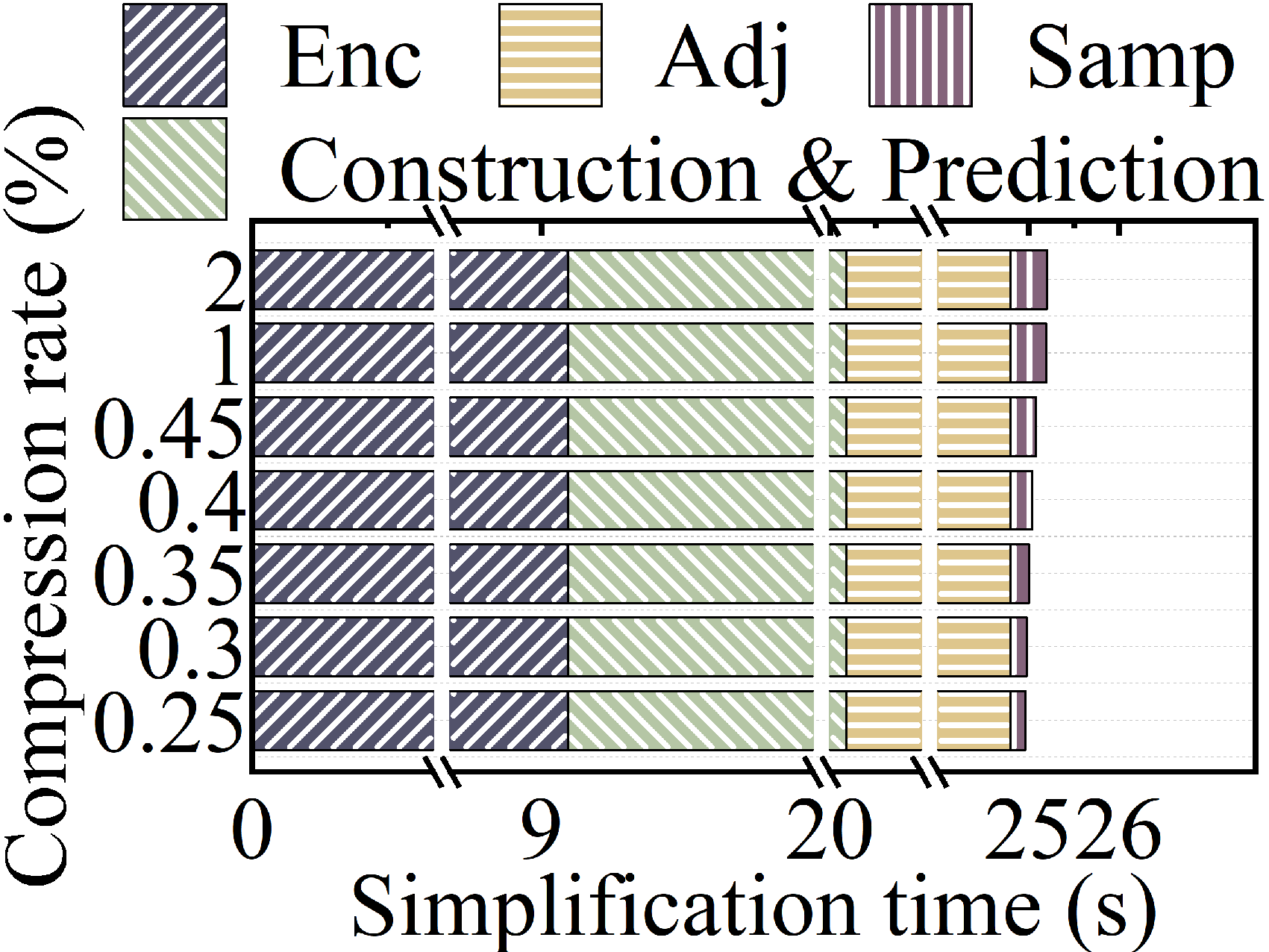}} 
            \quad
            \subfloat[ {\color{black}ML1 (OSM)}]{
            \includegraphics[width=2.5 cm]{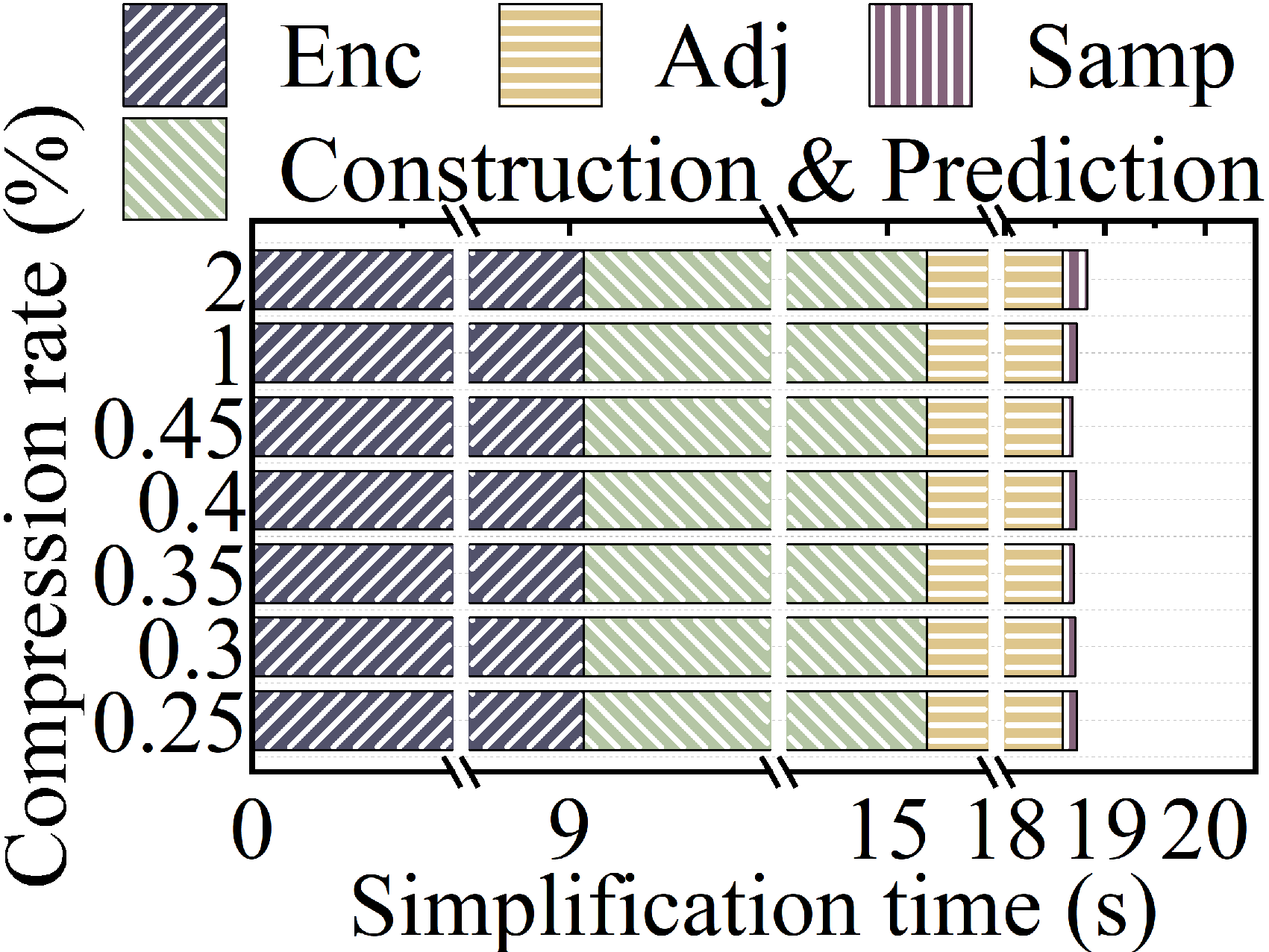}}  
              \vspace{-3mm}
              
                   \subfloat[ {\color{black}ML2 (Geolife)}]{
          \includegraphics[width=2.5 cm]{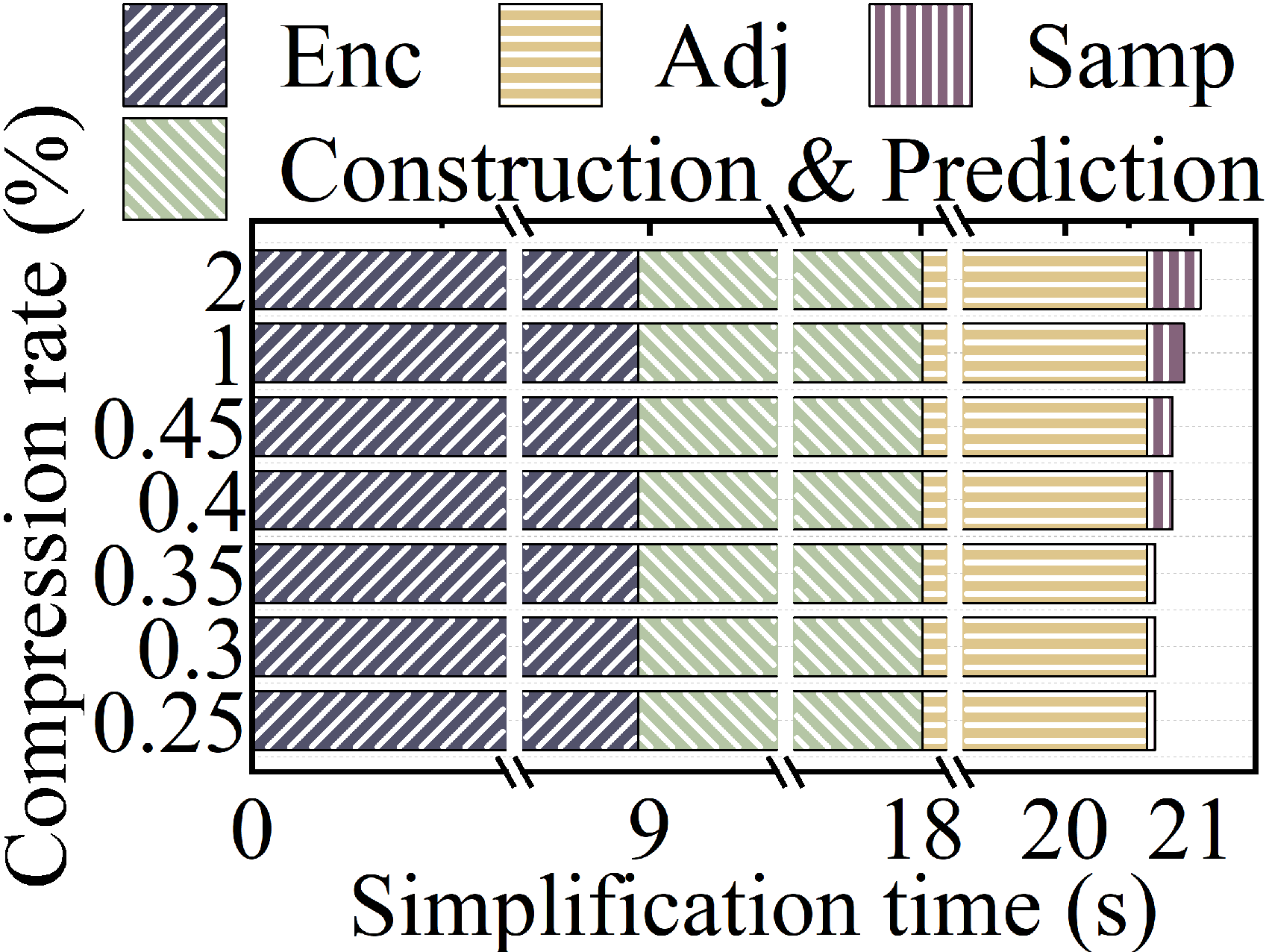}} 
          \quad
            \subfloat[ {\color{black}ML2 (T-Drive)}]{
            \includegraphics[width=2.5  cm]{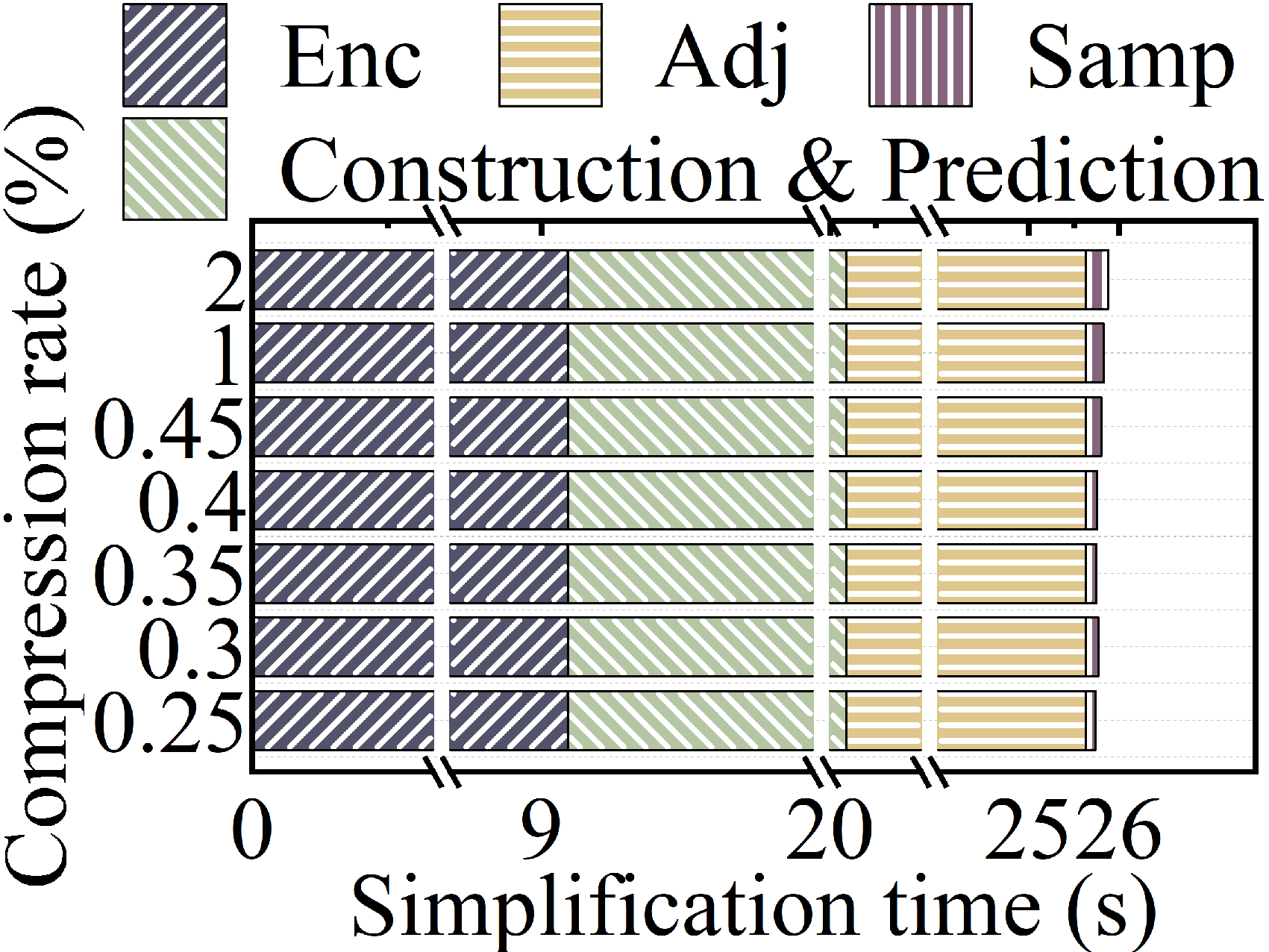}}  
            \quad
            \subfloat[ {\color{black}ML2 (OSM)}]{
            \includegraphics[width=2.5 cm]{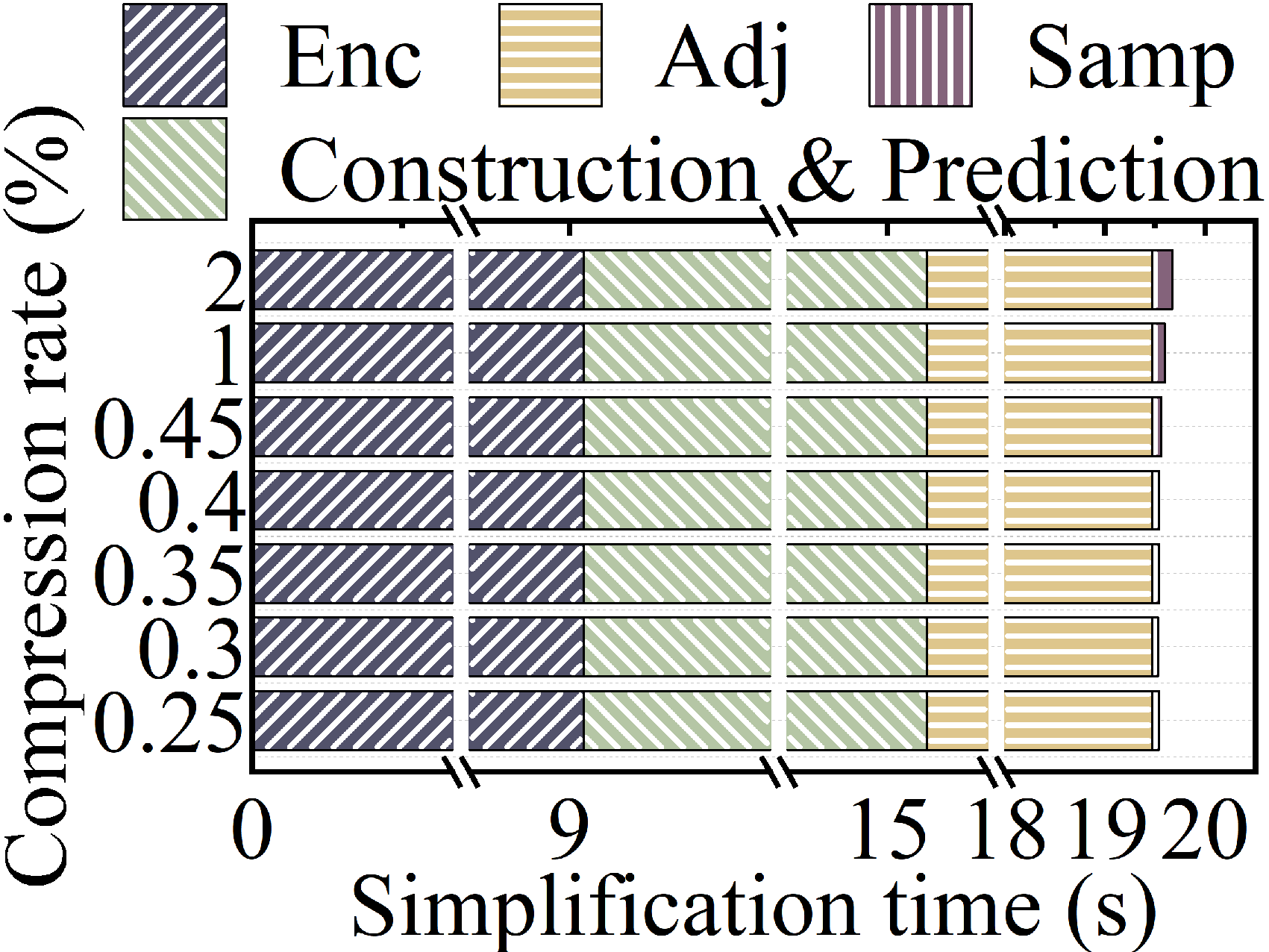}}

 \caption{ \color{black} Overall simplification time of \textsc{MLSimp}, including the time spent on each component.  
 ML1 and ML2 are as Fig.~\ref{efficiency}. 
 % ML1 and ML2 represent  \textsc{MLSimp} simplification for queries generated based on the data distribution and a Gaussian distribution, respectively.
 % Enc, Adj, Samp, Construction, and Prediction represent encoding, adjustment, sampling, graph construction, and importance prediction, respectively.
}
 \vspace{-6mm} 
   \label{component_time}
   \end{figure}

We study the simplification efficiency of \textsc{MLSimp}. Fig.~\ref{efficiency} reports the simplification time required by all methods discussed in Sec.~\ref{sec:5.2}. We report the overall simplification time rather than the time for simplifying individual trajectories. For RL4QDTS and \textsc{MLSimp}, we report the simplification time separately for queries generated based on the data distribution and queries generated based on a Gaussian distribution. Across the three datasets,  \textsc{MLSimp} generates simplified results in less than 60 seconds for all compression rates and datasets. Compared to the fastest Top-Down method, the average compression time is reduced by 42\%--70\%. 
% 44\%--70\%.
% This is because \textsc{MLSimp} quantifies the importance of all points without any iterations, and \textsc{MLSimp} inference uses only the lightweight GNN-TS model, resulting in much lower inference time compared to S3 with its Seq2Seq model. Additionally, unlike all baselines, our simplification time does not vary with the compression rate. This is because, as discussed in Sec.~\ref{sec:4.3.3}, its complexity is independent of the compression rate.

{\color{black}  Fig.~\ref{component_time}  further reports the overall simplification time of \textsc{MLSimp} on Geolife, T-drive, and OSM, including  the time for each component. 
% \textsc{MLSimp} is presented in two variants: \textsc{ML1} for data distribution-based queries and \textsc{ML2} for Gaussian distribution-based queries. 
The components measured include trajectory encoding, trajectory graph construction, importance prediction, importance adjustment, and trajectory sampling. 
% Trajectory encoding, graph construction, and importance prediction are part of GNN-TS. 
% Since the cost of trajectory graph construction is minimal and it is a prerequisite for importance prediction, the time for the graph construction is combined with the importance prediction time cost in Fig.~\ref{component_time}.
In the same dataset, the time cost of trajectory encoding, trajectory graph construction, and importance prediction in GNN-TS is unaffected by varying compression rates and query distributions.  This is because GNN-TS is only used to obtain global information of trajectory points and to predict importance based on this information. The main time cost during \textsc{MLSimp} simplification comes from the inference on the learned models T-Bert and GAT in the GNN-TS, as these models are multi-layered, with each layer involving multiple matrix computations. The second highest time cost is from the importance adjustment, influenced by simulated query results. The subsequent adjustment of trajectory points varies across different query distributions. Finally, the sampling time increases with the compression rate but remains minimal, as \textsc{MLSimp}  samples directly based on adjusted importance without requiring iterative sampling.}

% \subsection{Scalability  Test}

% We study the scalability of \textsc{MLSimp} on OSM. We first vary the trajectory database size from 200 million to 1 billion points, with a fixed compression rate of 0.25\%. The simplification time for all methods is shown in Fig.~\ref{scalability} (a). The  results show that the simplification time of \textsc{MLSimp} is consistently lower than those of other methods, and \textsc{MLSimp} also scales the best. This is because \textsc{MLSimp} does not require iterative calculations, multiple trajectories can be input simultaneously  for importance prediction, and both T-Bert and GNN models support parallel computation.

% We also conduct experiments on OSM with 100 million points with different compression rates, as shown in Fig.~\ref{scalability}  (b). As before, \textsc{MLSimp} exhibits the lowest simplification time and is not affected by the compression rate. Top-Down methods (including Top-Down and RL4QDTS) show an increase in simplification time as the compression rate increases, while  Bottom-Up methods (including Bottom-Up and RLTS) show a decrease in simplification time as the compression rate increases. This is because  the Top-Down methods select points to add to the database, while  the Bottom-Up methods select points to remove.

 \vspace{-2mm}
\subsection{Model Analysis}

   \begin{figure}
\centering
         \subfloat[ {Training time}]{
          \includegraphics[width=2.8 cm]{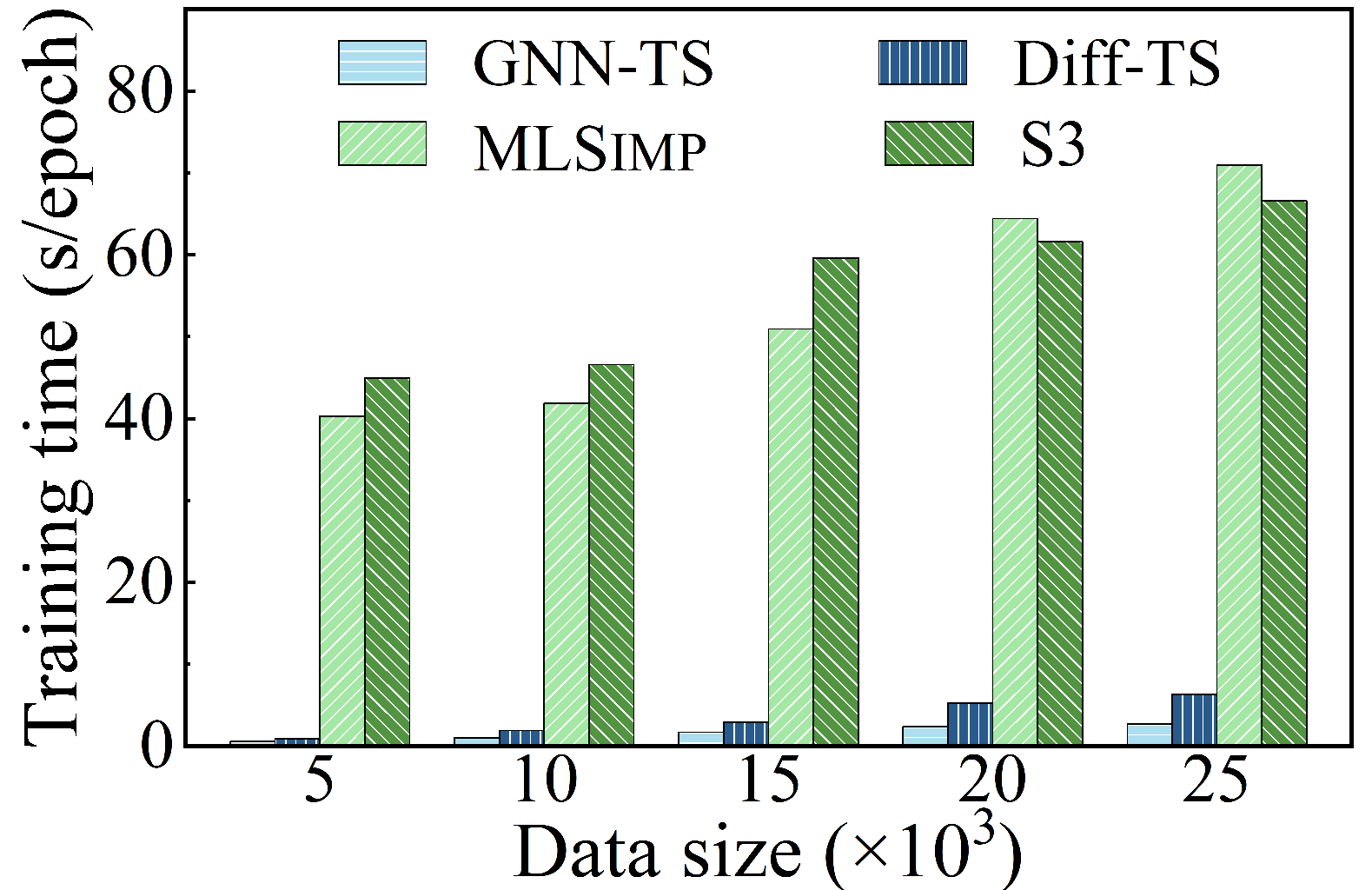}}   
          % \quad
                   \subfloat[ {\color{black}Warm-up epochs}]{
          \includegraphics[width=2.8cm]{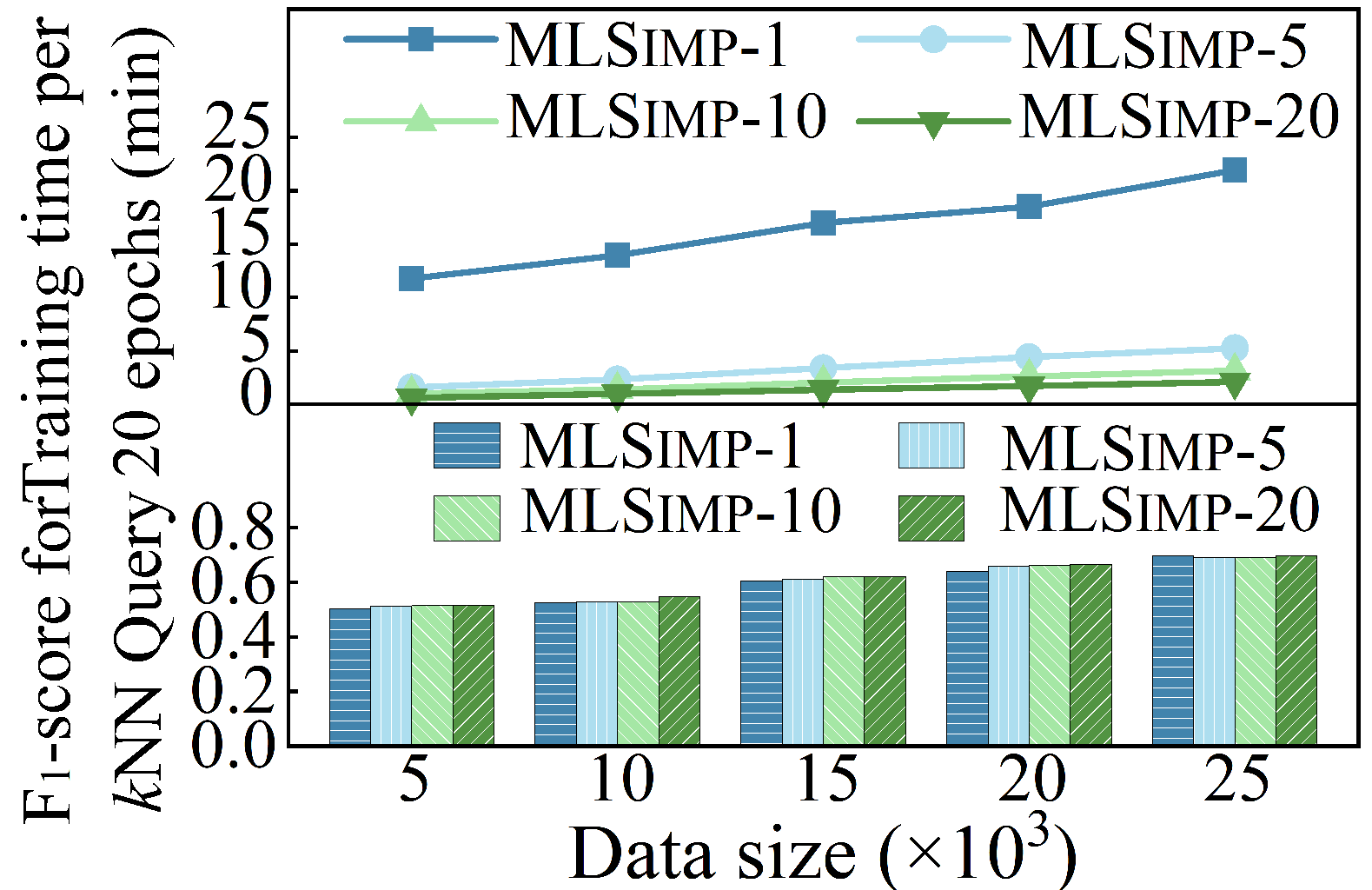}}
         \subfloat[ {Memory budget}]{
          \includegraphics[width=2.8cm]{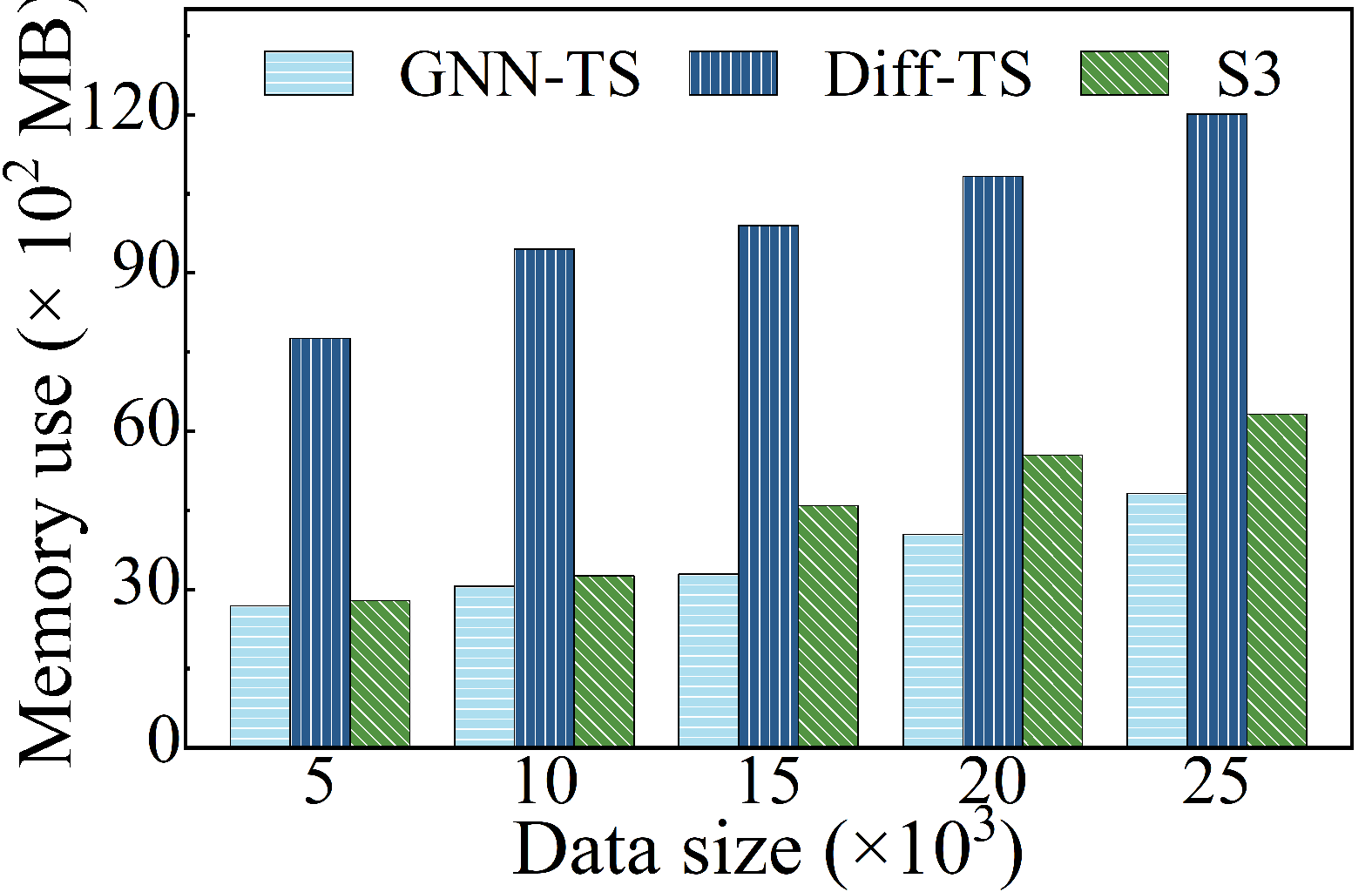}}
           \caption{Model analysis of \textsc{MLSimp}.}

       \vspace{-6mm} 
     \label{model analysis}
   \end{figure}
   
\begin{figure}
\centering

         \subfloat[ {The impact of $\lambda_3$}]{
          \includegraphics[width=2.8cm]{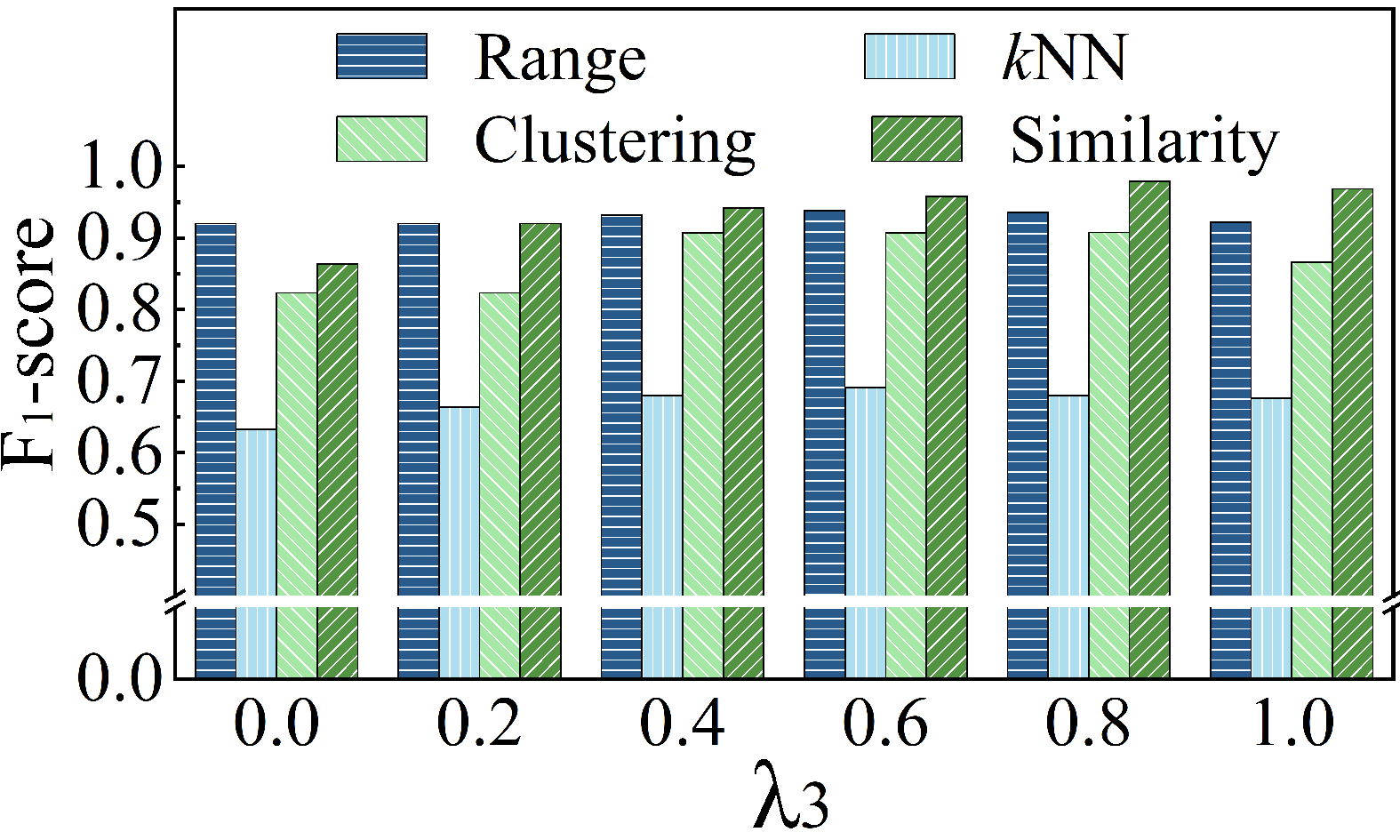}} 
         \subfloat[ {The impact of $\delta$}]{
          \includegraphics[width=2.8cm]{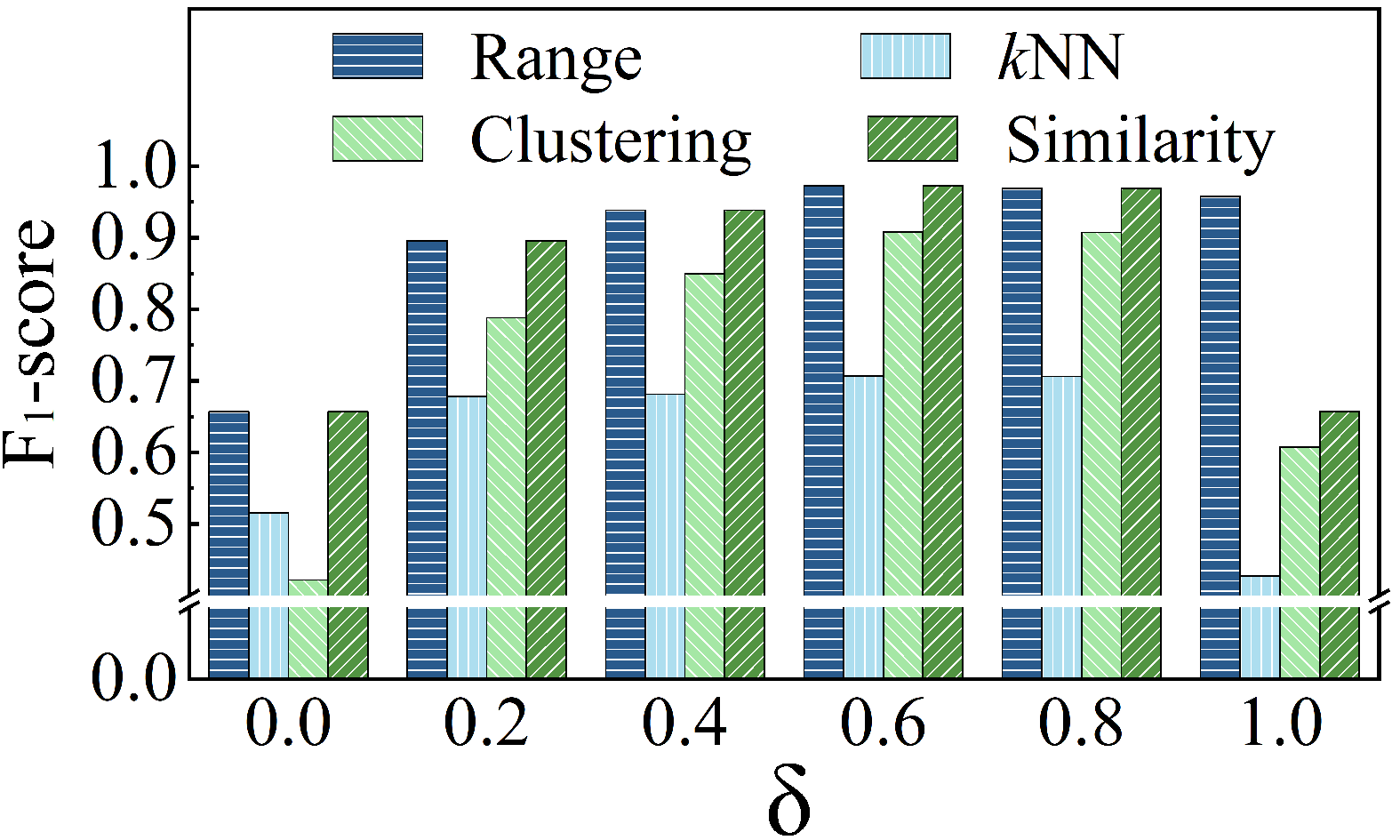}} 
                   \subfloat[ {\color{black}The impact of $|Q^w|$}]{
          \includegraphics[width=2.8cm]{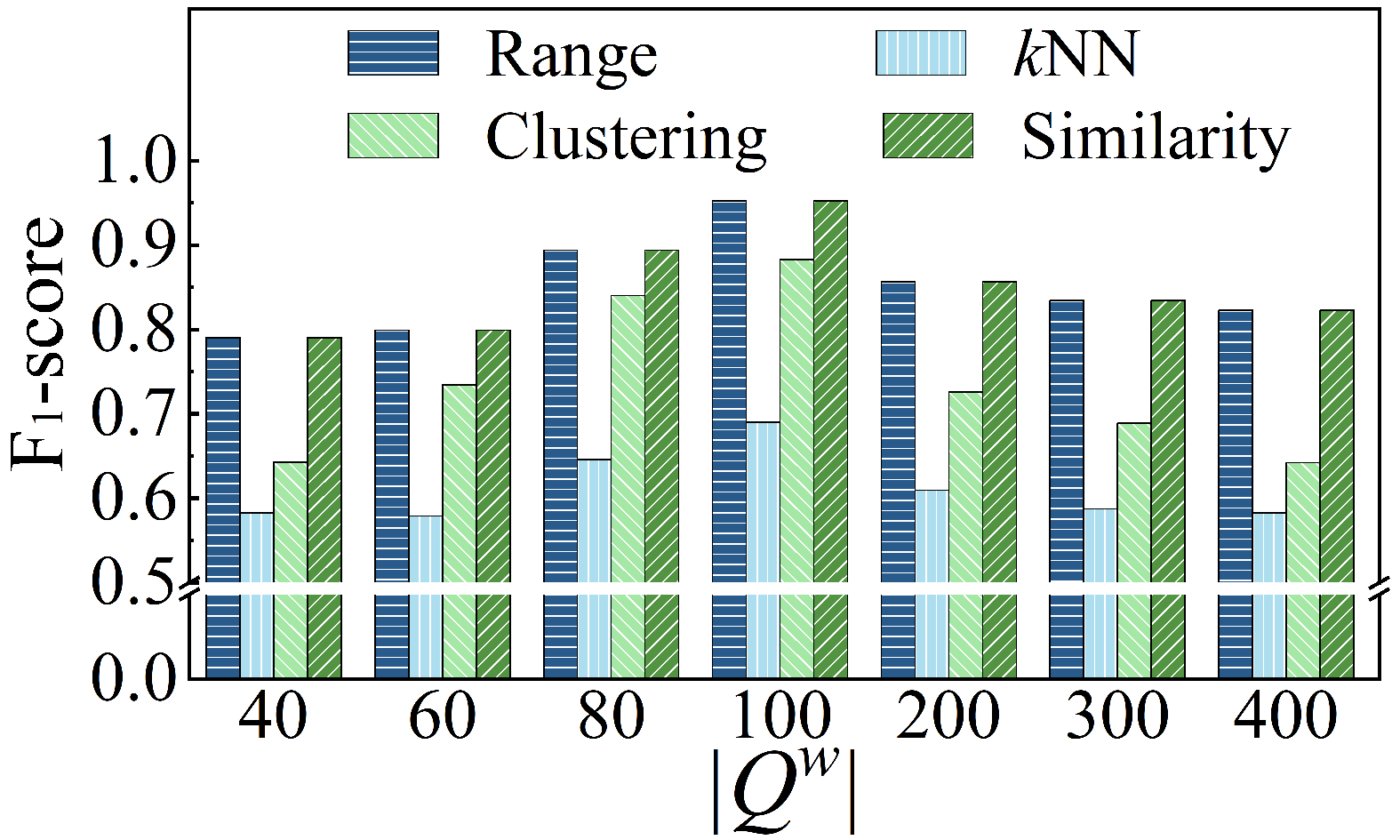}}

           \caption{\color{black}Impact of hyperparameters.}

           \vspace{-6mm}
     \label{hyperparameters}
   \end{figure}

We evaluate the training cost of \textsc{MLSimp} on Geolife. We compare \textsc{MLSimp} with the learning-based lightweight model S3. Since \textsc{MLSimp} iteratively trains GNN-TS and Diff-TS, we measure 
% the training time of each model for one epoch separately, as well as 
the overall training time for \textsc{MLSimp} for one epoch of the mutual learning. 
% We compare with the training time for one epoch of S3. 
The results are shown in Fig.~\ref{model analysis} (a). 
Since the simplification of S3 is based on the Seq2Seq model, which requires iterative encoding of each GPS point, the training time of S3 is substantially higher than those of GNN-TS and Diff-TS. 
% The time for one epoch of \textsc{MLSimp} training not only includes the time for one round of training for both models, but also includes the inference process of GNN-TS and Diff-TS for generating mutual learning information. 
However, the one-epoch training time of \textsc{MLSimp} is higher than that of S3 when the data size is large.  To save training costs, we only perform ML after multiple training epochs of GNN-TS and Diff-TS.
% , making the training time of \textsc{MLSimp}  manageable.

   {\color{black} Fig.~\ref{model analysis} (b)     presents (i) the training times for 20 epochs of GNN-TS and Diff-TS (upper part), and (ii) the query $F_1$-scores on the simplified datasets using models trained with different warm-up epochs before ML adjustments (lower part), where \textsc{MLSimp}-i (with $i \in {1, 5, 10, 15}$) indicates that GNN-TS and Diff-TS update the ML signal to each other after every $i$ warm-up epochs. At $i = 1$, the models update ML signals after each training epoch, eliminating the need for additional warm-up. 
   % Both tests vary the dataset size from 5,000 to 25,000 points.
As the data size increases, 
the training time and growth rate of \textsc{MLSimp}-1 are significantly higher than those of models with 5, 10, and 20 warm-up epochs. 
% \textsc{MLSimp}-20, which performs ML after 20 warm-up epochs, has the lowest training time, 90\% lower than  that of  \textsc{MLSimp}-1. 
This is because both GNN-TS and Diff-TS need to perform an inference to generate ML signals. As Diff-TS is a generative model requiring multiple iterations, each inference takes substantial time. 
% With 20 epochs of training, \textsc{MLSimp}-1 performs inferencing 20 times, while \textsc{MLSimp}-20 requires only one.
Although \textsc{MLSimp}-5, \textsc{MLSimp}-10, and \textsc{MLSimp}-20 reduce the frequency of updating ML information, the query performance of the trained \textsc{MLSimp}-i ($i \in \{1, 5, 10, 15\}$) models is almost identical. The reason may be that, despite the reduced frequency of ML information updates, the training objective remains unchanged, resulting in a similar distribution in the encoding space of the trajectory points in the final GNN-TS.
Therefore, in all experiments in Sec.~\ref{sec:5}, we use \textsc{MLSimp}-20 as the training strategy.}

Next, we evaluate the memory use of GNN-TS, Diff-TS, and S3. Since \textsc{MLSimp} uses \textcolor{black}{GNN-TS} for simplification, the memory use of \textcolor{black}{GNN-TS} is the same as the memory usage of \textsc{MLSimp} during simplification. Fig.~\ref{model analysis} (c) shows the results for \textcolor{black}{GNN-TS}, Diff-TS, and S3 on  Geolife with different data sizes. We observe that the memory use of GNN-TS is slightly lower than that of S3, while the memory use of Diff-TS is significantly higher than that of GNN-TS and S3. This is because Diff-TS contains stacked Transformer layers.

\vspace{-1mm}
\subsection{Hyperparameter Study}

\begin{figure}
\centering

         \subfloat[ {\color{black}Compression rate = 0.25\%}]{
          \includegraphics[width=3.9cm]{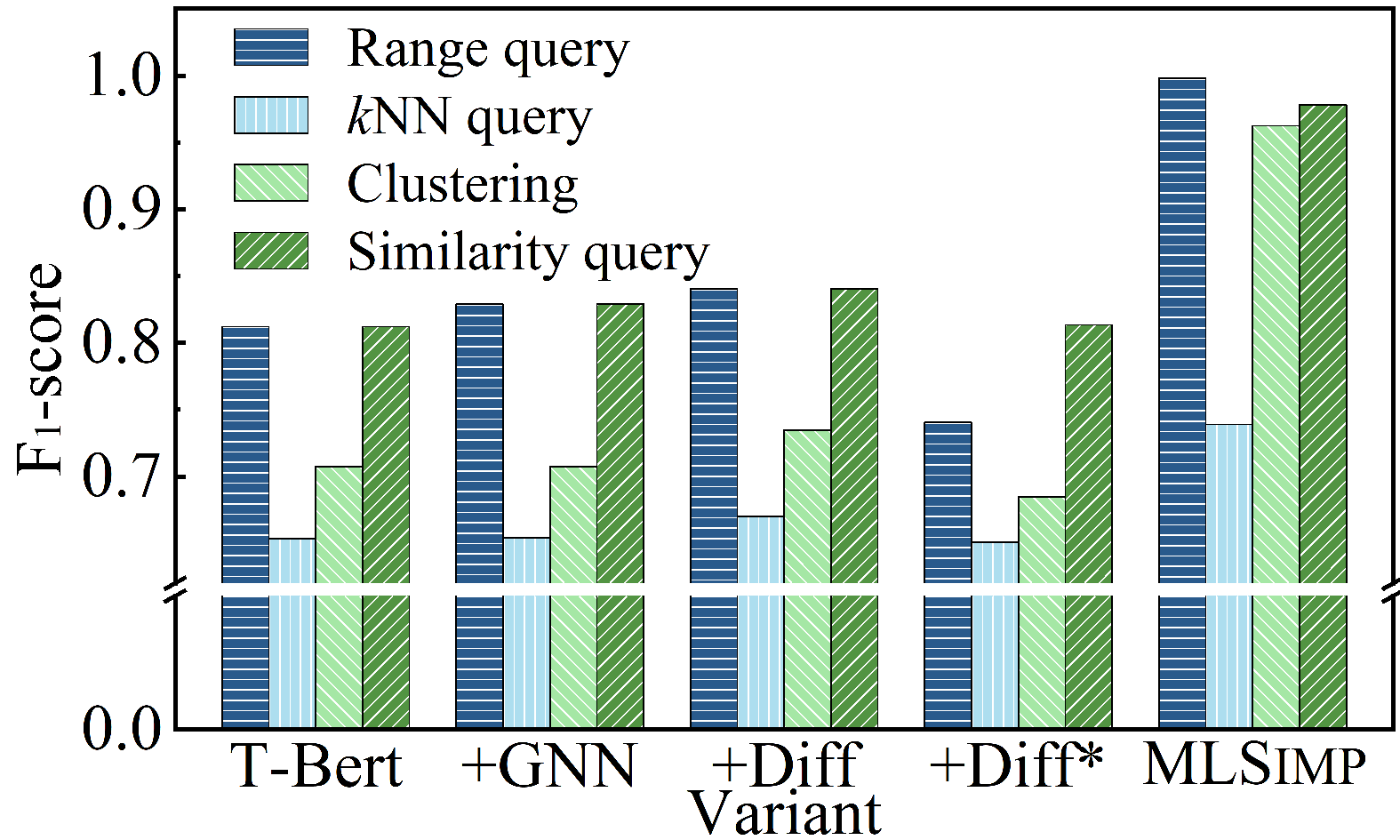}}
          \quad
         \subfloat[ {\color{black}Compression rate = 2\%}]{
          \includegraphics[width=3.9cm]{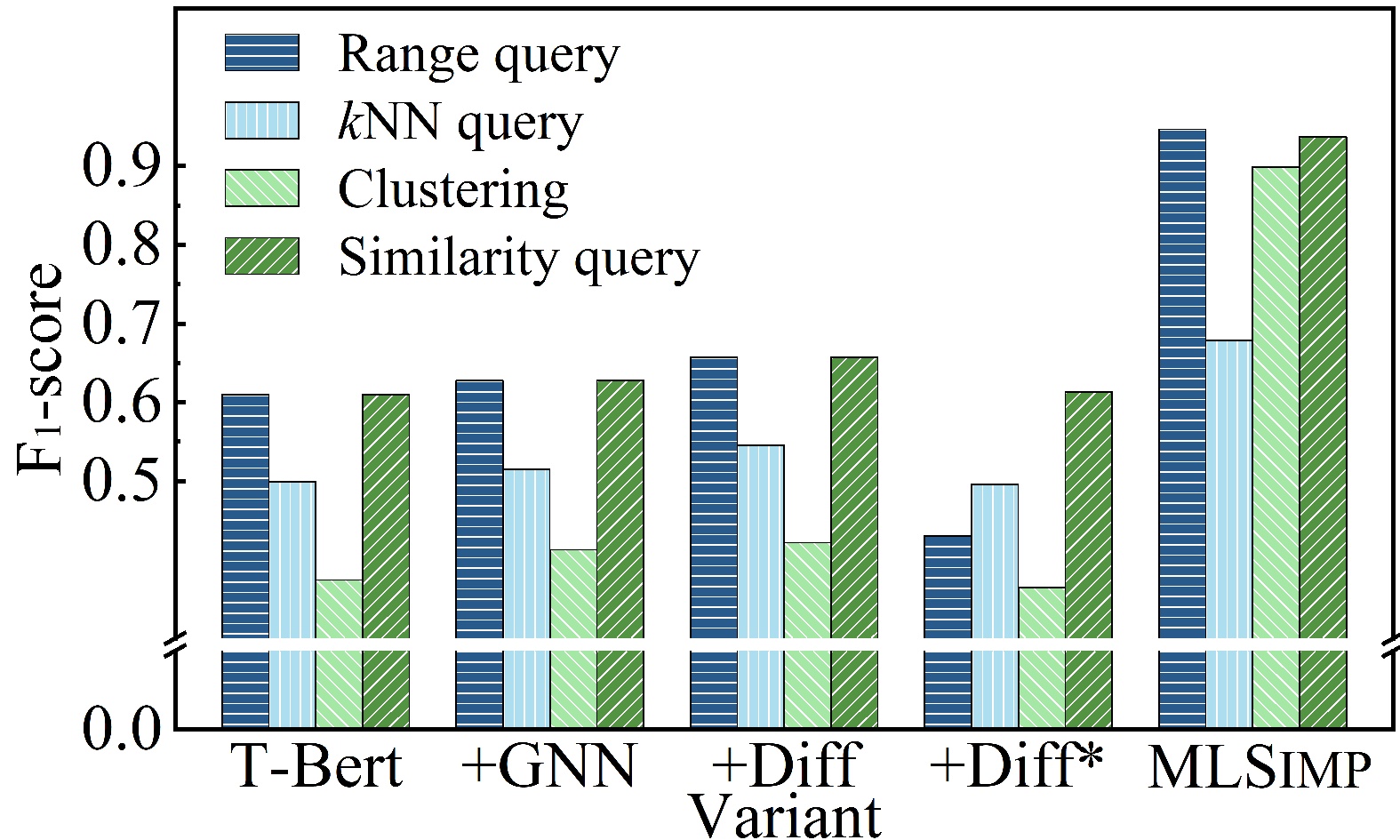}}

           \caption{\color{black}Ablation study results.}

    \vspace{-6mm}
     \label{Ablation Study}
   \end{figure}

Hyperparameter $\lambda_3$ is used to adjust the weight of the amplified label in the mutual training loss of GNN-TS. The $F_1$-scores of queries with different $\lambda_3$  are shown in Fig.~\ref{hyperparameters} (a), where $\lambda_3$ ranges from 0 to 1. \textsc{MLSimp} performs best across different queries when $0.6 \leq \lambda_3 \leq 0.8$. 
% This demonstrates that the amplified label benefits the training of GNN-TS. 
Additionally, the impact of $\lambda_3$ on range queries is less significant {\color{black}than} on the other three query types. This is because range queries have a lower correlation with the distribution of trajectories.
% , while the amplified label only emphasizes points containing more trajectory information.

Hyperparameter $\delta$ is the adjustment weight based on queries. The query performance for different $\delta$  is shown in Fig.~\ref{hyperparameters} (b), where  $\delta$ ranges from 0 to 1. When $\delta = 0$,  no adjustment is made. When $\delta = 1$, the importance of GPS points is only related to the queries. \textsc{MLSimp} achieves the best performance across different queries when $0.6 \leq \delta \leq 0.8$. This is because the query-based adjustment provides query information to \textsc{MLSimp}, allowing it to consider the importance of both the distribution of trajectories and the queries.

{\color{black}Hyperparameter $|Q^w|$ represents the number of queries  used for importance adjustment. The query performance for different values of $|Q^w|$ is shown in Fig.~\ref{hyperparameters} (c), where $|Q^w|$ varies from 40 to 400. When $|Q^w| = 100$, \textsc{MLSimp} achieves the best performance across different queries. 
When $|Q^w|$ exceeds 100, the $F_1$-scores of all queries decrease as the number of queries increases. Conversely, when $|Q^w|$ is below 100, the $F_1$-scores of all queries decrease as the number of queries decreases. This is because when too many queries are used for adjustment, the region involved in the adjustment increases, and the adjustment cannot focus on popular regions, potentially introducing noise. When too few queries are used, the importance adjustment lacks sufficient attention to popular regions, neglecting points that are truly important for the queries.}

%    \begin{figure}
% \centering
%          \subfloat[ {Data size}]{
%           \includegraphics[width=3.7cm]{figures/scalability_a_v2.png}}
%           \quad
%          \subfloat[ {Compression rate}]{
%           \includegraphics[width=3.7cm]{figures/scalability_a.png}}

%            \caption{Scalability test  on OSM.}

%            \vspace{-7mm}
%      \label{scalability}
%    \end{figure}

% However, since \textsc{MLSimp} only needs to retain GNN-TS as the compressor during simplification, it retains the lightweight model while improving the quality of simplified trajectory queries.

\vspace{-1mm}
\subsection{Ablation Study}

We conduct ablation experiments on Geolife and report the $F_1$-score of five variants in Fig.~\ref{Ablation Study}. The five variants are given as follows: (a) ``T-Bert'' represents a variant that only uses a pre-trained T-Bert to generate trajectory embeddings, followed by sampling based on the importance evaluation in Eq.~\ref{importance}; (b) ``+GNN'' is a variant that utilizes a GNN for self-supervised training on top of T-Bert to generate trajectory embeddings; (c) ``+Diff-TS'' is a variant where GNN-TS is trained with Diff-TS for mutual learning; {\color{black} (d) ``+Diff*''  is the variant where GNN-TS is trained using  ML and Diff-TS, but the loss function includes only the ML loss $\mathcal{L}_\textit{ml}$  and excludes the contrastive loss $\mathcal{L}_\textit{con}$ in Eq.~\ref{gnn-ts loss};}  (e) ``\textsc{MLSimp}'' represents the complete model with query-based importance adjustment.

Under both compression rates, GNN-TS, Diff-TS, the mutual learning paradigm, and importance adjustment enhance the performance of queries.
% The impact of importance adjustment on query performance is the most significant, as the query distribution can directly affect query accuracy. 
% Additionally, in Fig.~\ref{Ablation Study} (a), the improvement in performance by Diff-TS is slightly higher than in  Fig.~\ref{Ablation Study} (b), as Diff-TS amplifies signals generated for low compression rates, while the advantage of amplified labels becomes less apparent at higher compression rates.
{\color{black}Additionally,  the performance of the +Diff* variant is consistently lower than the performance of all other variants. The reason is that without the guidance of $\mathcal{L}_\textit{con}$, GNN-TS loses the ability to filter important trajectory points coarsely. 
% Diff-TS is trained using the important trajectory points predicted by GNN-TS as supervision information. 
The amplified signal generated by Diff-TS, influenced by GNN-TS, cannot guarantee the inclusion of important trajectory points.
% This amplified signal, when used as ML information for retraining GNN-TS, causes GNN-TS to deviate from our training objectives. 
% With multiple iterations of ML, the deviation is amplified, leading to reduced performance of GNN-TS.
}

   \begin{figure}
\centering   
         \subfloat[ {\color{black}Global simplification}]{
          \includegraphics[width=3.9cm]{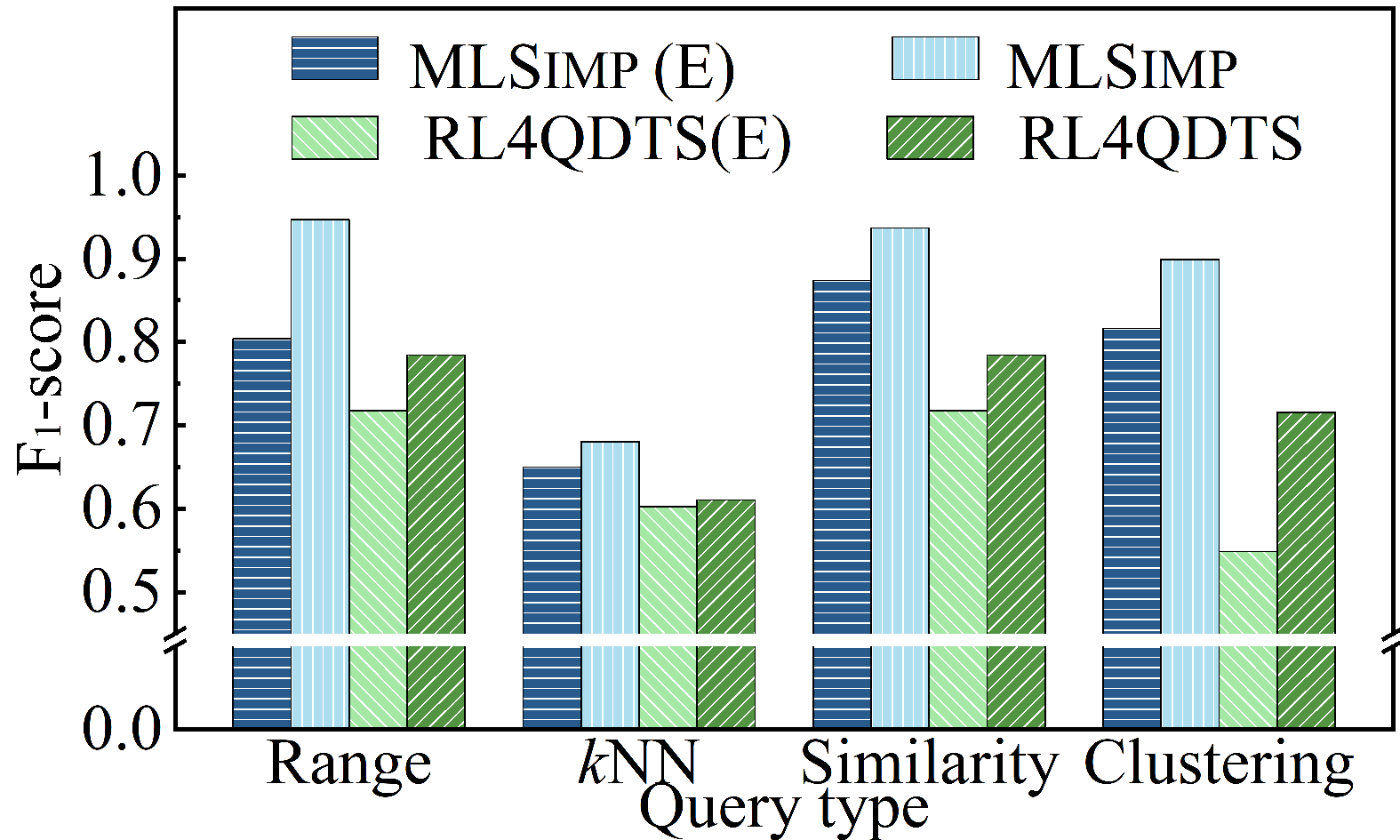}}
          \quad
         \subfloat[ {\color{black}Simulated queries}]{
          \includegraphics[width=3.9cm]{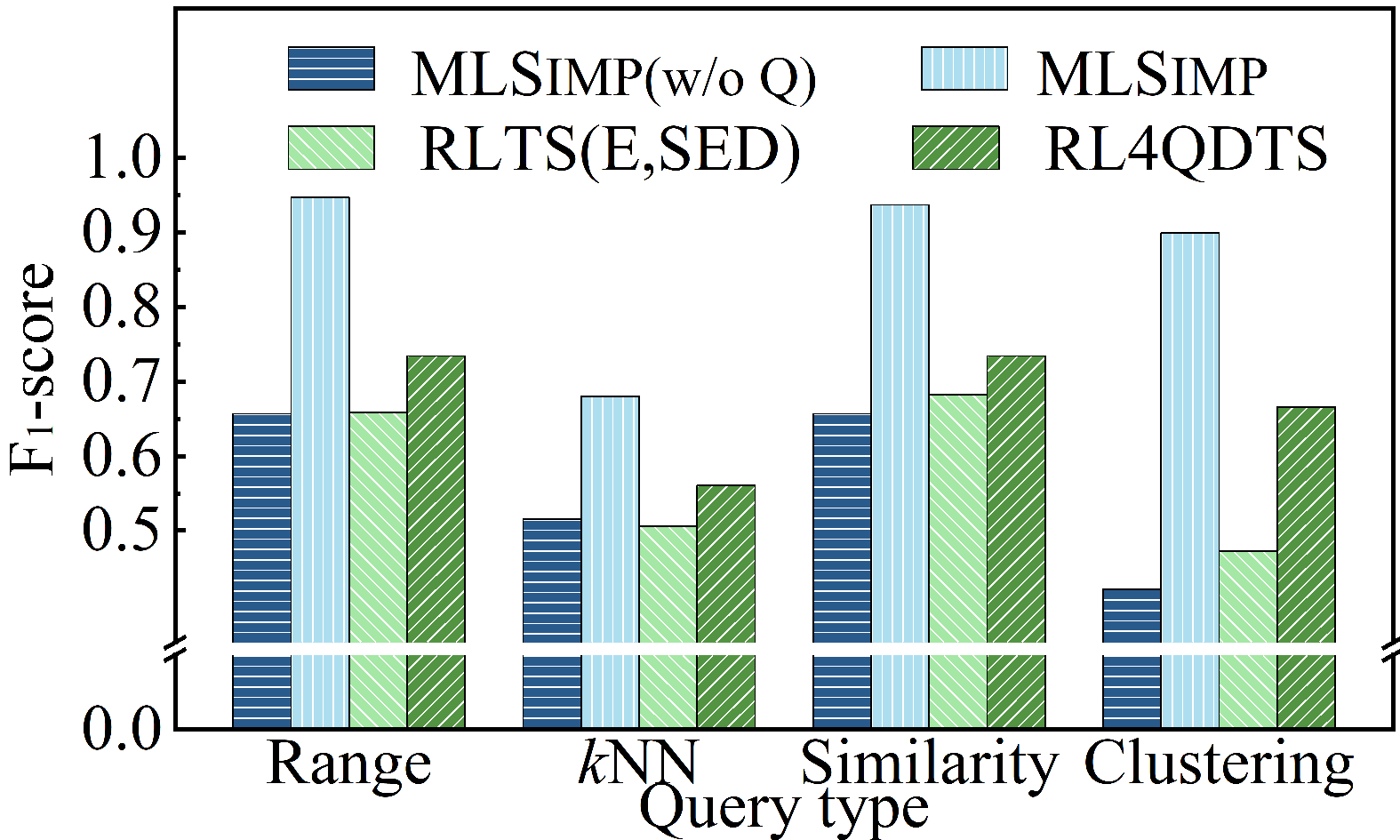}}

           \caption{\color{black}Case study results.}
           \label{fig:case_study}
                      \vspace{-6mm}
   \end{figure}

\vspace{-.7mm}
\subsection{Case Study}

{\color{black}
\noindent\textbf{Impact of global simplification.}  
 Fig.~\ref{fig:case_study} (a) shows  the performance of \textsc{MLSimp} and RL4QDTS on Geolife, comparing global simplification to their variants that simplify each trajectory independently (\textsc{MLSimp}(E) and RL4QDTS(E)). All methods are  tested across four types of queries at a compression rate of 0.25\%.
The global versions of \textsc{MLSimp} and RL4QDTS achieve higher $F_1$-scores than their independent counterparts. This is because global simplification adaptively selects the compression rate for each trajectory under a fixed database compression rate. 
% More points are retained for complex trajectories (higher compression rate) to preserve critical information, while fewer points are retained for simple trajectories (lower compression rate) to avoid redundancy. 
% In contrast, \textsc{MLSimp}(E) and RL4QDTS(E) apply the same compression rate to all trajectories. 
% Some simplified trajectories contain redundant data, while important points are discarded from complex trajectories, leading to an unnecessary loss of information. 
% Thus, global simplification benefits query result performance by being better at retaining important points while removing redundant points.

\vspace{.5mm}
\noindent\textit{\textbf{Discussion.}}  \textsc{MLSimp}(E) outperforms RL4QDTS(E) on similarity-based queries (e.g., $k$NN queries, similarity search, and clustering), even surpassing RL4QDTS. This is because the simplified trajectories of \textsc{MLSimp}(E) retain more global information than those of RL4QDTS(E) at the same trajectory length. \textsc{MLSimp}(E) effectively incorporates global information for each trajectory point during encoding through multiple iterations of information aggregation by GNN-TS, determining the importance of trajectory points based on their distribution. 
% Hence, \textsc{MLSimp}(E) can identify important points considering an entire trajectory, rather than just local information. 
RL4QDTS(E) and RL4QDTS, which only consider local information, retain points based on range queries, focusing on the specific distribution of individual locations and timestamps.
% Although RL4QDTS improves query performance with global simplification, the lack of global information reduces its performance on similarity-based queries.

\vspace{.5mm}
\noindent\textbf{Impact of simulated queries.} 
% To enhance the usefulness of simplified trajectories, both QDTS methods, \textsc{MLSimp} and RL4QDTS,  use simulated queries during simplification. 
% \textsc{MLSimp} uses simulated queries for adjustment (see Sec.~X), while RL4QDTS uses simulated queries to compute state vectors and select trajectory points via two agents. We study the impact of simulated queries on the performance of both QDTS methods. 
Fig.~\ref{fig:case_study} (b) reports the $F_1$-scores of \textsc{MLSimp} and RL4QDTS and their variants without using simulated queries at a compression rate of 0.25\% on four types of queries on Geolife. \textsc{MLSimp}(w/o Q) is a variant of \textsc{MLSimp} that does not use simulated queries for importance adjustment. 
% Since RL4QDTS's reinforcement learning framework inherently uses simulated queries for simplification, 
We select RLTS(E,SED)
% , a reinforcement learning framework similar to RL4QDTS  
as a reinforcement learning competitor that does not use simulated queries. The results of \textsc{MLSimp} and RL4QDTS with simulated queries consistently outperform \textsc{MLSimp}(w/o Q) and RLTS(E,SED) without simulated queries. This is because \textsc{MLSimp}(w/o Q) 
lacks adjustments based on query distribution during database simplification.
% resulting in an inability to adapt to the spatial range and time window of queries, even for similarity-based queries limited to a time window. 
RLTS(E,SED), cannot select trajectory points oriented toward queries. 
% RL4QDTS, using simulated queries to compute states and rewards, makes reinforcement learning selections more query-oriented.
}

\vspace{-2mm}
\subsection{Simplification Errors Evaluation}
\begin{figure}
\centering
         \subfloat[ {\color{black}SED error}]{
         \includegraphics[width=3.5cm]{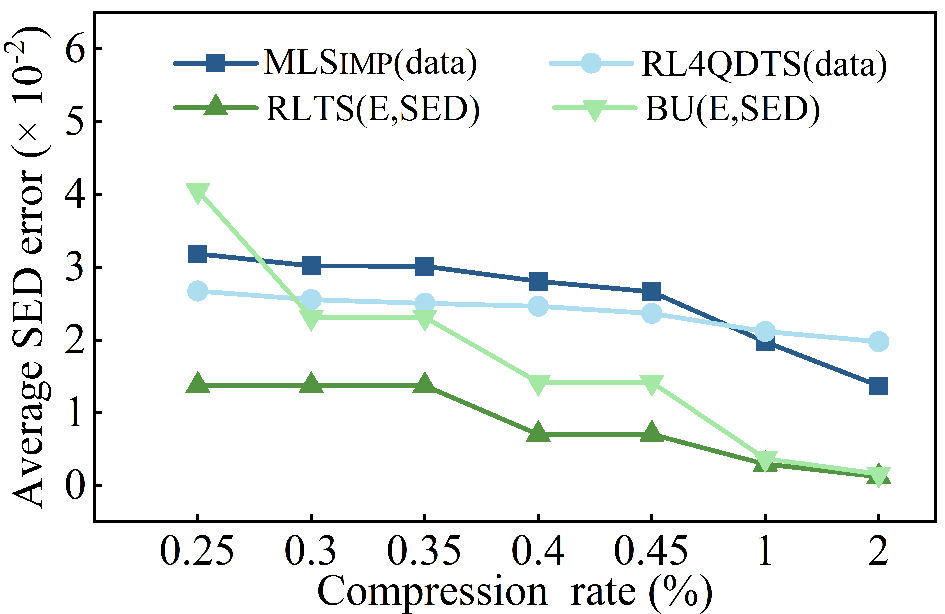}}
          \quad
         \subfloat[ {\color{black}PED error}]{
          \includegraphics[width=3.5cm]{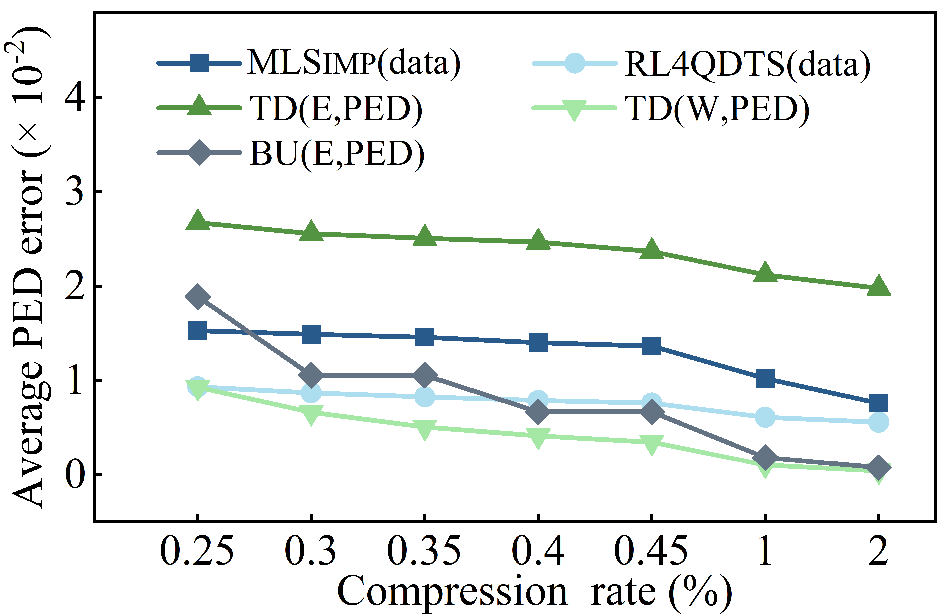}}

           \caption{\color{black}Simplification error evaluation on Geolife.}
           \label{fig:simerror}
 \vspace{-5.5mm}
   \end{figure}

 {\color{black}
 % We evaluate the simplification errors of the query-driven and error-driven methods on Geolife. 
 Fig.~\ref{fig:simerror} reports the average SED and PED on Geolife.  
 Since the distribution of queries is irrelevant to the error evaluation, the query-driven methods, \textsc{MLSimp} and RL4QDTS, report the simplification errors using queries generated based on the data distribution.
For the error-driven methods, we test the corresponding simplification error results according to the error metrics used by the baseline methods.
% For SED comparisons, we report the average error values of RLTS(E,SED) and BU(E,SED). For PED comparisons, we report the average error values of TD(E,PED), TD(W,PED), and BU(E,PED).
For both error metrics, the error-driven methods generally outperform the query-driven methods. RLTS(E,SED) achieves the lowest SED for simplified trajectories, while TD(W,PED) achieves the lowest PED for simplified trajectories. 
% This is because these methods aim to minimize trajectory errors and benefit from selecting the points that result in the lowest error during each iteration.
The simplification errors of the two query-driven methods are acceptable. \textsc{MLSimp}'s simplification error is at most 0.03 higher for SED and at most 0.01 higher for PED compared to the optimal error-driven method. This is because 
% the two query-driven methods aim to optimize query performance and retain the usability of trajectories for queries, such as maintaining the similarity between trajectories. 
the two query-driven methods generate similar simplified trajectories for similar trajectories, but do not fully support minimizing the errors between the simplified and original trajectories.
}

\vspace{-2mm}

\begin{figure}
\centering
          %  \subfloat[ {\color{black}Gaussian ($\mu\in \{0.5, 0.9\}$)}]{
          % \includegraphics[width=3.6cm]{figures/gau_mu_v2.png}}
          % \quad
          %                       \subfloat[ {\color{black}Gaussian ($\sigma\in \{0.25, 0,85\}$)}]{
          % \includegraphics[width=3.6cm]{figures/gau_sigma_v2.png}}
          % \vspace{-3mm}

         \subfloat[ {\color{black}The impact of $\mu$}]{
          \includegraphics[width=3.5cm]{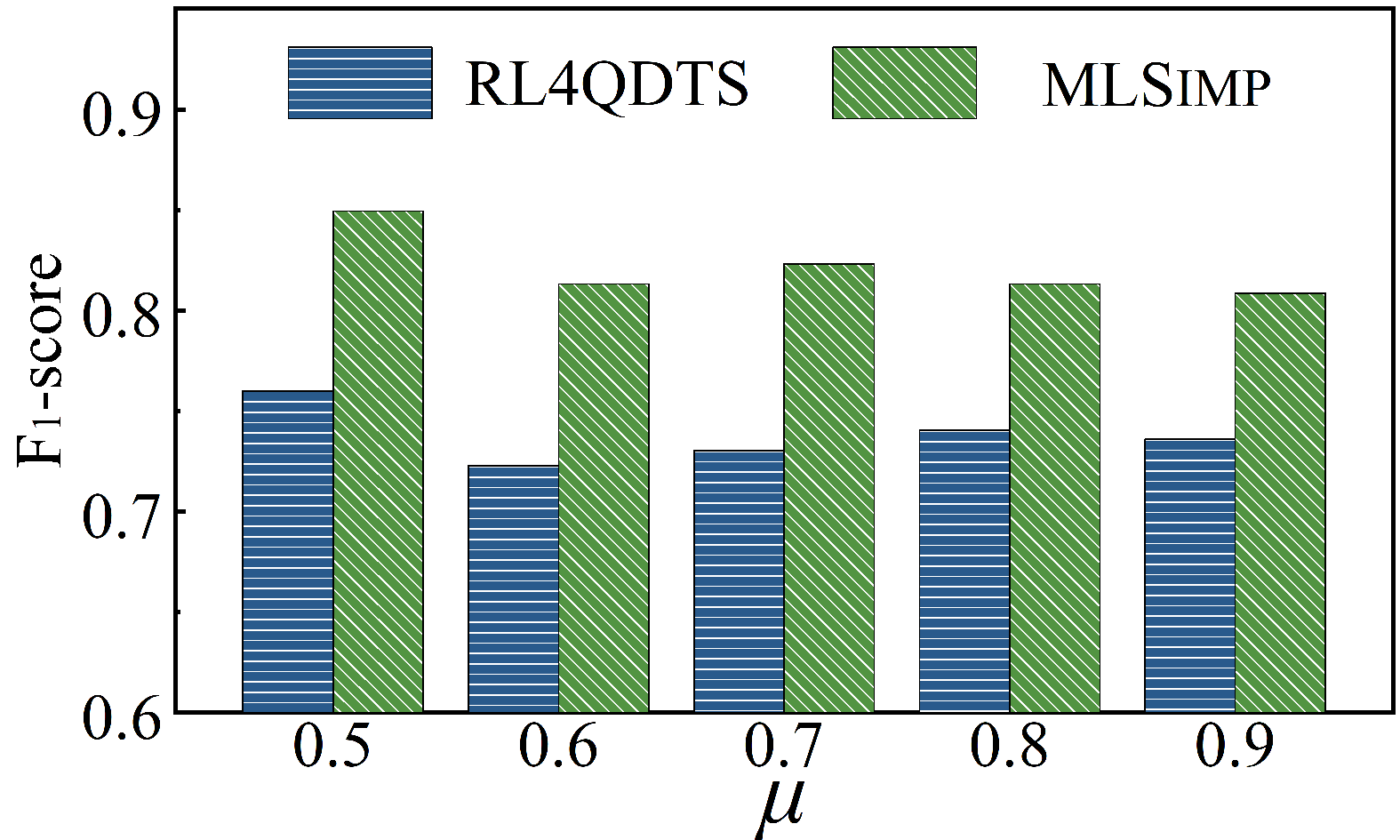}}
             \quad
         \subfloat[ {\color{black}The impact of $\sigma$}]{
          \includegraphics[width=3.5cm]{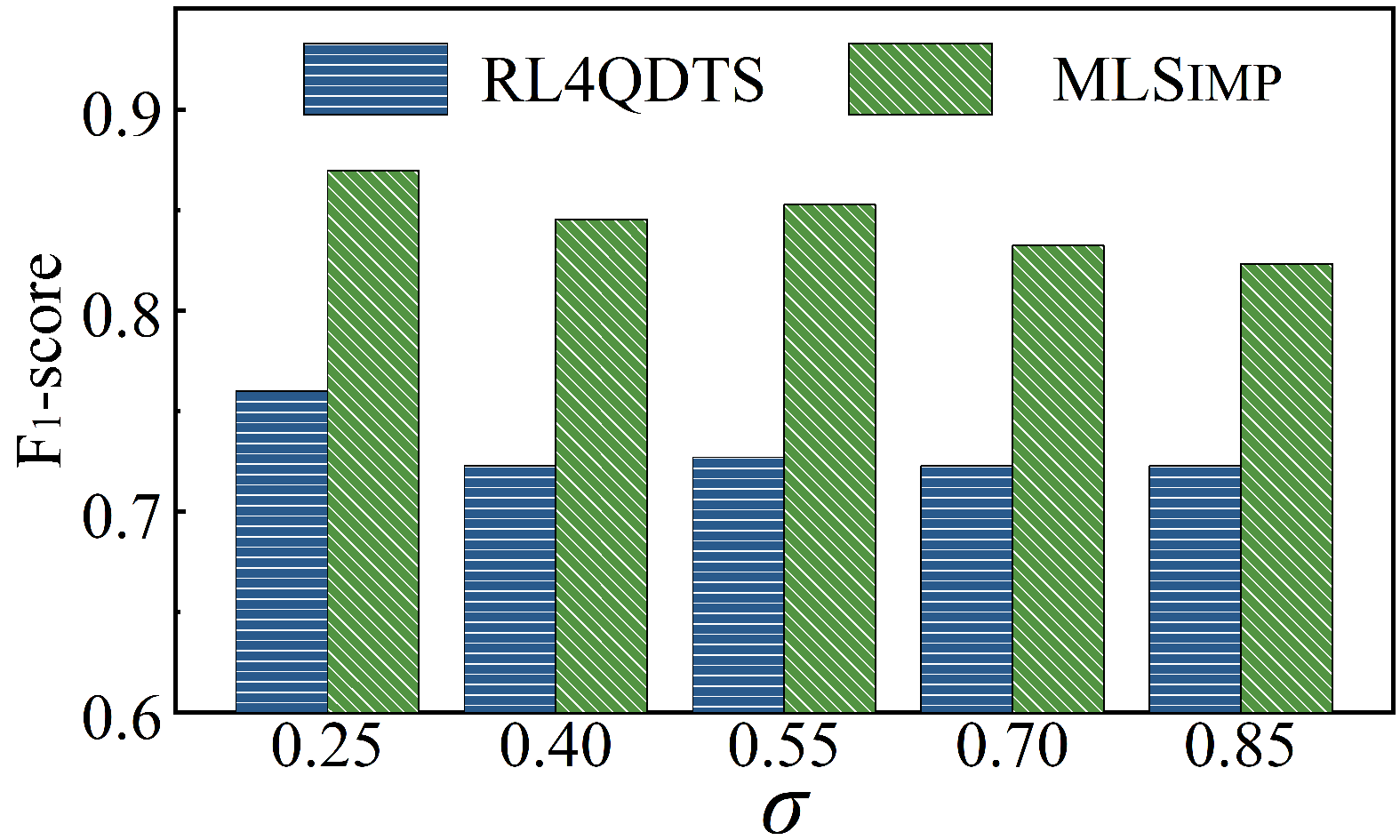}}
 \caption{\color{black}Transferability test results.}
 \label{fig:transfer}
           \vspace{-5.5mm}
   \end{figure}

\vspace{-3mm}
 \subsection{Transferability Test}  
{\color{black}We report the results for range queries with  distributions where $\mu$ varies from 0.5 to 0.9 and $\sigma$ from 0.25 to 0.85 in Figs.~\ref{fig:transfer} (a) and (b), respectively. 
% Prior to this, we show the query distributions for $N(\mu=0.5, \sigma=0.25)$, $N(\mu=0.9, \sigma=0.25)$, and $N(\mu=0.5, \sigma=0.85)$ in Figs.~\ref{fig:transfer} (a) and (b). As can be seen, the test query regions differ notably from the distribution used for simplification, i.e., $N(\mu=0.5,\sigma=0.25)$. 
% Specifically, with $\sigma=0.85$, the queries are no longer concentrated in one region but are distributed more evenly across the space.
They  show that although the performance of both RL4QDTS and \textsc{MLSimp} decline when tested with queries different from those used for simplification, \textsc{MLSimp} consistently outperforms RL4QDTS across different $\mu$ and $\sigma$ values.  This is due to \textsc{MLSimp}'s use of global information to eliminate semantically redundant points. Moreover, during adjustment, the hyperparameter $\delta$ controls the simplification result without relying solely on the generated queries for simplification, enhancing \textsc{MLSimp}'s ability to generalize across different query distributions.}
\vspace{-2mm}
\subsection{Visualization}
\begin{figure}
\centering
                   \subfloat[ {\color{black}Original trajecotries}]{
          \includegraphics[width=3.3 cm]{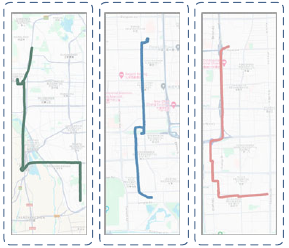}} 
          \quad\quad
         \subfloat[ {\color{black}GNN-TS without ML}]{
            \includegraphics[width=3.3 cm]{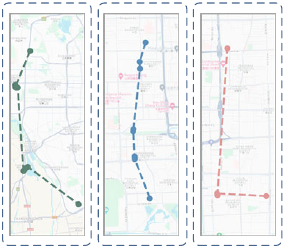}} 
\vspace{-3mm}
         
                   \subfloat[ {\color{black}Amplify signal by Diff-TS}]{
           \includegraphics[width=3.3 cm]{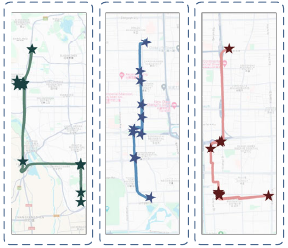}}  
           \quad\quad
                   \subfloat[ {\color{black}GNN-TS with ML}]{
           \includegraphics[width=3.3cm]{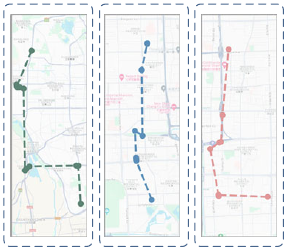}}  
       \caption{\color{black}Simplified trajectory visualization. Circles denote trajectory points, dashed lines connect adjacent points in the simplified trajectory, and stars denote points where the importance is amplified.}
\label{fig:visualization}
\vspace{-7mm}
   \end{figure}

{\color{black}Fig.~\ref{fig:visualization}   visualizes the original trajectories and simplified results (with a 2\% compression rate) of three trajectories from Geolife. Compared to~Fig.~\ref{fig:visualization} (a), the simplified trajectories of GNN-TS without ML (Fig.~\ref{fig:visualization} (b)) have multiple redundant nodes at some turning points and miss some critical turning points. In Fig.~\ref{fig:visualization} (c), Diff-TS identifies several important points in the trajectory during inference that are not retained in Fig.~\ref{fig:visualization} (b). Although the information in amplified signals still contains redundant and non-important points, the GNN-TS with ML incorporates the knowledge extracted by Diff-TS, generating simplified trajectories more similar in shape to the original trajectory, as shown in Fig.~\ref{fig:visualization} (d).  This is because the redundancy and non-important points in the GNN-TS without ML have low uniqueness or globality. Even when injected with the amplified signal, their importance remains lower than that of truly important points. Meanwhile, important points that are unsampled in Fig.~\ref{fig:visualization} (b) gain higher importance due to the amplification signal.}

%% file: src/conclusion.tex
\vspace{-1mm}
\section{Conclusion}
\label{sec:6}
We propose a trajectory simplification method called \textsc{MLSimp}, which retains trajectory information tailored for trajectory querying.
\textsc{MLSimp} combines two models, GNN-TS and Diff-TS, to simplify trajectories by predicting and adjusting the importance of trajectory points. GNN-TS uses a pre-trained Bert and a GNN to predict the importance of trajectory points, while Diff-TS utilizes a diffusion model to generate simplified trajectories. During training, the two models are used for mutual learning to enhance performance. Experiments on three real-world datasets offer concrete evidence of the effectiveness and efficiency of \textsc{MLSimp}.
% ---it is capable of consistently outperforming the state-of-the-art competitors, even at low compression ratios.  
In future research, it is of interest to extend  \textsc{MLSimp} to support online compression of streaming trajectories and road network trajectory compression. 